\documentclass[12pt]{article}
\usepackage{jheppub}
\usepackage{bbm}
\usepackage{epsfig, amsmath, amssymb, dsfont}
\usepackage{hyperref}

\hypersetup{
    colorlinks=true,
    linkcolor=blue,
    citecolor=blue,
    urlcolor=blue
}
\usepackage{graphicx,psfrag}
\usepackage{xcolor}
\usepackage{physics}
\usepackage[utf8]{inputenc}
\newcommand{\nn}{\nonumber}
\newcommand{\be}{\begin{equation}}
\newcommand{\ee}{\end{equation}}
\newcommand{\ben}{\begin{equation}}
\newcommand{\een}{\end{equation}}
\newcommand{\bea}{\begin{eqnarray}}
\newcommand{\eea}{\end{eqnarray}}
\newcommand{\bA}{\begin{array}}
\newcommand{\eA}{\end{array}}
\newcommand{\bc}{\begin{center}}
\newcommand{\ec}{\end{center}}
\newcommand{\al}{\alpha}

\newcommand{\ra}{\rightarrow}
\newcommand{\del}{\partial}

\newcommand{\ie}{{\it i.e.}}
\newcommand{\eg}{{\it e.g.}}

\def\BZ{{\mathbb Z}}

\newcommand{\lan}{\langle}
\newcommand{\ran}{\rangle}

\numberwithin{equation}{section}

\begin{document}


\begin{titlepage}

%

\bc

\hfill 
\\         [15mm]

{\Huge $dS$ extremal surfaces, replicas, boundary Renyi \\ [2mm]
   entropies in $dS/CFT$ and time entanglement} \\
\vspace{12mm}
{\large Kanhu Kishore Nanda,\ \ K.~Narayan,\ \ Somnath Porey,\ \ Gopal Yadav} \\
\vspace{3mm}
{\small \it Chennai Mathematical Institute, \\}
{\small \it H1 SIPCOT IT Park, Siruseri 603103, India.\\}
{\small Email: \ kanhukishore, narayan, sporey, gopalyadav at cmi ac in}\\

\ec
\vspace{30mm}

\begin{abstract}
~We develop further previous work on de Sitter extremal surfaces and
  time entanglement structures in quantum mechanics. In the first
  part, we first discuss explicit quotient geometries. Then we
  construct smooth bulk geometries with replica boundary conditions at
  the future boundary and evaluate boundary Renyi entropies in
  $dS/CFT$. The bulk calculation pertains to the semiclassical de
  Sitter Wavefunction and thus evaluates pseudo-Renyi entropies. In
  3-dimensions, the geometry in quotient variables is Schwarzschild de
  Sitter. The 4-dim $dS$ geometry involves hyperbolic foliations and
  is a complex geometry satisfying a regularity criterion that amounts
  to requiring a smooth Euclidean continuation.  Overall this puts on
  a firmer footing previous Lewkowycz-Maldacena replica arguments
  based on analytic continuation for the extremal surface areas via
  appropriate cosmic branes.\\ In the second part (independent of de
  Sitter), we study various aspects of time entanglement in quantum
  mechanics, in particular the reduced time evolution operator, weak
  values of operators localized to subregions, a transition matrix
  operator with two copies of the time evolution operator,
  autocorrelation functions for operators localized to subregions, and
  finally future-past entangled states and factorization. Based on
  these, we then give some comments on a cosmological transition
  matrix using the de Sitter Wavefunction.
\end{abstract}


\end{titlepage}

{\tiny 
\begin{tableofcontents}
\end{tableofcontents}
}



\section{Introduction}

Various fascinating generalizations of quantum information structures
arise in the context of holography
\cite{Maldacena:1997re,Gubser:1998bc,Witten:1998qj} towards de Sitter
space
\cite{Strominger:2001pn,Witten:2001kn,Maldacena:2002vr,Anninos:2011ui}.
Certain generalizations of the Ryu-Takayanagi formulation
\cite{Ryu:2006bv,Ryu:2006ef,Hubeny:2007xt,Rangamani:2016dms} of $AdS$
holographic entanglement to de Sitter space were studied in
\cite{Narayan:2015vda,Sato:2015tta,Narayan:2015oka,Narayan:2017xca,
  Narayan:2019pjl,Narayan:2020nsc} with more recent reinventions in
\cite{Doi:2022iyj,Narayan:2022afv} (also \cite{Doi:2023zaf,
  Narayan:2023ebn,Narayan:2023zen,Goswami:2024vfl}). These pertain to
RT/HRT extremal surfaces anchored at the $dS$ future boundary $I^+$
and amount to considering the bulk analog of setting up boundary
entanglement entropy in the dual CFT at $I^+$, in part towards
understanding de Sitter entropy \cite{Gibbons:1977mu} (reviewed in
\cite{Spradlin:2001pw}) as some sort of holographic entanglement
entropy. There are no $I^+\ra I^+$ turning points
\cite{Narayan:2015vda} so the surfaces necessarily have timelike
components and the bulk areas are complex-valued.  Conjectured
$dS/CFT$ dualities
\cite{Strominger:2001pn,Witten:2001kn,Maldacena:2002vr,Anninos:2011ui}
suggest Euclidean non-unitary ghost-like CFTs dual to de Sitter space
(note that on general grounds we expect that the gravity dual to an
ordinary Euclidean CFT is Euclidean $AdS$).
Studies in various toy models of ``entanglement entropy'' in
ghost-like theories explicitly reveal complex-valued entropies arising
naturally \cite{Narayan:2016xwq,Jatkar:2017jwz}, the negative norm
contributions here leading to imaginary components. Thus the timelike
components in the bulk areas are necessary to mirror this
complex-valued boundary entanglement entropy.

Recent investigations suggest that the areas of these extremal
surfaces are best interpreted as encoding pseudo-entropy or
time-entanglement \cite{Doi:2022iyj}, \cite{Narayan:2022afv},
entanglement-like structures involving timelike
separations. Pseudo-entropy \cite{Nakata:2020luh} is the entropy based
on the transition matrix $|f\ran\lan i|$ regarded as a generalized
density operator. In some sense this is perhaps the natural object
here since the absence of $I^+\ra I^+$ returns for extremal surfaces
suggests that extra data is required in the interior, somewhat
reminiscent of scattering amplitudes (equivalently the time evolution
operator), and of \cite{Witten:2001kn} viewing de Sitter space as a
collection of past-future amplitudes. This is also suggested by the
$dS/CFT$ dictionary $Z_{CFT}=\Psi_{dS}$\ \cite{Maldacena:2002vr}:
boundary entanglement entropy formulated via $Z_{CFT}$ translates to a
bulk object formulated via the Wavefunction $\Psi_{dS}$ (a single ket,
rather than a density matrix), leading (not surprisingly) to
non-hermitian structures. In \cite{Narayan:2023zen}, the $dS$ extremal
surfaces were recast via analytic continuation from $AdS$, amounting
to a space-time rotation: so they are analogous to $AdS$ RT ones on
constant time slices turned sideways. This then suggested a heuristic
version of the Lewkowycz-Maldacena argument for
the no-boundary $dS$ extremal surfaces for maximal (IR)
subregions. Roughly the boundary replica argument on $Z_{CFT}$
translates now to a bulk replica argument on the Wavefunction
$\Psi_{dS}$ which is essentially pseudo-entropy. The area now is
interpreted as the amplitude for creation of a cosmic brane that
localizes on the part Euclidean, part timelike no-boundary extremal
surface.\\
These have fuelled parallel developments in quantum mechanics and
CFT, independent of $dS$, pertaining to entanglement-like structures
involving timelike separations. Two aspects of this ``time-entanglement''
in simple toy models were described in \cite{Narayan:2022afv}, based on
(1) the time evolution operator and reduced transition amplitudes, 
dovetailing with pseudo-entropy \cite{Nakata:2020luh} (and some
discussions in \cite{Doi:2022iyj}), 
and (2) two copies of future-past entangled states and entirely
positive structures. In \cite{Narayan:2023ebn}, this was developed
further, with the time evolution operator regarded as a generalized
density operator. Partial traces lead to a reduced time evolution
operator for subregions and the corresponding complex-valued von
Neumann entropy, with sharp parallels with finite temperature
entanglement structures at imaginary temperature.
Entanglement structures for the time evolution operator along with a
projection operator onto some initial state towards isolating
components thereof ends up amounting to pseudo-entropy for this state
and its time-evolved final state.\
(Other partially related work that we found useful here include
\cite{Mollabashi:2020yie, Mollabashi:2021xsd, Hikida:2022ltr, Hikida:2021ese, Guo:2022jzs, Liu:2022ugc, Li:2022tsv, Cotler:2023xku, Jiang:2023ffu, Chen:2023gnh, Jiang:2023loq, Chu:2023zah, Chen:2023sry, Chen:2023eic, He:2023ubi, Omidi:2023env, Diaz:2023npx, He:2023syy, Grieninger:2023knz, Das:2023yyl, Guo:2024lrr, Basu:2024bal, Carignano:2024jxb, Fareghbal:2024lqa, Heller:2024whi, Bou-Comas:2024pxf, Xu:2024yvf, Caputa:2024gve, Afrasiar:2024ldn, Milekhin:2025ycm}.)

This paper contains explorations divided into two parts broadly.\\
In the first part, we put on a firmer footing the heuristic
analytic-continuation-based arguments in \cite{Narayan:2023zen} of the
Lewkowycz-Maldacena replica formulation (reviewed in
sec.~\ref{sec:dSnbRev}). We first discuss $n$-quotient geometries as a
tool to evaluate the extremal surface areas
(sec.~\ref{sec:singQuoGeom}, with more detail in $dS_3, dS_4$,
sections \ref{rcdS3}, \ref{rcdS4}, as well as future-past extremal
surfaces in entirely Lorentzian $dS$\ (sec.~\ref{sec:fpRep}), real
Euclidean extremal surfaces in the Euclidean sphere giving $dS$
entropy (sec.~\ref{sec:dS-sphere}), and no-boundary surfaces in
slow-roll inflation (sec.~\ref{sec:dSsr-rep})).  Then we construct
smooth bulk geometries (sec.~\ref{sec:dSreplica}) with replica
boundary conditions at the future boundary (reflecting 
$I^+$-anchored extremal surfaces), and evaluate
boundary Renyi entropies in $dS/CFT$. The bulk calculation pertains
to the semiclassical $dS$ Wavefunction and thus evaluates
pseudo-Renyi entropies.  Technically this uses the bulk action
alongwith the Gibbons-Hawking boundary term as well as appropriate
counterterms. The $dS_3$ case (sec.~\ref{sec:dS3replica}) is most
easily expressed in the static coordinatization which enable a simple
visualization of the way the replica copies are glued together for
these maximal subregions. This finally results in the boundary Renyi
entropies being pure imaginary, consistent with the imaginary central
charge in $dS_3/CFT_2$ \cite{Maldacena:2002vr}: the $n=1$ limit
recovers the boundary entanglement entropy matching the extremal
surface area. In terms of boundary quotiented variables, the cosmic
brane can be identified with the bulk conical singularity in
corresponding Schwarzschild $dS_3$ geometries with mass related to
$n$. This $dS_3$ analysis has close parallels with the $AdS_3$ replica
discussion in \cite{Lewkowycz:2013nqa}.  The 4-dim $dS$ geometry
(sec.~\ref{sec:dS4replica}) involves hyperbolic foliations, with
close parallels with the $AdS$ hyperbolic black holes in
\cite{Hung:2011nu}. Here this is a complex geometry which satisfies a
regularity criterion that amounts to requiring a smooth Euclidean
continuation. The embedding of the hyperbolic foliations into global
$dS_4$ is nontrivial and illustrates how the $n=1$ limit recovers the
known complex-valued no-boundary extremal surface area. This analysis
can be also understood via analytic continuation from a $-AdS$
framework, and is valid for de Sitter in any dimension
(sec.~\ref{sec:dSd+1replica}).  Overall this firms up the arguments
in \cite{Narayan:2023zen} for the extremal surface areas via
appropriate cosmic branes.

In the second part, we develop further various aspects of
time-entanglement structures in quantum mechanics, building on
\cite{Narayan:2022afv}, \cite{Narayan:2023ebn},
\cite{Narayan:2023zen}\ (reviewed in sec.~\ref{sec:tE-rev}). We relate
the reduced time evolution operator to weak values of operators
localized to subregions (sec.~\ref{sec:weak-rte}): this also relates
singularities in the corresponding pseudo-entropies to those in these
weak values. These quantities involve one copy of the time evolution
operator: we then introduce (sec.~\ref{mathdsT}) a related transition
matrix operator ${\mathds T}$ containing time-evolved versions of two
distinct states, which thus contains two copies of the time evolution
operator.  We then show how autocorrelation functions (especially for
operators localized to subregions) can be naturally recast via
${\mathds T}$ and partial traces thereof: this connects with some of
the discussion in \cite{Milekhin:2025ycm}. In sec.~\ref{sec:fp-fac},
we consider future-past entangled states and recast correlation
functions for operators localized to subregions in terms of these, and
comment on factorization aspects of these future-past states in
relation to that of correlation functions for long time
separations.

This suggests a synthesis of these quantum mechanical time
entanglement aspects with de Sitter space. In
sec.~\ref{sec:Synth-dSctm}, we consider a ``cosmological transition
matrix'' involving two copies of the de Sitter Wavefunction, at the
past and the future. We discuss how de Sitter entropy and the
future-past surface area arise from special cases here. We also
comment on some future work on related aspects in two copies of the
dual ghost-like CFT.

Lastly, we comment on certain maximal area aspects of these
no-boundary extremal surfaces containing timelike components in
sec.~\ref{sec:maximal}, and make some speculations on the sign of the
imaginary parts in these areas and their interpretations via cosmic
branes in sec.~\ref{impnba}.\ 
Sec.\ref{sec:Disc} contains a Discussion of various
aspects. The appendices provide various calculational details (in
particular App.~\ref{App:singQuoGeom} on quotient geometries,
App.~\ref{sec:-AdS} on $-AdS$,\ App.~\ref{adsds} on the $dS$ action
and App.~\ref{appnc} on examples with the ${\mathds T}$ operator).

\section{Review: no-boundary $dS$ extremal surfaces}\label{sec:dSnbRev}

We briefly review \cite{Narayan:2022afv,Narayan:2023zen} and previous
work here on generalizations of RT/HRT extremal surfaces to de Sitter
space (see also \cite{Doi:2022iyj}). These involve considering the
bulk analog of setting up entanglement entropy in the dual Euclidean
$CFT$ on the future boundary, restricting to some boundary Euclidean
time slice, defining subregions on these, and looking for extremal
surfaces anchored at $I^+$ dipping into the time (holographic)
direction: analysing this shows that there are no spacelike surfaces
connecting points on $I^+$ \cite{Narayan:2015vda}. In entirely Lorentzian
$dS$, there are future-past timelike surfaces stretching between
$I^\pm$ \cite{Narayan:2017xca,Narayan:2020nsc}, akin to rotated
analogs of the Hartman-Maldacena surfaces \cite{Hartman:2013qma} in
the eternal $AdS$ black hole \cite{Maldacena:2001kr}: these have pure
imaginary area, relative to $AdS$ spacelike RT/HRT surfaces.  With a
no-boundary type Hartle-Hawking boundary condition, the top half of
these timelike surfaces joins with a spacelike part on the hemisphere
giving a complex-valued area \cite{Doi:2022iyj},
\cite{Narayan:2022afv}\ (and \cite{Hikida:2022ltr,Hikida:2021ese} for
$dS_3/CFT_2$). The real part of the area arises from the hemisphere
and is precisely half de Sitter entropy. Due to the presence of
timelike components in these extremal surfaces, these areas are best
interpreted as pseudo-entropy, as we see below. From the dual side, in
various toy models of ``entanglement entropy'' in ghost-like theories
and ``ghost-spin'' lattice models, complex-valued entropies arise
naturally \cite{Narayan:2016xwq,Jatkar:2017jwz}: generic states, their
reduced density matrices via partial traces and the corresponding von
Neumann entropies show that the negative norm states here lead to
imaginary components of the boundary entanglement entropies (which are
also strictly speaking pseudo-entropies since the adjoints of states
are nontrivial). Thus the timelike components in the bulk areas are
necessary to mirror this complex-valued boundary entanglement entropy.
\begin{figure}[h] 
\hspace{0.25pc}
\includegraphics[width=6.7pc]{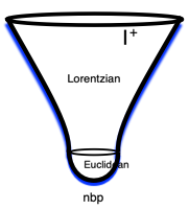}
\hspace{1pc}
\begin{minipage}[b]{29pc}
\caption{{ \label{fig1}
\footnotesize{
No-boundary de Sitter space, with the top Lorentzian region
continuing smoothly into the Euclidean hemisphere region
ending at the no-boundary point. Also shown are IR no-boundary
extremal surfaces (blue) anchored at the future boundary $I^+$
dipping into the time direction, timelike in the Lorentzian region
and going around the hemisphere. }}}
\end{minipage}
\end{figure}

Overall, directly analysing the bulk extremization and calculating the
no-boundary extremal surface areas for the IR (maximal) subregions at
the future boundary gives
\be\label{nbdS43rev}
(dS_4)\quad 
S_{nb} = -i\,{\pi l^2\over 2G_4} {R_c\over l} + {\pi l^2\over 2G_4}\,;
\qquad\quad
(dS_3)\quad 
S_{nb} = -i{l\over 2G_3}\log {2R_c\over l} 
+ {\pi\,l\over 4G_3}\,,
\ee
with the pure imaginary piece from the top Lorentzian part of $dS$ and
the real piece (precisely half de Sitter entropy) from the Euclidean
hemisphere. $R_c$ is the cutoff near $I^+$.

These can also be realized via analytic continuations from the $AdS$
RT surfaces which all lie on a constant time slice.
Under the\ $dS\leftrightarrow AdS$\ analytic continuation, this
$AdS$ constant time slice continues to $dS$ in the Lorentzian region
$r>l$ as
\be\label{nbdSstattoAdS2}
ds^2_{(r>L)} = {dr^2\over 1+{r^2\over L^2}} + r^2 d\Omega_{d-1}^2
\ \ \xrightarrow{\ L^2\ra -l^2\ } \ \
ds^2_{(r>l)} = -{dr^2\over {r^2\over l^2}-1} + r^2 d\Omega_{d-1}^2 \,.
\ee
The $AdS$ boundary at $r\ra\infty$ maps to the $dS$ future boundary
$I^+$ at $r\ra\infty$, and the $AdS$ region $r\in [L,\infty]$ maps
to the $dS$ future universe $F$ parametrized by $r\in [l,\infty]$\
(and $r$ is time here).
The $dS$ hemisphere $r<l$ is $\tau_E=-it=[0,{\pi\over 2}]$ where\
$-{dr^2\over {r^2\over l^2}-1}>0$\ is Euclidean.

The IR $dS$ surface starts at the boundary of the maximal subregion
of the $S^{d-1}$ (\ie\ hemisphere) so it is anchored on the equator
of the $S^{d-1}$ (blue shadow in Figure~\ref{fig1}) and wraps the
equatorial $S^{d-2}$. From (\ref{nbdSstattoAdS2}), it is clear that
the IR $dS$ extremal surface becomes a space-time rotation of that
in $AdS$.\ Its area continues as
\bea\label{IRsurfAdSdS}
{V_{S^{d-2}}\over 4G_{d+1}} \int_0^{R_c} {r^{d-2} dr\over \sqrt{1+{r^2\over L^2}}}
\ & \xrightarrow{\, L\ra -il\,} &\
{V_{S^{d-2}}\over 4G_{d+1}} \int_0^l {r^{d-2} dr\over \sqrt{1-{r^2\over l^2}}}
+ {V_{S^{d-2}}\over 4G_{d+1}}
\int_l^{R_c} {r^{d-2} \sqrt{dr^2\over -({r^2\over l^2}-1)}} \ , \nn\\
&& =\ {1\over 2} {l^{d-1}V_{S^{d-1}}\over 4G_{d+1}} - i\# {l^{d-1}\over 4G_{d+1}}
{R_c^{d-2}\over l^{d-2}} +\, \ldots
\eea
where the $\ldots$ are subleading imaginary terms.
For $AdS_4\ra dS_4$ and $AdS_3\ra dS_3$, we obtain
\be\label{IRsurfAdSdS4}
{V_{S^1}\over 4G_4}\int_0^{R_c}{rdr\over\sqrt{1+{r^2\over L^2}}} =
{\pi L^2\over 2G_4} \Big({R_c\over L} - 1\Big)\ \
\xrightarrow{\, L\ra -il\,}\ \
-i{\pi l^2\over 2G_4} {R_c\over l} + {\pi l^2\over 2G_4} = S^{IR}_{dS_4}\,,
\ee
\be\label{IRsurfAdSdS3}
{V_{S^0}\over 4G_3}\int_0^{R_c}{dr\over\sqrt{1+{r^2\over L^2}}} =
{2L\over 4G_3}\log{2R_c\over L}\ \ \xrightarrow{\, L\ra -il\,}\ \
-i{l\over 2G_3}\log{2R_c\over l} + {\pi l\over 4G_3} = S^{IR}_{dS_3}\,.
\ee
We continue to focus on these IR extremal surfaces pertaining
to maximal subregions here.

The fact that the $dS$ areas arise from analytic
continuations from $AdS$ suggests a natural way
to obtain the no-boundary de Sitter extremal surface areas through a
heuristic replica argument \cite{Narayan:2023zen} involving an
analytic continuation of the Lewkowycz-Maldacena formulation
\cite{Lewkowycz:2013nqa} in $AdS$ to derive RT entanglement entropy
(generalized in \cite{Dong:2016hjy}, \cite{Dong:2016fnf},
\cite{Dong:2013qoa}; see also \cite{Casini:2011kv} and the review
\cite{Rangamani:2016dms}).
In $AdS$, the boundary replica space $M_n$ extends into a smooth $n$-sheeted
bulk replica space ${\cal B}_n$ which is a smooth covering space with
replica boundary conditions. The quotient
${\tilde{\cal B}_n}$ of the bulk space by the $\BZ_n$ replica symmetry
(so its boundary is $\del{\tilde{\cal B}_n}=M_n/\BZ_n=M_1$, the
original boundary space) contains conical (orbifold) singularities
corresponding to $\BZ_n$ fixed points in the bulk for $n\neq 1$.
These reflect the fact that
the bulk quotient space is a solution to the bulk Einstein equations
only in the presence of a source with nontrivial backreaction.
The fixed points extend out from the subregion boundary so the
required source is a codim-2 (cosmic) brane with tension
${n-1\over n}{1\over 4G}$\ (giving deficit angle $2\pi-{2\pi\over n}$)
and area $A$ (wrapping all the transverse directions).
The smooth action is\
$I_n = n I_1 + I_{brane}$ with $I_{brane} = {n-1\over n} {A\over 4G}$.
Defining the bulk partition function $Z_n\equiv Z[{\tilde{\cal B}_n}]$
on the replica quotient space ${\tilde{\cal B}_n}$, the entropy via
replica is\
$S = -\lim_{n\ra 1} n\del_n (\log Z_n - n\log Z_1)
= -\lim_{n\ra 1} (1-n\del_n) I_n$\
in the semiclassical approximation where\ $Z_n\sim e^{-I_n}$, with
$I_n$ the action. Thus as $n\ra 1$, we
obtain\ $S={A\over 4G}$ which is the area of the extremal RT/HRT
entangling surface.
The analytic continuation from $AdS$ and the $dS/CFT$ dictionary
$Z_{CFT}=\Psi_{dS}$ gives
\be
Z_{bulk} \sim e^{-I_{bulk}}\ \ \longrightarrow\ \ \Psi_{dS}\sim e^{iS_{cl}}
\sim\ e^{iS^{(r>l)}}\,e^{S_E^{(r<l)}}
\ee
enabling a heuristic understanding of boundary entanglement entropy
\cite{Narayan:2023zen} in the Euclidean CFT dual to $dS$ via replica.
Semiclassically, the Wavefunction is given by the action, the top
Lorentzian part (with real $S^{(r>l)}$) being a pure phase while the
bottom Euclidean hemisphere has real action.
We pick
boundary Euclidean time slices as $S^d$ equatorial planes at $I^+$
(\ie\ some $S^{d-1}$) and consider the IR extremal surface obtained
from the maximal subregion (half the $S^{d-1}$). We construct $n$
copies and appropriately glue them cyclically, with replica boundary
conditions. The $n$-copy replica Wavefunction $\Psi_n$ on the
quotient bulk replica space satisfies, semiclassically,\
$Z_n\sim e^{-I_n}\ra \Psi_n\sim e^{iS_n}$\ with
$-I_n\ \ra\ iS_n = iS_n^{(r>l)} + S_E^{(r<l)}$\,.
The codim-2 brane source generating the conical singularities
for $n\neq 1$
has nontrivial (part Euclidean, part Lorentzian) time evolution:
in the $n\ra 1$ limit it satisfies the no-boundary condition and
wraps the will-be no-boundary $dS$ extremal surface. Now
continuing from above gives $-I_{brane}\ra -{n-1\over n} {A_{brane}\over 4G}$
and\
$S_t = \lim_{n\ra 1} (1-n\del_n) \log \Psi_n = {A^{dS}_{brane}\over 4G}$\
as the area/entropy in $dS$, 
giving boundary entanglement entropy in the dual ghost-like Euclidean CFT.
A crucial point here is that since the analytic continuation maps
$Z_{bulk}^{AdS}$ to the de Sitter Wavefunction $\Psi_{dS}$, this is
now a replica formulation on $\Psi_{dS}$, considering
the $dS/CFT$ dictionary $Z_{CFT}=\Psi_{dS}$\ \cite{Maldacena:2002vr}.
With the Wavefunction $\Psi_{dS}$ regarded as an amplitude (or
transition matrix from ``nothing''), this gives pseudo-entropy.
In particular the codim-2 brane that smooths out bulk (orbifold)
singularities is now a time-evolving, part Euclidean, part timelike,
brane: this gives a complex area semiclassically. In this
Lewkowycz-Maldacena formulation, the area of these
no-boundary $dS$ extremal surfaces arises as the amplitude for cosmic
brane creation. So it is important that the divergent pieces of the
area arising from near the future boundary are pure imaginary, since
otherwise this amplitude would diverge. As it is, there is a finite
probability: the real part arises from the maximal hemisphere, with size
set by $dS$ entropy.
In \cite{Goswami:2024vfl}, no-boundary extremal surfaces in slow-roll
inflation models were studied (see App.~\ref{sec:dSsr-rep}): there are
new features in the area integrals which must be defined carefully as
complex-time-plane integrals with appropriate time contours
(Figure~\ref{fig3}). In the end the cosmic brane probability matches
the corresponding parts in the Wavefunction, vindicating the above
picture.

The above Lewkowycz-Maldacena replica is fuelled by the analytic
continuation. In what follows, we will flesh this out more
elaborately, without recourse to analytic continuation.

\section{de Sitter quotient geometries}\label{sec:singQuoGeom}

In this section we will study explicit quotient metrics near $n=1$ for
various de Sitter like spaces (which are similar to Fursaev's
constructions \cite{Fursaev:2006ih}) and recover boundary entanglement
entropy by evaluating the areas of no-boundary $dS$ extremal surfaces
in \cite{Narayan:2022afv,Narayan:2023zen} (reviewed in
sec.~\ref{sec:dSnbRev}).  In subsequent sections, we will construct
smooth bulk replica geometries for various $dS$ spaces, which we will
then use to construct boundary Renyi entropies via $dS/CFT$.


Denoting the bulk quotient space as $B_n/Z_n=O_n$,
near $n=1$ we have $O_{1+\epsilon}=O_1=B_1$ which is the original
space (\eg\ $dS_{d+1}$). Including the cosmic brane as explicit
source gives a smooth description of the bulk+brane space: the 
cosmic brane action is $I_{brane} = {n-1\over n}\,{A_{brane}\over 4G}$
with $A_{brane}$ the area given by the action over the singular locus.
Near $n=1$ we expand as $n=1+\epsilon$ and obtain
$I[B_{1+\epsilon}] = (1+\epsilon) I[B_1] + \epsilon {A_{brane}\over 4G}$,
which gives the extremal surface area as the entropy
\be\label{I[Bn]I[brane]}
S = \lim_{n\ra 1}\ n\del_n \big(I[B_n]-nI[B_1]\big)
= \del_\epsilon \big(I[B_{1+\epsilon}] - (1+\epsilon) I[B_1]\big)
= {A_{brane}\over 4G}\,.
\ee
We thus recover the entropy or extremal surface area through the
$n\ra 1$ limit of a family of quotient spaces.\
Thus the boundary entanglement entropy in these no-boundary $dS$-like
spaces 
is obtained using the above arguments and the $dS/CFT$ dictionary
$Z_{CFT}=\Psi_{dS}$ with
\be
\log Z^{CFT}_n = \log\Psi_n\equiv -I[B_n]\,;\qquad
\log\Psi_n=n\log\Psi_1 - nI_{brane}\,,
\ee
with $I[B_n]$ comprising the Euclidean part (from the hemisphere) and
the Lorentzian part (and $\Psi_1$ indicates a single copy).
We obtain
\begin{equation}
\mathcal{S} = -n \del_n (\log Z^{CFT}_n - n \log Z^{CFT}_1)|_{n=1}
= -n \del_n (\log\Psi_n - n \log\Psi_1)|_{n=1}  
= \frac{A_{brane}}{4G}. \label{sent}
\end{equation}
We will now display quotient metrics, valid near $n=1$, in various
$dS$-like spaces and use the above.

\subsection{$dS_3$ quotient and boundary EE} \label{rcdS3}

We first consider the calculation of no-boundary extremal surfaces and
the associated boundary entanglement entropy (or bulk pseudo-entropy)
in $dS_3$. We will compute the pseudo-entropy in two different but
equivalent ways. First, we will consider $dS_3$ and an  appropriate
quotient geometry, and thereby obtain the pseudo-entropy via
cosmic branes.  Second, we will follow \cite{Donnelly:2018bef} to
obtain the pseudo-entropy directly from the on-shell action
(App.~\ref{adsds}).

\begin{figure}[h] 
  \includegraphics[width=25pc]{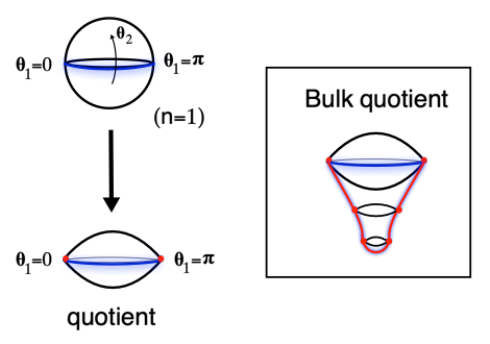} 
\hspace{0.2pc}
\begin{minipage}[b]{13.5pc}
  \caption{{\label{figreplica}\!\! \footnotesize{The 
        boundary $S^2$ with the IR subregion $\theta_1=[0,\pi]$ on
        the  boundary Euclidean time slice given by the equatorial
        plane $\theta_2=0$. The quotient space arises from the
        replica-like space with $n$ copies, and contains conical
        singularities at the North/South Pole endpoints. The box
        shows the bulk quotient space and the cosmic brane (red).} \\ }}
\end{minipage}
\end{figure}

As reviewed in sec.~\ref{sec:dSnbRev}, recall that we had defined
boundary Euclidean time slices as equatorial ($S^1$) planes of the
$S^2$ at $I^+$ and then defined the IR boundary subregion as the half
circle on that $S^1$. The extremal surface then stretches out into the
bulk $r$-direction from the semicircle endpoints giving the areas
(\ref{nbdS43rev}), (\ref{IRsurfAdSdS3}).
In the present context, we consider the boundary Euclidean time
slice as some $\theta_2=const$ slice: the IR subregion is then
the interval defined by $\theta_1=[0,\pi]$ on that slice, and the
endpoints are the North/South Poles (Figure~\ref{figreplica}).

Thus consider the quotient metric
\be\label{metnds2}
ds^2 = -\frac{dr^2}{\frac{r^2}{l^2}-1} + r^2 (d\theta_1^2
+ \frac{1}{n^2} \sin^2\theta_1 {d\theta_2^2})\,.\qquad
\ee
For $n=1$, this is global no-boundary $dS_3$. It is Lorentzian for
$r>l$ where $r$ is a timelike direction and describes the Euclidean
hemisphere for $r<l$. The sphere smoothly shrinks to zero size at
$r=0$ which is the no-boundary point, implementing Hartle-Hawking
boundary conditions.
For $n\gtrsim 1$, this is akin to the singular spaces in
\cite{Fursaev:2006ih}, adequate near the $n=1$ limit to calculate
boundary entanglement entropy (but not for Renyi entropies).

The $S^2$ at $n=1$ is defined by the latitude angle $\theta_1$ and the
azimuthal angle $\theta_2$, with ranges $\theta_1=[0,\pi]$
and $\theta_2=[0,2\pi]$. In (\ref{metnds2}), roughly speaking, the
azimuthal $\theta_2$-direction is squashed so the equator of original
size $2\pi$ now has size ${2\pi\over n}$: thus we have the (approximate)
shape of a rugby ball (or American football), with
conical singularities at the North and South poles (Figure~\ref{figreplica}),
$\theta_1=0$ and $\theta_1=\pi$. These singularities apart, the
metric is locally $dS_3$ with $R=6/l^2$; at $n=1$ the space is $dS_3$.
The bulk metric in the vicinity of the singular locus is
$ds^2 = -\frac{dr^2}{\frac{r^2}{l^2}-1} + r^2 (d\theta_1^2
+ \frac{\theta_1^2}{n^2} {d\theta_2^2})$.
The singular locus runs along the $r$-direction, starting at the
future boundary $I^+$ at the North Pole ($\theta_1=0$), down the
Lorentzian $r$-direction to the complexification point $r=1$ and
then along the Euclidean hemisphere to the no-boundary point $r=0$
and back to $r=1$, up the Lorentzian $r$-direction at the South
Pole ($\theta_1=\pi$) to $I^+$. At $n=1$, this traces out
the no-boundary extremal surface. With $n\gtrsim 1$ this family
of metrics represent the conical spacetime geometry (with deficit
angle $2\pi{n-1\over n}$ from the $\theta_2$-periodicity
${2\pi\over n}$) sourced by cosmic
branes of tension ${n-1\over n}\,{1\over 4G}$ at the singular loci above.

We will focus on
this quotient space (\ref{metnds2}) analytically continuing $n$ to
non-integer values near $n=1$. Then the space is almost smooth with
mild conical singularities. Taking $n=1+\epsilon$ the deficit
angle is $2\pi\epsilon$ upto $O(\epsilon)$.
Following \cite{Fursaev:1995ef}, we can calculate curvature
invariants due to the conical singularity: for \eqref{metnds2} we have
\begin{equation}
\int d^3x \sqrt{g}\, \textbf{R} = 4 \pi \left(1-\frac{1}{n}\right) \int dr \sqrt{g_{rr}} +\frac{1}{n} \int d^3 x \sqrt{g}\, \textbf{R}_{dS}, \label{req}
\end{equation} 
where $\textbf{R}_{dS}$ is the Ricci scalar $\frac{6}{l^2}$ of the
smooth $dS_3$ space. Note that this is no-boundary $dS_3$, with
$\sqrt{g_{rr}}={1\over\sqrt{1-r^2/l^2}}$ defined in the Euclidean
hemisphere region $r<l$ and then continued to the Lorentzian region
$r>l$. The above equation is schematically of the form\
$I[B_n/\BZ_n]=I_{brane}+{1\over n}I[B_1]$\ (and $O_1=B_1$), so multiplying
by $n$ essentially gives the analog of (\ref{I[Bn]I[brane]}). Then
employing the arguments there, we recover the associated entropy
which gives the extremal surface area\ (one could have instead
equivalently used a version of the replica formula for the quotient
space object $I[O_n]\equiv I[B_n/\BZ_n]$ without the extra $n$-factor).
The area of this 1-dimensional surface in the entropy (\ref{sent})
is then as given in (\ref{nbdS43rev}), (\ref{IRsurfAdSdS3}).
The factor of 2 in (\ref{IRsurfAdSdS}) is $V_{S^0}$ which can be
interpreted as the two halves of the cosmic brane curve in the
$r$-direction, along $R_c\ra l\ra 0\ra l\ra R_c$.

We have thus obtained boundary entanglement entropy in $dS_3/CFT_2$
from the quotient $dS_3$ via $Z_{CFT}=\Psi_{dS}$.
We now derive this result in an alternative way, generalizing the
arguments in \cite{Donnelly:2018bef}.
The semiclassical Wavefunction $\Psi_1 \sim e^{iI}$ of no-boundary
$dS_3$ is (App.~\ref{adsds})
\begin{equation} \label{Z1}
	\log \Psi_1 = {1\over 16\pi G_3} \left(4 \pi^2 l -8 i \pi l \log(\frac{2 R_c}{l}) + 8 i \pi \frac{R_c^2}{l} - 4 i \pi l\right) .
\end{equation}
$\Psi_1$ indicates a single copy here.
The spherical symmetry of the boundary $S^2$ implies that the boundary
stress tensor is symmetric: this can then be used to trade the $\del_n$
derivative of $\log Z_{CFT}$ in the Weyl anomaly of the $CFT$ on $S^2$
for $\del_{R_c}$, following the arguments in \cite{Donnelly:2018bef}
(which do not depend on unitarity of the CFT). Thus, using the
$dS/CFT$ dictionary $Z^{CFT}_1=\Psi^{dS}_1$, we obtain the boundary
entanglement entropy  
\begin{equation}
	\mathcal{S} = \left(1 - \frac{R_c}{2} \del_{R_c}\right) \log \Psi_1
	= {\pi l\over 4 G_3} - i {l \over 2 G_3} \log(\frac{2 R_c}{l})\,.
	\label{dels}
\end{equation}
It is important to note that in this calculation, the $AdS$ analytic
continuation (\ref{AdSdSL-il}) employed renders the sign of the
leading divergence to be the opposite of that in (\ref{dSact-LG}),
(\ref{ehactdS3L}). This in turn ensures that the first subleading term
in (\ref{Z1}) then has the same sign as the extremal surface area
(\ref{dels}). Relatedly, the first subleading term in (\ref{Z1})
differs by a factor of $l^2$ from the leading term, which thus
contributes one relative minus sign under the $AdS\ra dS$ analytic
continuation: thus $L\ra \pm il$ lead to $\pm i$ in the imaginary
part in the extremal surface area (\ref{dels}) above.
Further comments appear in sec.~\ref{impnba}.

\subsection{$dS_4/CFT_3$ boundary EE via replica quotient} \label{rcdS4}

In this subsection we discuss a $dS_4$ quotient geometry along the
same lines as for $dS_3$ above.
The metric of the $n$-quotient $dS_4$ space is
\begin{equation}
ds^2= -\frac{dr^2}{\frac{r^2}{l^2}-1} + r^2 \left( d\theta_1^2 + \frac{1}{n^2} \sin^2\theta_1 d\theta_2^2 + \cos^2\theta_1 d\theta_3^2 \right) . \label{metnds3}
\end{equation} 
The $n$ factor here is similar to that in (\ref{metnds2}) for the
$dS_3$ quotient.
The geometry near $\theta_1=0$ is
$ds^2 = -\frac{dr^2}{\frac{r^2}{l^2}-1} + r^2 \big(d\theta_1^2
+ \frac{\theta_1^2}{n^2} {d\theta_2^2} + d\theta_3^2 \big)$.
Thus the angles $\theta_1,\theta_2$ define the 2-dim cone at the tip
of which the cosmic brane is localized, generating the conical
singularity. The $S^3$ geometry in (\ref{metnds3}) for $n=1$ is the
Hopf fibration: this description dovetails with the boundary
subregion as reviewed in sec.~\ref{sec:dSnbRev} as follows. We
define the boundary Euclidean time slice as some $\theta_2=const$
slice, giving $d\theta_1^2+\cos^2\theta_1d\theta_3^2$ defining an
$S^2$. The equator here is $\theta_1=0$ while the North/South Poles
are $\theta_1={\pi\over 2}$\,. The boundary of the maximal
(hemispherical) subregion on this $S^2$ is the equator at
$\theta_1=0$. The cosmic brane thus extends out from $I^+$ from
this subregion boundary, wrapping the equatorial $S^1$ parametrized
by $\theta_3$ and going along the bulk $r$-direction over
$0\ra l\ra R_c$: its action/area is
\be\label{Abraneds4}
\mathcal{S} \equiv {A_{brane}\over 4G_{4}} =
{V_{S^{1}}\over 4G_{4}} \int_0^l {r dr\over \sqrt{1-{r^2\over l^2}}}
+ {V_{S^{1}}\over 4G_{4}}
\int_l^{R_c} {r\,dr\over\sqrt{-({r^2\over l^2}-1)}}
= -i\frac{\pi l^2}{2 G_4}\frac{R_c}{l}+\frac{\pi l^2}{2 G_4}\ ,
\ee
which gives the $dS_4/CFT_3$ boundary entanglement entropy as the
no-boundary $dS_4$ IR extremal surface area, using \eqref{sent},
(\ref{ehact}), (\ref{ehactdS4L}), (\ref{ehactdSE}): this matches
(\ref{IRsurfAdSdS4}).
The analog of (\ref{req}) useful for the action calculation here is\
$\int {\bf R} = 4\pi (1-{1\over n}) A_{brane} + {1\over n}\int {\bf R}_{dS}$
with the conical singularity giving a $\delta$-function which when
integrated over the $S^3$ reduces to $A_{brane}$.

We can also recover this from the semiclassical Wavefunction for a
single copy
\begin{equation}
	\log \Psi_1 = {\pi l^2\over 2 G_4} - i {\pi R_c^3\over 8 G_4}
        \left(-\frac{4}{l} +\frac{6 l}{R_c^2}\right) .
\end{equation}
Then following \cite{Donnelly:2018bef} and using the boundary $S^3$
spherical symmetry, we can obtain the boundary entanglement entropy
in $dS_4/CFT_3$ as
\begin{align}\label{DS-dS4replica}
	\mathcal{S} &= \left(1- \frac{R_c}{3} \del_{R_c}\right) \log \Psi_1 = 
	{\pi l^2\over 2 G_4} -i\frac{\pi l^2}{2 G_4}\frac{R_c}{l}\,,
\end{align}
along the lines of (\ref{dels}).
The above answer matches (\ref{Abraneds4}) obtained from the explicit
quotient metric and cosmic brane.

The expression for $\log\Psi_1$ above can also be obtained from the
Euclidean $AdS_4$ partition function with the analytic continuation
(\ref{AdSdSL-il}). Thus, as in the comments after (\ref{dels}), we
note that relative to (\ref{ehactdS4L}), the signs of
the imaginary terms in $\log\Psi_1$ here have been reversed by the
analytic continuation $L\ra -il$, which then ensures that the
divergent term in the entropy above (which differs by one $l^2$-factor
from the leading divergence in $\log\Psi$) matches that in the
extremal surface area.

\medskip

\noindent {\bf $dS_{d+1}$ quotient:}\ \
Considering a general $dS_{d+1}$ space ($d>2$), the $n$-quotient
metric near $n=1$ is
\begin{equation}
ds^2= -\frac{dr^2}{\frac{r^2}{l^2}-1} + r^2 \left( d\theta_1^2 + \frac{1}{n^2} \sin^2\theta_1 d\theta_2^2 + \cos^2\theta_1 d\Omega_{d-2}^2 \right), \label{dSd+1-nreplica}
\end{equation}
where $d\Omega_{d-2}^2$ is the metric of $S^{d-2}$. As in the earlier
analysis, $\theta_1,\theta_2$ form a cone. This gives 
\begin{equation}
I = \frac{n-1}{n} \frac{V_{S^{d-2}}}{4 G_{d+1}} \int_0^{R_c} dr\, r^{d-2} \frac{1}{\sqrt{1-\frac{r^2}{l^2}}}\,,
\end{equation}
as the cosmic brane action and thereby the entropy via (\ref{sent})
(matching (\ref{IRsurfAdSdS})).
We can also obtain the entanglement entropy by continuing the $AdS$ action to $dS$ with $L \rightarrow -i l$ and then using the formula,
\begin{equation}
\mathcal{S} = \left(1 - \frac{R_c}{d-1} \del_{R_c}\right) \log \Psi_1\,,
\end{equation}
with $R_c$ the UV cutoff near the future boundary.

Various generalizations of these quotient constructions can be
analysed: see App.~\ref{sec:fpRep} for future-past extremal surface
areas (see also \cite{Arias:2019pzy} for related discussions in a
slightly different context), App.~\ref{sec:dS-sphere} for real
extremal surfaces in the Euclidean sphere giving $dS$ entropy, and
App.~\ref{sec:dSsr-rep} for slow-roll no-boundary extremal surfaces.

\section{de Sitter replicas: boundary Renyi via $dS/CFT$}\label{sec:dSreplica}

In general we imagine constructing a bulk replica (covering) space $B_n$
comprising $n$ copies of the original space $B_1$ glued together with
replica boundary conditions: this is a branched cover branched over
the subregion boundary, and we go from the $k$th copy to the next
cyclically. If we now quotient this space by the $\BZ_n$ replica
symmetry which permutes the $n$ copies, then we obtain an
orbifold-like space with conical singularities localized at the 
$\BZ_n$ fixed points. The quotient space can be interpreted as
the backreacted geometry of codim-2 cosmic branes which are
localized at the locus that becomes the extremal surface.
As $n\ra 1$, there is no backreaction: the tensionless
cosmic brane becomes the extremal surface.

At finite $n$ however, this requires new bulk geometries that satisfy
replica boundary conditions at the holographic boundary and are smooth
in the bulk with no singularities. This requirement of a smooth bulk
replica space is in accord with $AdS/CFT$ intuition as pointed out in
\cite{Headrick:2010zt}, where we require bulk perturbations to satisfy
regularity conditions in the interior. This logic, implemented by
Lewkowycz-Maldacena \cite{Lewkowycz:2013nqa}, and implicit in
\cite{Casini:2011kv} for entanglement entropy and developed in
detail in \cite{Hung:2011nu} for Renyi entropies (see also
\cite{Camps:2016gfs,Horowitz:2017ifu}), is conceptually and
technically distinct from the quotient spaces (akin to
\cite{Fursaev:2006ih}) discussed in the previous section.

In this section we will describe smooth bulk $dS$-like replica spaces
which we will then use to calculate boundary Renyi entropies using the
$dS/CFT$ dictionary $Z^{CFT}_1=\Psi^{dS}_1$ \cite{Maldacena:2002vr}
where the subscript $1$ denotes a single copy ($n=1$). The boundary
Renyi entropy for the replica-$n$ geometry with
$Z^{CFT}_n=\Psi^{dS}_n$ is defined by
\be\label{dSrenyi}
S_n = \frac{1}{1-n} \log({Z_n\over Z_1^n}) = 
\frac{1}{1-n} \log(\frac{\Psi_n}{\Psi_1^n})
\sim i\,\frac{I_n- n I_1}{1-n}\,,
\ee
where the Wavefunction in the semiclassical regime is\
$\Psi_n\sim e^{iI_n}$, with $I_n$ the action for the replica-$n$
space.
Our $dS$ analysis has close parallels with the $AdS$ studies in
\cite{Lewkowycz:2013nqa} and \cite{Hung:2011nu}.
We describe $dS_3$ first, and then $dS_4$ and higher dimensional $dS$.

It is worth noting that the broad logic of our replica construction in
these cosmological cases is along the lines of the generalization of
\cite{Lewkowycz:2013nqa} in \cite{Dong:2016hjy} to the general
time-dependent $AdS$ context.  However several central aspects here
are really features in de Sitter with the spacelike future boundary
and $dS/CFT$ in mind, and so have no analog in $AdS$ (which pertain to
spacelike surfaces anchored on the timelike boundary, even when there
is nontrivial time dependence). For instance, bulk time is not
boundary (Euclidean) time here: the maximal subregions at $I^+$
considered throughout this work lie on boundary Euclidean time slices
(well-defined given the existence of boundary spatial isometries),
which are essentially ``vertical'' slices in the bulk.  The extremal
surfaces lie on these slices of the bulk geometry and dip into the
bulk time direction (Figure~\ref{fig1}), with timelike components
necessarily arising (due to the absence of $I^+\ra I^+$ turning points
\cite{Narayan:2015vda}).  Overall the picture here is a space-time
rotation of the $AdS$ story, and so is more analogous to the $AdS$ RT
formulation on constant time slices turned sideways (akin
to \cite{Narayan:2017xca,Narayan:2020nsc}), than HRT with
time dependence. Thus in an essential sense, we expect the full
richness of \cite{Dong:2016hjy} to only enter for generic subregions
(and their correspondingly more complicated extremal surfaces).
Likewise, the replica space also exhibits new geometric features as we
will see: for instance, the cosmic brane locus is not spacelike and
relatedly, a boundary replica on $Z_{CFT}$ translates to a bulk
replica on $\Psi_{dS}$ as we already saw in sec.~\ref{sec:dSnbRev},
reviewing \cite{Narayan:2023zen}. Here this results in
(\ref{dSrenyi}), a complex-valued pseudo-Renyi-entropy in general.  In
the end, it will turn out that our formulations here amount to
analytic continuation from $AdS$ (which executes the space-time
rotation), so in some essential sense there are of course close
interrelations between our analysis here and that in
\cite{Lewkowycz:2013nqa}, \cite{Hung:2011nu}, \cite{Dong:2016hjy}.  At
various places in the text, we will try to highlight specific novel
features, from an intrinsically $dS$ perspective.

\subsection{$dS_3$ replica geometry: boundary Renyi}\label{sec:dS3replica}

\begin{figure}[h] 
  \includegraphics[width=21.5pc]{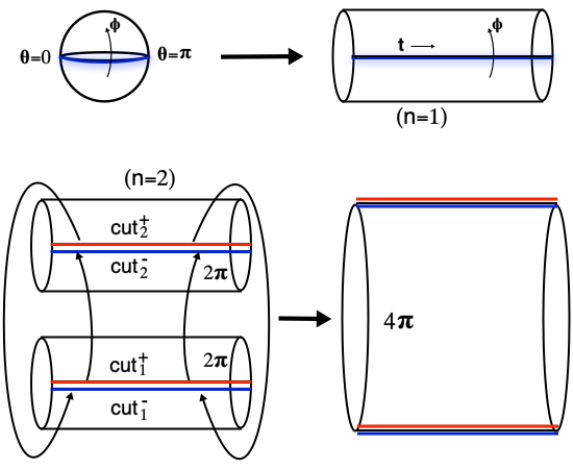} 
\hspace{0.5pc}
\begin{minipage}[b]{16pc}
  \caption{{\label{figreplicadS3}\!\! \footnotesize{Depicted (top) is
        the boundary $S^2$ with a conformal transformation to the
        cylinder. Also shown is the maximal (IR) subregion as the
        semicircle $\theta=[0,\pi]$ on the  boundary Euclidean time
        slice given by the equatorial plane $\phi=0$, which maps
        to a line along the entire cylinder length. The bottom shows
        the $n=2$ replica space obtained by gluing two copies of the
        cylinder cyclically along the cuts as $1^+\ra 2^-\ra 2^+\ra 1^-$
        and back to $1^+$. Since the gluing is along the entire
        cylinder length, this gives a fatter cylinder with
        $\phi$-periodicity $4\pi$.}  }}
\end{minipage}
\end{figure}

Consider the 3-dim geometry described by
\be\label{dS3replica}
ds^2 = - {dr^2\over {r^2\over l^2} - {1\over n^2}} +
\Big({r^2\over l^2} - {1\over n^2}\Big) dt^2 + r^2 d\phi^2\,,
\qquad \phi\equiv \phi+2\pi n\,.
\ee
This has $R={6\over l^2}$\,.
For $n=1$, this is $dS_3$ in the static coordinatization. The $r>{l\over n}$
region describes the future and past universes with $g_{rr}<0$ and
$r$ is bulk time. The $r<{l\over n}$ region is the static patch where
$g_{tt}<0$ and $t$ is the time coordinate.\\
For large $r\ra\infty$, the metric approaches\
$ds^2\sim -l^2{dr^2\over r^2} + r^2({dt^2\over l^2}+d\phi^2)$\ so the
boundary space has replica boundary conditions with $t$ noncompact
but $\phi\equiv \phi+2\pi n$. One way to understand this as a replica
space is as follows: start with the boundary space as a 2-sphere $S^2$
and then make a boundary conformal transformation to the cylinder,
with the boundary metrics 
\be\label{S2-cyl-confmap}
ds^2=\sin^2\theta\, \Big({d\theta^2\over\sin^2\theta}+ d\phi^2\Big)
\quad\ra\quad ds'^2=dt^2+d\phi^2\,,
\ee
where $t=\int {d\theta\over\sin\theta}$ is now the noncompact cylinder
length direction. This is essentially the sphere-to-cylinder
map familiar from radial quantization around the North pole of the
$S^2$, the latitudes $\theta=const$ now becoming $t=const$ slices.
Now consider a $\phi=0$ slice regarded as a boundary Euclidean time
slice of the $S^2$ and the (maximal) subregion defined by a semicircle
$\theta=[0,\pi]$: in terms of the cylinder picture, this is a
``longitudinal'' slice $\phi=0$ with the subregion defined by the
entire line $t=[-\infty,\infty]$ on this slice. A replica space
corresponding to $n$ copies of this subregion can be constructed
by introducing a cut along this $t$-line in each cylinder copy and
joining the cylinder copies $k$ to $k+1$ along the cuts cyclically.
Since the cylinder copies are being glued along their entire length
directions, the effective cross-section enlarges thus making a
``fatter'' replica-cylinder with cross-section angle $\phi$-periodicity
$2\pi n$. For $n=2$ this is depicted in Figure~\ref{figreplicadS3}.

The bulk geometry above is then constructed as a smooth geometry whose
future boundary satisfies these replica boundary conditions. Thus 
the $n$-factors in (\ref{dS3replica}) have been fixed by demanding
regularity of the full geometry. 
For instance starting with
\be
ds^2 = - {dr^2\over f(r)} +  
f(r) dt^2 + r^2 d\phi^2\,,\qquad
f(r) = {r^2\over l^2} - r_h^2\,,
\ee
satisfying replica boundary conditions at the future boundary
$r\ra\infty$, and requiring that the metric near $r=0$ is smooth
with no conical singularities gives
\be
ds^2 \sim - r_h^2 dt^2 + {dr^2\over r_h^2} + n^2 r^2 d{\tilde\phi}^2
\quad \Rightarrow\quad r_h={1\over n}\,,
\ee
where we have redefined ${\tilde\phi}\equiv {\phi\over n}$
which has periodicity $2\pi$.
Near $r\gtrsim r_h$ this metric becomes\
$ds^2 \sim - {dr^2\over 2r_h ({r\over l} - r_h)}
+ 2r_h ({r\over l} - r_h) dt^2 + r_h^2l^2 d\phi^2
\equiv - d\rho^2 + {\rho^2\over n^2l^2}dt^2 + l^2 d{\tilde\phi}^2$
which is ordinary regular Milne in the $(\rho,t)$-subplane (since
$t$ is noncompact) and in the $(\rho,{\tilde\phi})$-subplane.
Further since $t$ is a noncompact direction, there are no issues
with closed timelike curve pathologies (unlike if the blackening
$f(r)$ factor appeared in $g_{{\tilde\phi}{\tilde\phi}}$).
Thus we recover the smooth $dS_3$-like replica space (\ref{dS3replica}).
As an interesting aside, it is worth noting that this geometry can
be realized as an analytic continuation $l^2\ra -L^2$ from the
corresponding $AdS_3$-like geometry with replica boundary conditions
studied by Lewkowycz-Maldacena \cite{Lewkowycz:2013nqa}.

\medskip

\noindent {\underline {\bf Boundary Renyi entropies:}}\ \ 
We will now evaluate the boundary Renyi entropies in $dS/CFT$ using
(\ref{dSrenyi}). Towards this, we use the (bare) action
(which is (\ref{ehact}))
\be\label{tildeI-dS3}
16\pi G_3\,{\tilde I} = \int d^3x \sqrt{-g} \left(R-\frac{2}{l^2}\right) - 2\int d^2x \sqrt{\gamma} K\,,
\ee
with the extrinsic curvature at the future boundary $r=R_c$ given by
\be
K = \frac{1}{\sqrt{-g}} \del_r(\sqrt{-g} n^r) = \frac{1}{l r} \del_r \left(l r \sqrt{\frac{r^2}{l^2}-n^{-2}}\right)\ \ \xrightarrow{r\ra R_c}\ \ 
\frac{2 R_c^2 - l^2 n^{-2}}{l^2 R_c \sqrt{\frac{R_c^2}{l^2}-n^{-2}}}\,.
\ee
Then the action for the replica geometry (\ref{dS3replica}), using
$R= \frac{6}{l^2}$, is
\be
{\tilde I}_n = {1\over 16\pi G_3} \int r d\phi dt dr \frac{4}{l^2}
- {2\over 16\pi G_3} \int d\phi dt\, R_c
\sqrt{\frac{R_c^2}{l^2}-{1\over n^{2}}}\, K
= \frac{4\pi n}{16\pi G_3\,l^2} L_t (n^{-2} l^2 - R_c^2)\,, \label{osact2}
\ee
where $L_t=\int dt$ which we will evaluate below.
We now add a counterterm to remove divergences and render the action
finite:
\be
I_{ct} = {\Lambda\over 16\pi G_3} \int d^2x \sqrt{\gamma}
= {\Lambda\over 16\pi G_3} \int d\phi dt\, R_c \sqrt{\frac{R_c^2}{l^2}-n^{-2}}
\sim {\Lambda 2\pi n\over 16\pi G_3}\, {L_t\over l} \left(R_c^2 - \frac{l^2}{2 n^2}\right)\,.
\label{Icteq2}
\ee
Choosing $\Lambda=\frac{2}{l}$ gives the regulated action
\be
I_n = {\tilde I}_n + I_{ct} = \frac{2\pi n}{16\pi G_3\,l^2}\, L_t n^{-2} l^2\,.
\label{regact2}
\ee
Now we note that $L_t=\int dt = \int_0^\pi {d\theta\over\sin\theta}$
is the length of the subregion given by the semicircle interval
stated above. This however diverges over $\theta \in [0,\pi]$, so
we regulate as $\theta=[\delta, \pi-\delta]$. This gives
\be
L_t = 2 \log(\frac{2}{\delta}) \equiv
2 \log(\frac{2 R_c}{l})\,, \label{Lxeq} 
\ee
where we have identified $\delta=\frac{l}{R_c}$ to match the known
boundary entanglement entropy at $n=1$ with the extremal surface area.
Thus using the regulated actions, we finally obtain the boundary Renyi
entropy (\ref{dSrenyi}) as
\be
S_n = {i\over 1-n}\, \frac{2\pi n}{16\pi G_3\,l}\, L_t l (n^{-2}-1)
= \Big(1+{1\over n}\Big)\, i {l\over 4G_3} \log {2R_c\over l} \equiv
\Big(1+{1\over n}\Big)\, {c\over 6} \log {2R_c\over l}
\,,
\ee
where $c=i{3l\over 2G_3}$ is the $dS_3/CFT_2$ central charge
\cite{Maldacena:2002vr}.
The $n\ra 1$ limit gives the boundary entanglement entropy\
${c\over 3} \log {2R_c\over l}$\,, which is half the future-past
extremal surface area \cite{Narayan:2017xca,Narayan:2022afv}.

The Euclidean part of the $dS_3$ replica space can be obtained from
the $r<{l\over n}$ part of (\ref{dS3replica}) with Euclidean time
$t\ra i\tau_E$ and $\tau_E=[0,{\pi\over 2}]$. The Renyi entropy
calculation with the Euclidean gravity action then gives
\be
S_n^E = \Big(1+{1\over n}\Big)\,{l\over 4G_3}\,{L_{\tau_E}\over 2} =
\Big(1+{1\over n}\Big)\,{l\over 4G_3}\,{\pi\over 2}\,,
\ee
with $L_{\tau_E}=2\int_0^{\pi/2} d\tau_E=\pi$ arising from the
great circle path in the hemisphere.
Thus the boundary Renyi entropy for no-boundary $dS_3$ is given
by appending the Lorentzian and Euclidean contributions as
\be\label{dS3renyi}
S_n^{nb} = {1+n\over 2n}\, \Big({\pi l\over 4G_3}
+ {i l\over 2G_3}\log{ 2R_c\over l} \Big)\,,
\ee
and the $n\ra 1$ limit gives the boundary entanglement entropy
which matches the no-boundary extremal surface area
(\ref{IRsurfAdSdS3}), but with the imaginary part being $+i$,
obtained via the analytic continuation $L\ra il$ in
(\ref{IRsurfAdSdS3}).
In the present calculation we have used the expanding branch
Wavefunction (\ref{dSact-LG}), (\ref{ehactdS3L}), in contrast
with (\ref{Z1}) used in (\ref{dels}): see also the comments there.

\medskip

\noindent {\underline {\bf Cosmic brane:}}\ \ 
While we have used the smooth replicated geometry (\ref{dS3replica})
to evaluate the Renyi entropies, it is interesting to understand
how this geometry is recast in terms of the boundary variables
quotiented by the $\BZ_n$ replica symmetry. This is most easily seen
by considering the same geometry (\ref{dS3replica}) but featuring
the quotient angle variable ${\tilde\phi}={\phi\over n}$\,, \ie\
\be\label{dS3replicaQuotient}
ds^2 = - {dr^2\over {r^2\over l^2} - {1\over n^2}} +
\Big({r^2\over l^2} - {1\over n^2}\Big) dt^2 + r^2 d{\tilde\phi}^2\,,
\qquad {\tilde\phi}\equiv {\tilde\phi}+2\pi\,.
\ee
Now we see that there is a singularity at $r=0$, with
\be
ds^2 \sim -{dt^2\over n^2} + n^2dr^2+ r^2d{\tilde\phi}^2
\ee
revealing a conical deficit angle. The location $r=0$ in fact
corresponds to the North and South poles of the $S^2$ that we began
with (before conformal transforming to the cylinder). In fact this
quotient metric (\ref{dS3replicaQuotient}) can be recognized as
Schwarzschild $dS_3$ with metric in conventional form written as
\be\label{dS3replicaQuotientSdS3}
ds^2 = -\Big(1-8G_3E - {r^2\over l^2}\Big) dt^2
+ {dr^2\over 1-8G_3E - {r^2\over l^2} } + r^2 d{\tilde\phi}^2\,,
\qquad 8G_3E=1-{1\over n^2}\,,
\ee
with the black hole mass $E$ encoding $n$. In this quotiented
space, the singular locus at $r=0$ which is timelike in the Lorentzian
region (and Euclidean in the hemisphere after Euclideanizing
$t\ra i\tau_E$) can be recognized as the location of the cosmic
brane for $n\neq 1$. In terms of the global structure of the geometry,
the cosmic brane is localized as a timelike curve at the North and
South poles in the Lorentzian part and wraps the great semicircle
in the Euclidean hemisphere. In the $n\ra 1$ limit, this is the
no-boundary extremal surface in $dS_3$ extending out from the
boundary of the subregion along the $t$-direction at the future
boundary.

We have thus recovered the boundary Renyi entropies (\ref{dSrenyi}),
(\ref{dS3renyi}), from a bulk replica calculation via $dS/CFT$. As can
be seen the answer is structurally similar to those in ordinary
unitary 2-dim CFTs with $AdS_3$-like gravity duals: as such, our
analysis can also be realized as an analytic continuation from $AdS_3$
which is not surprising since these are boundary quantities (analogous
to dual operator correlation functions being analytic continuations
\cite{Maldacena:2002vr}). In particular the $n$-dependence appears as
a simple multiplicative factor. This might appear surprising since the
dual CFT here is exotic with pure imaginary central charge $c=i{3
  l\over 2G_3}$, so in this sense our result here is a prediction for
the dual CFT per se. However since we are considering fairly simple
subregions, the boundary replica space is relatively simple (simply a
fatter cylinder as we have seen, Figure~\ref{figreplicadS3}) and in
effect the Renyi entropies are only sensitive to the CFT central
charge and not more intricate CFT data. It would be interesting to
understand this better intrinsically in the CFT.

A more basic question is whether the $dS_3$ replica geometry we have
employed is the dominant one. In the current 3-dim case, it is very
likely that this is in fact the unique geometry with these boundary
conditions pertaining to the maximal subregion. In 3-dimensions, the
only maximally symmetric geometries are de Sitter or Schwarzschild de
Sitter.  We have seen that the replica geometry (\ref{dS3replica}) is
mappable to ordinary $dS_3$ after appropriately reabsorbing the
$n$-factors into the coordinates so the replica space is just the
smooth bulk completion of the boundary replica space. Likewise the
boundary quotiented geometry is precisely Schwarzschild $dS_3$ with
the mass encoding the replica-$n$. For generic non-maximal subregions
or multiple subregions, the replica geometry is likely to be rather
nontrivial, and would be interesting to understand.  In this regard,
it would appear that generalizations of the $AdS_3/CFT_2$ studies in
\cite{Hartman:2013mia}, \cite{Faulkner:2013yia}, might be of relevance
here: naively these suggest that the Renyi entropies would depend on
detailed properties of the 2-dim CFT, nonunitary in the current $dS_3$
case.

\subsection{$dS_4$ replica geometry: boundary Renyi}\label{sec:dS4replica}

Now we will consider replica geometries for $dS_4$-like spaces. The
boundary here is the 3-sphere $S^3$: we consider some equatorial
plane as the boundary Euclidean time slice (which gives an $S^2$)
and consider the maximal subregion comprising say the Northern
hemisphere so its boundary is the $S^1$ interface between the
Northern/Southern hemispheres. Towards constructing the replicated
geometries, we will find it convenient to instead consider hyperbolic
boundary foliations of de Sitter since the corresponding $AdS$
versions are known \cite{Hung:2011nu} and serve as a useful crutch.
Pulling out a conformal factor in the $S^3$ Hopf fibration gives\
$\sin^2\theta\, \Big( d\phi^2 +
{d\theta^2 + \cos^2\theta d\psi^2\over\sin^2\theta} \Big)
\equiv \sin^2\theta ( d\phi^2 + dH_2^2 )$: this is analogous to
the sphere-to-cylinder conformal transformation
(\ref{S2-cyl-confmap}) in the $dS_3$ case.
However we will find it more convenient to consider the boundary
space in the form below (instead of the $\theta$-coordinate)
\be
ds_3^2 = d\phi^2 + dH_2^2 = d\phi^2 + d\chi^2 + \sinh^2\chi\, d\psi^2
\ee
\ie\ as $S^1\times H_2$, with $H_2$ the above 2-dim hyperbolic metric.
Pure $dS_4$ in this hyperbolic boundary conformal frame is
\be\label{dS4-S1xH2}
ds^2 = -\frac{dr^2}{f(r)} + l^2 f(r) d\phi^2 + r^2 dH_2^2\,,\qquad
f(r) = \frac{r^2}{l^2}+1\,,\qquad r=[il,0]\cup[0,R_c]\,.
\ee
Here $r$ is bulk time which is defined along a complex time-contour
comprising a Lorentzian part over $[0,R_c]$\ (with $R_c$ the future
boundary cutoff) and a Euclidean part over $[il,0]$. In the Euclidean
part, the geometry can be mapped to $S^4$ by also Wick rotating
$\chi=i\varphi$ as
\be\label{dS4-S1xH2Eucl}
r=i\rho\,,\ \ \chi=i\varphi\,:\qquad
ds^2 = {d\rho^2\over 1-{\rho^2\over l^2}} +
l^2 \Big(1-{\rho^2\over l^2}\Big) d\phi^2 +
\rho^2 (d\varphi^2+\sin^2\varphi\, d\psi^2)\,,
\ee
which is the Hopf fibration, the $\phi$-circle shrinking
smoothly at $\rho=l$ which is a complex zero of $f(r)$ with $r=il$.

The analytic continuation $l^2\ra -L^2$ applied to (\ref{dS4-S1xH2})
gives
\be
ds^2 = {dr^2\over {r^2\over L^2}-1} + L^2 \Big({r^2\over L^2}-1\Big) d\phi^2
+ r^2 dH_2^2
\ee
which is Euclidean $AdS_4$ (with $\phi\equiv it$ Wick-rotated from 
Lorentzian time $t$) in hyperbolic foliations studied fruitfully in
\cite{Hung:2011nu}: our analysis below of the Renyi story has close
parallels wth the $AdS$ discussions there.

Consider now the $\phi=0$ slice and the maximal subregion defined
by half the $H_2$, its boundary being the $S^1$ parametrized by $\psi$.
We now consider the replica space obtained by gluing $n$ copies of
the 3-dim space along the cuts at the subregions in the $H_2$: since
the subregions run along the entire $H_2$ $\chi$-direction (wrapping
the $\psi$-circle) on the $\phi=0$ slice, the effective geometry
enlarges in the $\phi$-direction which puffs up to have periodicity
$2\pi n$. This is similar to the fatter replica-cylinder described
in the paragraph after (\ref{S2-cyl-confmap}).  Visualizing the
present 3-dim $S^1\times H_2$ replica geometry in detail appears
difficult unfortunately!

We want to construct a bulk replica space with these replica boundary
conditions where $\phi\equiv \phi+2\pi n$\,: so consider (with
$c_1$ a parameter)
\be\label{dS4H2n-rep}
ds^2 = -\frac{dr^2}{f(r)} + l^2 f(r) d\phi^2 + r^2 dH_2^2\,,\qquad
f(r) = \frac{r^2}{l^2}+1+{c_1\over r}\,,\qquad \phi\equiv \phi+2\pi n\,.
\ee
We want to require that this be a smooth geometry: towards this, we
define an auxiliary Euclidean metric by Wick-rotating $r\ra i\rho$
and demand that this Euclidean metric satisfy regularity conditions
in the interior. This criterion for regularity of a de Sitter like
geometry was used effectively in \cite{Das:2013mfa} for a $dS$-brane,
which ended up giving a complex geometry in the $dS_4$ case and we
will see some parallels here: similar arguments appear in
\cite{Maldacena:2019cbz}.\ Wick-rotating gives a $-AdS$ type space
with metric
\be
ds^2 = -\left(\frac{d \rho^2}{-f(\rho)} + l^2 (-f(\rho)) d\phi^2
+ \rho^2 dH_2^2 \right)\,,\qquad
-f(\rho) = \frac{\rho^2}{l^2} -1 + \frac{i c_1}{\rho}\,.
\ee
Now note that $f(r)$ develops a complex zero at $r_h=i\rho_h$ so
\be
f(\rho_h) = 0 \quad\ra\quad c_1 = -i\rho_h \frac{l^2-\rho_h^2}{l^2}\,.
\ee
Now we demand that the $-AdS$ geometry at the complex zero of $f(r)$
is regular with no conical deficit: with
$f(\rho)\sim f'(\rho_h) (\rho-\rho_h)$ near $\rho_h$, regularity
with $\phi$-periodicity $2\pi n$ requires
\be\label{-AdSrhoh}
{l\over 2} f'(\rho_h)\,2\pi n = 2\pi
\quad\Rightarrow\quad \rho_h = \frac{l}{3n} (\sqrt{1+3n^2}+1)\,.
\ee
For $n=1$, this gives $\rho_h=l$ and $c_1=0$, which is pure $dS_4$
in (\ref{dS4-S1xH2}) above.
In terms of the original $r$-variable in (\ref{dS4H2n-rep}), we have
pure imaginary $r_h$ and $c_1$ 
\be\label{dS4n-c1rh}
c_1 = r_h {l^2+r_h^2\over l^2}\,,\quad r_h=i \frac{l}{3n} (\sqrt{1+3n^2}+1)\,,
\qquad f(r_h)=0\,,
\ee
and for $n=1$ we have $r_h=il$ and $c_1=0$.

We would like to evaluate the boundary Renyi entropies (\ref{dSrenyi})
for these $dS_4$-like geometries using the semiclassical $dS/CFT$
dictionary as in the $dS_3$ case. The action is given by
\be
\tilde{I}_n = {1\over 16\pi G_4} \int d^4 x \sqrt{-g} (R-2\Lambda)
- {1\over 8\pi G_4} \int d^3x \sqrt{\gamma} K\,,
\ee
with
\be
  R = \frac{12}{l^2}\,,\quad
  \sqrt{-g} d^4x= l dH_2 d\phi r^2 dr\,, \quad
  \sqrt{\gamma}K d^3x = \left(\frac{3 R_c^3}{l} + 2 l R_c
  +\frac{3l}{2} c_1 \right) dH_2 d\phi\,,
\ee
where we have used the unit normal $n^\mu=(n^r,0,0,0)$ to the $r=R_c$
boundary and the extrinsic curvature $K = \nabla_\mu n^\mu$ as
\be
n^r = \sqrt{\frac{r^2}{l^2}+1 + \frac{c_1}{r}}\,,\quad
K = \frac{1}{\sqrt{-g}} \del_r(\sqrt{-g} n^r) = \frac{3 \frac{r^3}{l^2} +2r + \frac{3 c_1}{2}}{r^2 (f(r))^{\frac{1}{2}}}\,,\quad
\sqrt{\gamma} = l R_c^2 \sqrt{f(R_c)}\,\sinh\chi\,.
\ee
In evaluating the bulk term, we take the bulk (complex) time range as
$r=[R_c,r_h]$, \ie\ from the future boundary cutoff to the complex
zero at $r_h$. This gives
\be
\tilde{I}_n = -\frac{4\pi A n}{16\pi G_4\,l} \left(2 R_c^3 +2 l^2 R_c
+ \frac{3l^2}{2} c_1 + r_h^3\right)\,.
\ee
To remove the cutoff-dependent divergent terms, we add the counterterm
(following well-known holographic renormalization procedures
\cite{Balasubramanian:1999re,Myers:1999psa,deHaro:2000vlm,
  Skenderis:2002wp})
\be
16\pi G_4 I_{ct} = \int \sqrt{\gamma} \left(\alpha R_b + \beta\right),\quad
R_b = -\frac{2}{R_c^2}\,,\quad
\beta = \frac{4}{l}\,,\ \ \alpha = -l\,.
\ee
Adding these, the full regulated action finally becomes (using
(\ref{dS4n-c1rh}))
\be
I_n = \tilde{I}_n + I_{ct} = -\frac{2\pi A n}{16\pi G_4\,l} r_h (r_h^2-l^2)
\ee
where $A=\int dH_2$ is the area of the hyperboloid (which has
interesting features as we will see in detail below). The
semiclassical Wavefunction of the replica space then becomes
\be
\log\Psi_n = i I_n = i \frac{2\pi A n}{16\pi G_4 l}\, r_h (-r_h^2+l^2)\,,
\ee
giving the boundary Renyi entropy (\ref{dSrenyi}) as\
(with $r_h$ in (\ref{dS4n-c1rh}))
\be\label{dS4renyi}
S_n = {i\over 1-n} \frac{2\pi A n}{16\pi G_4 l}\, \Big( r_h (-r_h^2+l^2)
- 2il^3 \Big)\,.
\ee
In the $n\ra 1$ limit, it can be seen that\
$I_n\sim I_1+{I_1\over 2}(n-1)^2$\ so we obtain the boundary
entanglement entropy as
\be\label{dS4renyiEE}
S_1 = -{l^2\over 4G_4} A\ \equiv\
{\pi l^2\over 2G_4} \Big( {iR_c\over l} + 1\Big)\,,
\ee
where the expression on the right is the no-boundary $dS_4$
extremal surface area
(\ref{IRsurfAdSdS4}), but with the imaginary part being $+i$,
obtained via the analytic continuation $L\ra il$\ (analogous to
the comments after (\ref{dS3renyi})).

Thus we see that the area $A$ of the hyperboloid is complex-valued! Of
course this is the analytic continuation of the entanglement entropy
from the $AdS$ hyperbolic black hole \cite{Hung:2011nu}.\
To see how this arises directly in the present de Sitter context, it
is instructive to look at the way $dS_4$ in the hyperbolic foliation
(\ref{dS4-S1xH2}) arises via a nontrivial cutoff surface embedded in
global $dS_4$ for the simplest $n=1$ case. Towards this, we recall
that $dS_4$ is defined \cite{Spradlin:2001pw} as the hyperboloid
surface
\be\label{emmet}
X_1^2+X_2^2+X_3^2 +X_4^2-T^2=l^2 
\ee
in 5-dim Minkowski space. Then global $dS_4$ is given by the
parametrization
\begin{align} 
  &	X_1=l\cosh(\eta) \sin(\alpha) \cos(\beta),\ \ T=l \sinh(\eta),\ \
  X_2 = l\cosh(\eta) \cos(\alpha) \cos(\gamma), \nonumber \\
  & 	X_3 = l\cosh(\eta) \cos(\alpha) \sin(\gamma),\ \
  X_4= l\cosh(\eta) \sin(\alpha) \sin(\beta)\,, \label{pareq}
\end{align}
where we have the Hopf fibration form of the $S^3$ metric.
Thus the metric becomes
\begin{equation}
  ds^2 = -l^2 d\eta^2 + l^2 \cosh[2](\eta) d\Omega_3^2
  = -\frac{d\hat{r}^2}{\frac{\hat{r}^2}{l^2}-1}+ \hat{r}^2 d\Omega_3^2\,,\qquad
  \hat{r} = l \cosh(\eta)\,,
  \label{gds}
\end{equation}	
where the second form is the no-boundary $dS_4$ metric and the
coordinate relation above is in the Lorentzian region.

On the other hand, the hyperbolic $dS_4$ (\ref{dS4-S1xH2}) arises
from the parametrization
\begin{align} 
  & X_1 = \sqrt{r^2+l^2} \cos(\phi),\ \  X_4=  \sqrt{r^2+l^2} \sin(\phi),\ \
  X_2 = r \sinh(\chi) \cos(\psi), \nonumber \\ 
  & X_3 = r \sinh(\chi)  \sin(\psi),\ \  T= r \cosh(\chi) . \label{pareq2}
\end{align}
In the overlapping charts, we match the coordinates obtaining
relations. For instance, matching $T$ in (\ref{pareq}) and (\ref{pareq2})
using (\ref{gds}), and likewise matching $X_1/X_4$ and $X_2/X_3$,
and using those and matching $X_1$ and $X_2$ gives
\bea\label{rhatr-betaphiEtc}
\sqrt{\hat{r}^2-l^2} = r \cosh\chi\,,\qquad
\beta =\phi\,,\quad \gamma=\psi\,,\nn\\
\hat{r} \sin\alpha= \sqrt{r^2+l^2}, \qquad
\hat{r} \cos\alpha = r\sinh\chi\,.
\eea
These then give
\be
\cot(\alpha) = \frac{r}{\sqrt{r^2+l^2}} \sinh\chi\,. \label{req2}
\ee
Using (\ref{rhatr-betaphiEtc}) and (\ref{req2}) we obtain
\be\label{chi-hatr-al}
{1\over\cosh\chi} = \sqrt{\frac{\hat{r}^2 \sin^2\alpha-l^2}{\hat{r}^2-l^2}}\ ,
\qquad r=\sqrt{\hat{r}^2 \sin^2\alpha-l^2}\ .
\ee
These are nontrivial relations for the embedding of the hyperbolic angle
$\chi(\hat{r},\alpha)$ and the radial coordinate $r(\hat{r},\alpha)$
as functions of the global coordinates $\hat{r}, \alpha$.
The full range of $\alpha$ (which is real-valued)
is $\alpha \in [0,\frac{\pi}{2}]$:  this however translates to
complex values for $\chi$ via the embedding relation above in
the Lorentzian part $\hat{r}>l$ of no-boundary $dS_4$. In particular
on a fixed cutoff slice $\hat{r}=R_c$ of no-boundary $dS_4$
(\ref{gds}), we find
\be\label{chimax0}
\sin\al < {l\over R_c} \quad\ra\quad {1\over\cosh\chi}
= i\,\sqrt{\frac{l^2-R_c^2 \sin^2\alpha}{R_c^2-l^2}}\,,
\ee
so that $\chi$ acquires imaginary values. In particular, as $\al\ra 0$,
we have 
\be\label{chimax}
{1\over\cosh\chi_{max}} \sim \frac{i l}{R_c}\,,
\ee
with $\chi_{max}$ the limiting (imaginary) value of $\chi$. On the
other hand, as $\al\ra {\pi\over 2}$ we see that $\chi\ra 0$.
Thus the area of the hyperboloid $H_2$ in the inherited $\chi$-range
here becomes
\be\label{H2areaA}
A = 2\pi \int_{0}^{\chi_{max}} \sinh\chi\, d\chi
= 2 \pi (\cosh\chi_{max}-1) = 2 \pi \left(\frac{-i R_c}{l}-1\right)\,.
\ee
It is worth noting that (\ref{chi-hatr-al}) gives real values
on the strict future boundary $R_c\ra\infty$ (at fixed $\al$) since
${1\over\cosh\chi}\ra\sin\al$\,: the feature (\ref{chimax}) arises
for a fixed boundary screen $R_c$ as we vary $\al$. It is also
interesting to note from (\ref{rhatr-betaphiEtc}) that for fixed
$\hat{r}=R_c\gg l$ we have $r\cosh\chi\sim R_c$ so that the
region (\ref{chimax0}) with imaginary $\cosh\chi$ requires $r$ to
also be pure imaginary, \ie\ we are in the Euclidean region
(\ref{dS4-S1xH2Eucl}). This is also directly seen from the
second relation in (\ref{chi-hatr-al}), when $\sin\al < {l\over R_c}$\,.
In other words, the hyperbolic future boundary screen $r=r_c$
is a nontrivial embedding (\ref{chi-hatr-al}) into global $dS_4$.
This is expected: in the global foliation, we have finite size
$S^3$ slices at $\hat{r}=const$, so hyperbolic cross-sections can
only be obtained with the $S^1\times H_2$ foliations embedded as
nontrivially curved slices.

Using the area $A$ in (\ref{H2areaA}) then corroborates the $n=1$
limit as matching the extremal surface areas (\ref{dS4renyiEE}): thus
finally the boundary Renyi entropies we have evaluated are
(\ref{dS4renyi}) with the area $A$ above. The real part of the Renyi
entropies (\ref{dS4renyi}) using $A$ above is
\be
Re\,S_n = {n\over 1-n} {\pi\over 4G_4\,l}
\left(\rho_h (\rho_h^2+l^2)-2l^3\right) ,
\ee
with $\rho_h$ in (\ref{-AdSrhoh}). In the $n\ra 1$ limit it can be
seen that this becomes half $dS_4$ entropy.

\subsubsection{Real $dS_4$-like hyperbolic cosmologies}

The asymptotically $dS_4$ spacetimes (\ref{dS4H2n-rep}) regarded
as real geometries with $c_1$ real (and $l \phi\ra t$ taken noncompact)
are interesting in their own right, independent of $dS/CFT$: we have
\be\label{dS4H2-realc1}
ds^2 = -\frac{dr^2}{f(r)} + f(r) dt^2 + r^2 dH_2^2\,,\qquad
f(r) = \frac{r^2}{l^2}+1+{c_1\over r}\,.
\ee
These are analogous to the real $dS_4$ bluewall geometry in
\cite{Das:2013mfa} (distinct from the complex $dS_4$ brane there),
and are in a different class from the complex geometry
(\ref{dS4H2n-rep}) relevant for the $dS/CFT$ Renyi analysis.

At large $r$, the spacetime (\ref{dS4H2-realc1}) approaches $dS_4$.
However the interior structure differs depending on $c_1$. For $c_1<0$,
we find a single real zero 
\be
f(r_h)=\frac{r_h^2}{l^2}+1-{|c_1|\over r_h}=0\,.
\ee
This is identical structurally to the blackening factor in the
$AdS_4$ Schwarzschild black hole. However now the singularity at
$r=0$ inside the horizon is timelike, since
\be
r\ra 0:\qquad ds^2 \sim  + {r dr^2\over |c_1|} - {|c_1|\over r} dt^2
+ r^2 dH_2^2\,.
\ee
The extended Penrose diagram is similar to that in the $dS_4$
bluewall, with asymptotic $dS_4$ universes flanked by horizons that
cloak the timelike singularity. As in that case, this resembles the
interior region (within the inner horizons) of the Reissner-Nordstrom
black hole.

On the other hand, $c_1>0$ implies there is no real zero
to the function $f(r)$ so
\be
c_1>0,\ \ r\ra 0:\qquad ds^2 \sim  -{r dr^2\over c_1} + {c_1\over r} dt^2
+ r^2 dH_2^2\,,
\ee
and the $r=0$ locus is now a spacelike, naked singularity.

\subsection{Higher dimensional $dS$ and boundary Renyi}\label{sec:dSd+1replica}

Various features of our analysis for $dS_3, dS_4$ carry over to higher
dimensional $dS$ as well: in an essential way, this amounts to analytic
continuation from an auxiliary $-AdS$ calculation as we describe in
App.~\ref{sec:-AdS}. There are several parallels with the Renyi
analyses in \cite{Hung:2011nu} via $AdS$ hyperbolic black holes.
We describe some aspects directly in $dS$ here.

The embedding formalism that we employed to find the range of $\chi$ in $dS_4$ can be extended to higher dimensional $dS$. Consider a replicated $dS_{d+1}$ with a boundary sphere $S^d$ with the following metric 
\begin{equation}
	d\Omega_d^2 = d\alpha^2 + \sin[2](\alpha) d\phi^2 + \cos[2](\alpha) d\Omega_{d-2}^2
\end{equation}
with $\phi \equiv \phi+2\pi n$. Using the standard arguments as before the above metric is conformally equivalent to 
$S^1 \times H_{d-1}$ where
\begin{equation}
	d H_{d-1}^2 = d\chi^2 + \sinh[2](\chi) d\Omega_{d-2}^2
\end{equation}
So the $dS$ metric here is
\begin{equation}\label{dshighd}
  ds^2 = -\frac{dr^2}{f(r)} + l^2 f(r) d\phi^2 + r^2 dH_{d-1}^2\,,\qquad
  f(r) =\frac{r^2}{l^2} + 1 +\frac{c_1}{r^{d-2}}\,.
\end{equation}
When we embed this metric \eqref{dshighd} into the global $dS$ metric
\begin{equation}
	ds^2 = -\frac{d\hat{r}^2}{\frac{\hat{r}^2}{l^2}-1}+ \hat{r}^2 d\Omega_d^2 \label{Gds}
\end{equation}
we can use the $U(1) \times SO(d-1)$ symmetry of \eqref{dshighd} to fix $d-1$ parameters out of a possible $d+1$. That then leaves us with $2$ parameters in each metric and as a result (\ref{rhatr-betaphiEtc}) will always be valid for any $d$. Consequently the maximum value $\chi_{max}$ will always be given by (\ref{chimax}).

The on-shell action can then be evaluated as before (see also
App.~\ref{sec:-AdS}), yielding the boundary Renyi entropy (\ref{dSrenyi})
\begin{equation}
	S_n = -\frac{i}{1-n} \frac{2\pi n}{l} V_{d-1} \frac{1}{16\pi G_{d+1}} \left(r_h^{d-2} (r_h^2- l^2) - i^d 2l^d\right)\,,
\end{equation}
generalizing (\ref{dS4renyi}) for $dS_4$.
The boundary entanglement entropy in the limit $n \rightarrow 1$
is given by
\begin{equation}
	S_{EE} = -\frac{2\pi}{l} V_{d-1} 2 i\, i^d l^d \frac{1}{16\pi G_{d+1}} = \frac{1}{4G_{d+1}} i^{d-1} l^{d-1} V_{d-1}\,, \label{EEhighd}
\end{equation}
where
\begin{equation}
	V_{d-1} = \Omega_{d-2} \int \sinh[d-2](\chi) d\chi
\end{equation}
is the hyperboloid volume.
As a check of the above, we can verify that this recovers the earlier
$dS_3, dS_4$ results.  Now for $d=4$, \eqref{EEhighd} gives
\begin{align}
	V_3 = -2\pi \left(\frac{R_c^2}{l^2} + \log(\frac{2 R_c}{l}) + \frac{1}{2} - \frac{i \pi}{2}\right) .
\end{align}
So, analogous to (\ref{dS4renyiEE}), we get
\begin{equation}
	S_{EE} = \frac{\pi^2 l^3}{4 G} + i \frac{\pi l^3}{2 G} \left(\frac{R_c^2}{l^2} + \log(\frac{2 R_c}{l})\right) ,
\end{equation}
(with $+i$ in the imaginary parts) and the real part is exactly
half $dS_5$ entropy, matching that in the extremal surface area
(\ref{IRsurfAdSdS}).


\section{Time entanglement in quantum mechanics: review}\label{sec:tE-rev}

We first briefly review various entanglement-like structures in
quantum mechanics discussed in
\cite{Narayan:2022afv,Narayan:2023ebn,Narayan:2023zen}.  The fact that
extremal surfaces anchored at the $dS$ future boundary do not return
suggests that extra data is required in the far past, somewhat
reminiscent of scattering amplitudes or equivalently time evolution
(\ie\ final states from initial states). With this analogy, we
consider the normalized time evolution operator as a generalized
density operator, normalized at any time $t$, in quantum mechanics per
se, independent of $dS$: these lead to various interesting
entanglement-like structures with timelike separations in their own
right. Then based on these studies of toy models without gravity, we
will return to de Sitter space in sec.~\ref{sec:Synth-dSctm}. Then
partial traces lead to a reduced time evolution operator and the
corresponding von Neumann entropy:
\be\label{Ut-rhotA}
{\cal U}(t)=e^{-iHt}\ \ra\
      \rho_t(t) \equiv {{\cal U}(t)\over {\rm Tr}\,{\cal U}(t)}
\quad\ra\quad \rho_t^A = tr_B\,\rho_t\quad\ra\quad
S_A = -tr (\rho_t^A\log\rho_t^A)\,.
\ee
There are sharp parallels with ordinary finite temperature entanglement
structures, except with imaginary temperature $\beta=it$.

A closely related quantity is pseudo-entropy, the entropy of the
reduced transition matrix \cite{Nakata:2020luh}.
Starting with the time evolution operator and including projection
operators onto initial states $|I\ran$ leads to transition
matrices between these initial states and their time-evolved final
states $|F\ran$: then taking partial traces leads to reduced
transition amplitudes and the associated pseudo-entropies: we have
\be\label{tE2-tEpPE}
\rho_{t,|I\ran} = \Big({{\cal U}(t)\,|I\ran\lan I|\over
  {\rm Tr}\, ({\cal U}(t) |I\ran\lan I|)}\Big)
\ \ \xrightarrow{\, {\rm Tr}_B\,}\ \ \rho_{t,|I\ran}^A \equiv
{\cal T}_{F|I}^A = {\rm Tr}_B
\left({|F\ran\lan I|\over {\rm Tr}(|F\ran\lan I|)} \right),\ \
|F\ran={\cal U}(t)|I\ran\,.
\ee
Thus the reduced time evolution operator with projection $\rho_{t,|I\ran}^A$
is equivalent to the pseudo-entropy reduced transition matrix
${\cal T}_{F|I}^A$ when the final state is the time-evolved initial
state, \ie\ $|F\ran={\cal U}(t)|I\ran$.

These can be evaluated explicitly for bipartite systems with the
Hilbert space being characterized by Hamiltonian eigenstates
$|i,i'\ran$ with energies $E_{i,i'}$.
For instance, consider a 2-qubit system with Hamiltonian
$H=E_{ij}|ij\ran$ where $i,j=1,2$, label the energy eigenstate basis
(with $E_{21}=E_{12}$).
Then, with
$\theta_1\equiv -(E_{22}-E_{11})t$ and $\theta_2\equiv -(E_{12}-E_{11})t$,
the normalized time evolution operator and the reduced time evolution
operator obtained by a partial trace over the second qubit are
\be\label{rhotrhotA-2qubit}
\rho_t = \sum_{i,j}
          {e^{-iE_{ij}t}\over \sum_{kl} e^{-iE_{kl}t}}\, |ij\ran\lan ij| 
\quad \ra\quad
\rho_t^A = {\big(1+e^{i\theta_2}\big)
|1\ran\lan 1| + \big(e^{i\theta_1}+e^{i\theta_2}\big) |2\ran\lan 2|
  \over 1+e^{i\theta_1}+2e^{i\theta_2}}\,.
\ee
This generically has complex-valued von Neumann entropy.

Including projections onto states $|I\ran$,\ using the normalization
$|c_{11}|^2+|c_{22}|^2=1$ and redefining\ $|c_{11}|^2=x$,\ we have
from (\ref{tE2-tEpPE}),
\bea\label{rhot12St12}
&& |I\ran=c_{11}|11\ran+c_{22}|22\ran\,,\quad
|F\ran={\cal U}(t)|I\ran=c_{11}e^{-iE_{11}t}|11\ran+c_{22}e^{-iE_{22}t}|22\ran\,,
\nn\\
&& \rho_{t,|I\ran}^1 = {1\over x+(1-x)e^{i\theta}}
\Big(x|1\ran\lan 1|+ (1-x)e^{i\theta}|2\ran\lan 2| \Big)\,,\qquad
\theta=-(E_{22}-E_{11})t=-\Delta E\,t\,,\quad \nn\\
&& S_{t,|I\ran}^1 = -{x\over x+(1-x)e^{i\theta}}\log {x\over x+(1-x)e^{i\theta}}
- {(1-x)e^{i\theta}\over x+(1-x)e^{i\theta}}
\log {(1-x)e^{i\theta}\over x+(1-x)e^{i\theta}}\,.\qquad
\eea
Here $S_{t,|I\ran}^1$ is the pseudo-entropy of the reduced transition
matrix corresponding to states $|I\ran, |F\ran$.
This is a real-valued entropy, periodic in $t$, exhibiting a
divergent value at\ $e^{i\theta}=-{x\over 1-x}$.\ \ 
The maximally entangled thermofield-double (Bell-pair) type state
with $x={1\over 2}$ has a singularity at $e^{i\theta}=-1$,\ 
\ie\ $t={(2n+1)\pi\over E_{22}-E_{11}}$\,.

Now consider the entirely timelike future-past surfaces
\cite{Narayan:2017xca,Narayan:2020nsc}: these suggest some sort of
generalized entanglement between $I^\pm$. By analogy, consider
future-past states, entangling states between a past time slice $P$
and a future time slice $F$,
\be\label{psifpTFD}
|\psi\rangle_{fp} = \sum \psi^{i_n^F,i_n^P} |i_n\ran_F |i_n\ran_P\ .
\ee
The corresponding future-past density matrix
$\rho_{fp} = |\psi\ran_{fp}\lan\psi|_{fp}$ after a partial trace
over the second ($P$) copy then gives a reduced density
matrix with nontrivial entanglement entropy. To see how this works, let
us consider a very simple toy example of a 2-state system in ordinary
quantum mechanics. The action of the Hamiltonian $H$ on these
(orthogonal basis) eigenstates and the resulting (simple) time
evolution are\
$H|k\ran = E_k|k\ran$ with $k=1,2$ and\
$|k\ran_F \equiv |k(t)\ran = e^{-iE_kt}|k\ran_P$.
We consider the $F$ and $P$ slices to be separated by time $t$
and obtain the $F$ state from the $P$ state by time evolution through
$t$.
The future-past TFD state (\ref{psifpTFD}) in this toy case and the
corresponding future-past density matrix are
\bea
&& |\psi\ran_{fp} = {1\over \sqrt{2}} |1\ran_F |1\ran_P +
{1\over \sqrt{2}} |2\ran_F |2\ran_P
= {1\over \sqrt{2}} e^{-iE_1t} |1\ran_P |1\ran_P
+ {1\over \sqrt{2}} e^{-iE_2t} |2\ran_P |2\ran_P\ , \nn\\
&& \rho_{f} = {\rm Tr}_P |\psi\ran_{fp}\lan\psi|_{fp}
= {1\over 2} |1\ran_F\lan 1|_F + {1\over 2} |2\ran_F\lan 2|_F\ .
\eea
We have normalized the coefficients for maximal entanglement at $t=0$.
For nonzero $t$, there are extra phases due to the time evolution but
they cancel in the reduced density matrix obtained by tracing
$|\psi\ran_{fp}\lan\psi|_{fp}$ over the entire second copy as\
$\delta_{ij}\psi_{fp}^{ki}(\psi_{fp}^*)^{lj}$, so we have obtained
an entirely positive structure. This is true for general future-past
states, via the doubling and partial trace. We will explore this
further later.

\section{Time evolution operator, partial traces:
  weak values, autocorrelation functions, future-past entanglement}

In what follows, we will find various relations between the time
evolution operator and partial traces thereof with weak values of
operators localized to subregions, as well as autocorrelation
functions thereof.
We have in mind a bipartite decomposition of the full system: we will
use notation specific to a 2-qubit system for illustration but the
indices will also suffice more generally.

\subsection{Weak values and the reduced time evolution operator}
\label{sec:weak-rte}

\begin{figure}[h] 
\hspace{1pc}
\begin{minipage}[b]{14pc}
  \includegraphics[width=13pc]{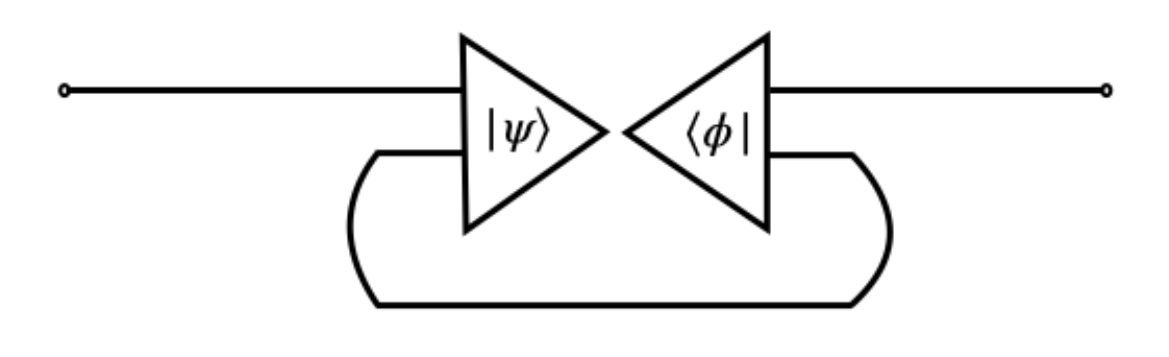} 
  \caption{{\label{figredtrmat} \footnotesize{The reduced transition
  matrix for states $|\psi\ran$ and $|\phi\ran$.} }}
\end{minipage}
\hspace{4pc}
\begin{minipage}[b]{18pc}
\includegraphics[width=17pc]{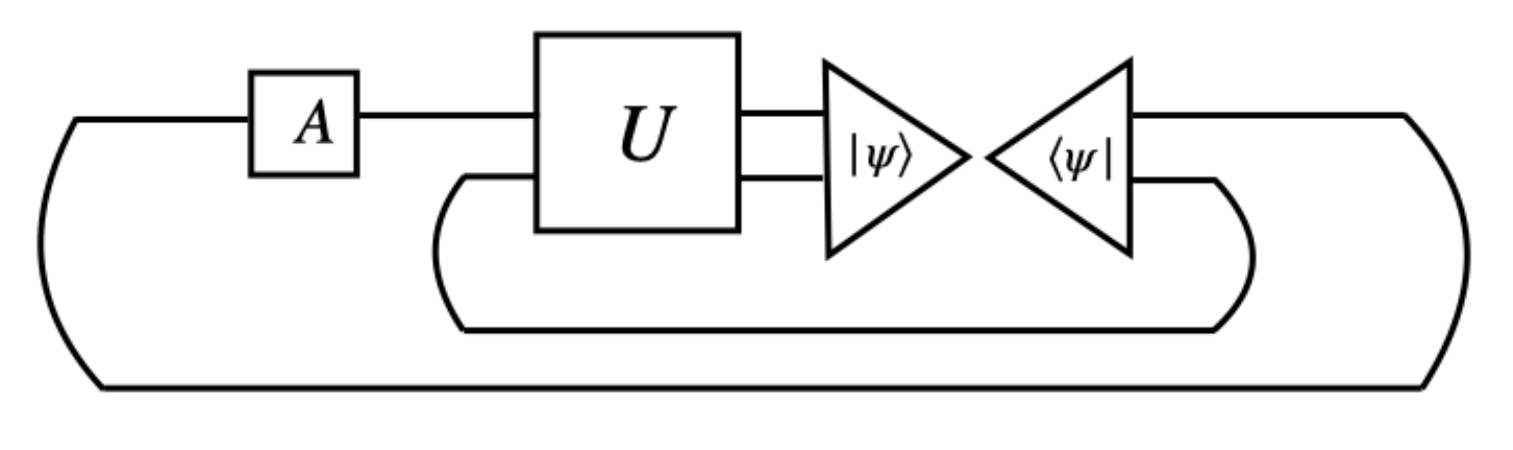} 
\caption{{\label{figweakApsipsi} \footnotesize{The weak value of $A$
      localized to subregion-1 with state $|\psi\ran$ and its
      time-evolution. }}}
\end{minipage}
\end{figure}

Consider two generic states $|\psi\ran$, $|\phi\ran$, and an operator
$A$ localized to subregion-1 (first index):
\be\label{psiphi-basisdecomp}
|\psi\ran = c_{ij} |ij\ran\,,\qquad
|\phi\ran = c'_{ij} |ij\ran\,,\qquad A=A_{\al\beta}|\al\ran\lan \beta|\,.
\ee
The weak value of $A$ in the states $|\psi\ran\,\ |\phi\ran$ is the
(normalized) amplitude 
\be\label{A-weakvalue}
{\lan \phi| A |\psi\ran\over \lan \phi|\psi\ran}
= {\sum_{kl,\ ij} \lan kl| c'^*_{kl}\
A_{\al\beta}|\al\ran\lan\beta|\ c_{ij}|ij\ran \over
\sum_{kl,\ ij} \lan kl| c'^*_{kl}\ c_{ij}|ij\ran}
=  {\sum_{\al,\beta,j}\  A_{\al\beta}\ c'^*_{\al\,j}\ c_{\beta\,j}
    \over \sum_{ij} c'^*_{ij}\ c_{ij}}
= {\rm Tr} (A\, {\cal T}_{\psi|\phi}^1)\,,
\ee
where the index contractions $k=\al,\ l=j,\ \beta=i$, arise from
the fact that $A$ is localized to subregion-1 (first index).\
In the last expression, we have recast the weak value in terms of
\be
{\mathcal T}_{\psi|\phi}^1 = {\rm Tr}_2 \left({|\psi\ran\lan \phi|\over {\rm Tr}(|\psi\ran\lan \phi|)} \right),\qquad
\big({\cal T}_{\psi|\phi}^1\big)_{\al\beta}=
{\sum_{j}\   c'^*_{\al\,j}\ c_{\beta\,j}
    \over \sum_{ij} c'^*_{ij}\ c_{ij}}\,,
\ee
where ${\cal T}_{f|i}^1$ is the reduced transition matrix for two
arbitrary states $|i\ran, |f\ran$ obtained after a partial trace over
subregion-2. Diagrammatically, the reduced transition matrix and the
weak value are represented in Figure~\ref{figredtrmat} and
Figure~\ref{figweakApsipsi}.

The expectation value of $A$ in $|\psi\ran$ is then obtainable by
thinking of $|\phi\ran$ as a post-selected state and summing over
all post-selected states weighted by the amplitude for post-selection:
\be
{\lan\psi| A |\psi\ran\over \lan\psi|\psi\ran}
= \sum_\phi {\lan\psi|\phi\ran\, \lan\phi| A |\psi\ran\over
  {\rm Tr} (|\phi\ran\lan\phi|)\ \lan\psi|\psi\ran}
= \sum_\phi {\lan\psi|\phi\ran\, \lan\phi| A |\psi\ran\over
  \lan\phi|\phi\ran\, \lan\psi|\psi\ran}\,.
\ee

We see that the weak value can be interpreted as the amplitude for
an initial state\ $|\psi\ran$\ ``kicked'' by an operator $A$ to
land in a final state\ $|\phi\ran$.\ It is then interesting to
evaluate this weak value or kick amplitude when the final state
is the time evolution of the initial state, \ie\
$|\phi\ran={\cal U}(t)|\psi\ran = \sum c_{ij} e^{-iE_{ij}t}|ij\ran$.\
A closely related object, the conjugate of the above, is
\be\label{A-weakvalue-psi(t)}
{\lan \psi| A |\psi(t)\ran\over \lan \psi|\psi(t)\ran}
= {\sum_{kl,\,ij} \lan kl| c^*_{kl}\
A_{\al\beta}|\al\ran\lan\beta|\ c_{ij}|ij\ran\, e^{-iE_{i j} t} \over
\sum_{kl,\,ij} \lan kl| c^*_{kl}\ c_{ij}|ij\ran\,e^{-iE_{i j} t} }\
= {\sum A_{\al\beta}\, c^*_{\al j} c_{\beta j}\, e^{-iE_{\beta j} t}\over
  \sum c^*_{i j} c_{i j}\, e^{-iE_{i j} t}} = {\rm Tr} (A\, \rho_{t,|\psi\ran}^1)\,,
\ee
where $\rho_{t,|\psi\ran}^1$ is the reduced time evolution operator
with projection onto the state $|\psi\ran$\ (\ref{tE2-tEpPE}),
obtained by a partial trace over subregion-2 (second index),
\be\label{redTE1}
\rho_{t,|\psi\ran}^1 
= {\rm Tr}_2 (\rho_{t,|\psi\ran})
    = {\sum_j c^*_{\al j}\, c_{\beta j}\, e^{-iE_{\beta j} t}\over
      \sum c^*_{i j}\, c_{i j}\, e^{-iE_{i j} t}}\, |\al\ran\lan\beta|
\equiv {\rm Tr}_2 \Big({|\psi(t)\ran\lan\psi|\over
  {\rm Tr}\, (|\psi(t)\ran\lan\psi|)}\Big)\,.
\ee
The last expression is written to make manifest the fact that the time
evolution operator with projection onto $|\psi\ran$ amounts to the
transition matrix between the initial state $|\psi\ran$ and its time
evolved final state $|\psi(t)\ran={\cal U}(t) |\psi\ran$.

Let us now recall that the time-entanglement (or pseudo) entropy of
this reduced time evolution
operator with projection exhibits singularities at certain specific
time locations\ (\ref{rhot12St12}).
These singularities can be traced to the zeroes of the normalization
factor in the denominator: for instance the generic initial state
$|I\ran=c_{11}|11\ran+c_{22}|22\ran$ gives the inner product in the
normalization
\be\label{normzn-<I|F>-gen}
   {\rm Tr} (|F\ran\lan I|) = \lan I|F\ran =  {1\over 2}
   \big( x\,e^{-iE_{11}t} + (1-x) e^{-iE_{22}t} \big)
= 0 \quad\ra\quad e^{-i(E_{22}-E_{11})t}=-{x\over 1-x}\,.
\ee
Thus the singularity in (\ref{rhot12St12}) essentially stems from the
fact that a generic initial state becomes orthogonal to its
time-evolved final state at specific time locations. Note that it is
necessary that the initial state is not an energy eigenstate:
otherwise the normalization gives a single time evolution phase which
cannot vanish.

We now see that at precisely these time locations, the weak value 
(\ref{A-weakvalue-psi(t)}) also diverges:
\be
\lan A\ran_t \equiv {\lan \psi| A |\psi(t)\ran\over \lan \psi|\psi(t)\ran}
= {\rm Tr} (A\, \rho_{t,|\psi\ran}^1)
= {1\over x+(1-x)e^{i\theta}}
\big( x A_{11} + (1-x) e^{i\theta} A_{22} \big)\,.
\ee
The overall factor leading to the divergence (matching with
(\ref{rhot12St12})) traces back to the inner product
$\lan \psi|\psi(t)\ran$ vanishing, as in (\ref{normzn-<I|F>-gen}).
At $t=0$ this becomes\ $\lan A\ran= x A_{11} + (1-x) A_{22}$,\
the expectation value, while at
$e^{-i(E_{22}-E_{11})t}=-{x\over 1-x}$
the weak value acquires a
singularity, reflecting the singularity in the entropy at these
time locations. Since this holds for generic hermitian operators $A$
(distinct from the identity operator),
this singularity is an intrinsic property of the state and its time
evolution defined by the quantum system.

Along the same lines, it is interesting to note a more general weak
value, involving the time-evolved state $|\psi(t)\ran$ and a distinct
state $|\phi\ran$\,:\ consider
\bea\label{A-weak-psiphi}
&& {\lan \phi|\,A\,|\psi(t)\ran\over \lan \phi|\psi(t)\ran}
= {\rm Tr} (A\, \rho_{t,|\psi\ran,|\phi\ran}^1)\,,\\
&& \rho_{t,|\psi\ran,|\phi\ran}^1 = 
{\rm Tr}_2 \Big({{\cal U}(t)\,|\psi\ran\lan\phi|\over
  {\rm Tr}\, ({\cal U}(t) |\psi\ran\lan\phi|)}\Big)
= {\rm Tr}_2 \Big({|\psi(t)\ran\lan\phi|\over
  {\rm Tr}\, (|\psi(t)\ran\lan\phi|)}\Big)
= {\sum_j c'^*_{\al j}\, c_{\beta j}\, e^{-iE_{\beta j} t}\over
  \sum c'^*_{i j}\, c_{i j}\, e^{-iE_{i j} t}}\, |\beta\ran\lan\al|\,,\nn
\eea
Here we have the time evolution operator along with the transition
matrix $|\psi\ran\lan\phi|$ from the state $|\phi\ran$ to the state
$|\psi\ran$. The last expression is expressed in component form,
using the basis decomposition (\ref{psiphi-basisdecomp}), similar to
the expressions in (\ref{A-weakvalue}), (\ref{A-weakvalue-psi(t)}).

It is of course straightforward to see that at the initial time
slice $t=0$, the weak value (\ref{A-weakvalue-psi(t)}) collapses down
to the expectation value in the initial state $|\psi\ran$,
\be
{\lan \psi|\,A\,|\psi(t)\ran\over \lan \psi|\psi(t)\ran}\Big\vert_{t=0}
= \lan A\ran\,,\qquad \rho_{t,|\psi\ran}^1 = \rho^1_{|\psi\ran}
= {\rm Tr}_2 \Big({|\psi\ran\lan\psi|\over
  {\rm Tr}\, (|\psi\ran\lan\psi|)}\Big)\,,
\ee
and the reduced time evolution operator collapses to the reduced
density matrix for $|\psi\ran$.

\noindent
\underline{\bf Evolution of subregions and the reduced time evolution
  operator}:\ \ It is natural to ask if time evolution for a subregion
is governed by the reduced time evolution operator (\ref{Ut-rhotA})
pertaining to that subregion. To address this, we note that
subregion-1 states can be constructed by starting with states of the
full system but deleting access to information about the subregion-2
components of the states. So amplitudes or inner products of states
with information localized to subregion-1 are obtained by partial
traces over subregion-2, \ie\ with states in (\ref{psiphi-basisdecomp}),
we have
\be
{\rm Tr}_2 \big( \lan \phi|\psi(t)\ran \big) =
{\rm Tr}_2 \big( \lan kj|\, c'^*_{kj}\, c_{ij}\, e^{-iE_{ij}t}\,|ij\ran \big)
\ee
Including the normalization, we see that this is in fact closely
related to the matrix elements of the reduced time evolution
operator appended with the transition matrix projector
$\rho_{t,|\psi\ran,|\phi\ran}^1$ in (\ref{A-weak-psiphi}), \ie\
\be
{\sum_j c'^*_{kj}\, c_{ij}\, e^{-iE_{ij}t}\over
  \ \sum_{ij} c'^*_{ij}\, c_{ij}\, e^{-iE_{ij}t}\ }\ =\
\big( \rho_{t,|\psi\ran,|\phi\ran}^1 \big)_{ik}\,.
\ee
Likewise the matrix elements of the reduced time evolution operator
with projection onto state $|\psi\ran$ are related to the inner
products
\be
{\rm Tr}_2 \big( \lan \psi|\psi(t)\ran \big) \quad\longleftrightarrow\quad
{\sum_j c^*_{kj}\, c_{ij}\, e^{-iE_{ij}t}\over
  \ \sum_{ij} c^*_{ij}\, c_{ij}\, e^{-iE_{ij}t}\ }\ =\
\big( \rho_{t,|\psi\ran}^1 \big)_{ik}\,.
\ee   
In this sense, the reduced time evolution operator controls the
evolution of the subregion, through its matrix elements after partial
traces over subregion-2: we have
\be
{\rm Tr}_2 \big( \lan ij|\, \Big({{\cal U}(t)\over
  {\rm Tr}\,{\cal U}(t)}\Big)\, |kl\ran \big) =
\big( \rho_t^1 \big)_{ik}\,.
\ee

\subsection{Autocorrelation functions: the ${\mathds T}$ operator}
\label{mathdsT}

Noting $|\psi(t)\ran={\cal U}(t) |\psi\ran$ above, we have\
$\lan\psi(t)| \psi(t)\ran = \lan\psi| \psi\ran$. Now consider the
expectation value of the time-dependent Heisenberg picture operator\
$A(t) = {\cal U}(t)^\dag A\ {\cal U}(t)$\ (with $A\equiv A(0)$)
\be\label{A(t)-expecvalue}
{\lan\psi| A(t) |\psi\ran \over \lan\psi|\psi\ran}
= {\lan\psi(t)|\, A\, |\psi(t)\ran \over \lan\psi|\psi\ran}\
=\ {\rm Tr} (A\,{\mathds T}^\rho)\,,\qquad
{\mathds T}^\rho \equiv { |\psi(t)\ran \lan\psi(t)|\over
  {\rm Tr} (|\psi\ran \lan\psi|)}\,.
\ee
This ${\mathds T}^\rho$ operator is of course just the density operator
at general time $t$: writing it in this manner with open legs on
both sides is just convenient to see how expectation values
work for operators localized on subregions.\\
It is worth noting that the operator ${\mathds T}^\rho$ coincides
with the operator $\rho_{t,|\psi\ran,|\phi\ran}$ before the partial trace
in (\ref{A-weak-psiphi}) when $|\phi\ran={\cal U}|\psi\ran$, \ie\
when $|\phi\ran$ is time-evolved from the state $|\psi\ran$. This
is evident directly from (\ref{A(t)-expecvalue}) and the
transition matrix operator inside the ${\rm Tr}_2$-bracket in
(\ref{A-weak-psiphi}), \ie\
\be
{\mathds T}^\rho = { |\psi(t)\ran \lan\psi(t)|\over
  {\rm Tr} (|\psi\ran \lan\psi|)}
\quad\longleftrightarrow\quad
{{\cal U}(t)\,|\psi\ran\lan\phi|\over
  {\rm Tr}\, ({\cal U}(t) |\psi\ran\lan\phi|)} \equiv
\rho_{t,|\psi\ran,|\phi\ran}\,,\qquad 
|\phi\ran={\cal U}(t)|\psi\ran\,.
\ee
Along the lines of the more general weak values in 
(\ref{A-weakvalue}), (\ref{A-weak-psiphi}), we can construct a
more general version of this ${\mathds T}^\rho$ operator as
\be\label{mathds-Tg}
{\mathds T}\equiv { |\psi(t)\ran \lan\phi(t)|\over
  {\rm Tr} (|\psi\ran \lan\phi|)}
=    {{\cal U}(t)\,|\psi\ran \lan\phi|\,{\cal U}(t)^\dag\over
     {\rm Tr} (|\psi\ran \lan\phi|)}\,.
\ee
This is a transition matrix involving time-evolved copies of two
distinct states $|\psi\ran$ and $|\phi\ran$. When the states are
the same, ${\mathds T}={\mathds T}^\rho$,\ \ie\ we obtain the
density matrix at time $t$, but more generally this is a new
object, useful in constructing 2-point autocorrelation functions
and connects with related structures studied in
\cite{Milekhin:2025ycm}.  This ${\mathds T}$-operator is
essentially similar to the transition matrices for pseudo-entropy
\cite{Nakata:2020luh}, except with two copies of the time
evolution operator explicitly stuck in.

\begin{figure}[h] 
\hspace{0.5pc}
\begin{minipage}[b]{9pc}
  \includegraphics[width=8pc]{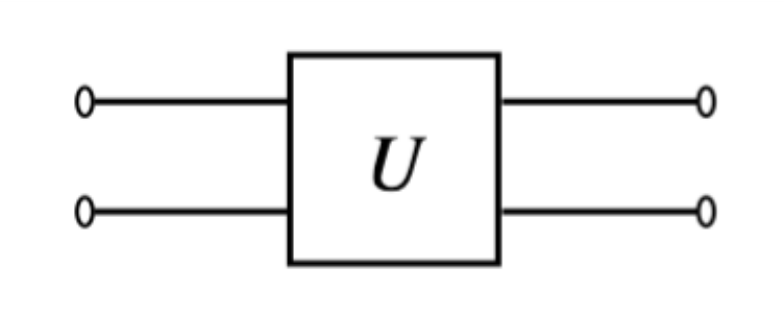} 
  \caption{{\label{figTimeEvOp} \footnotesize{The time evolution
        operator ${\cal U}(t)$.} }}
\end{minipage}
\hspace{4pc}
\begin{minipage}[b]{25pc}
\includegraphics[width=25pc]{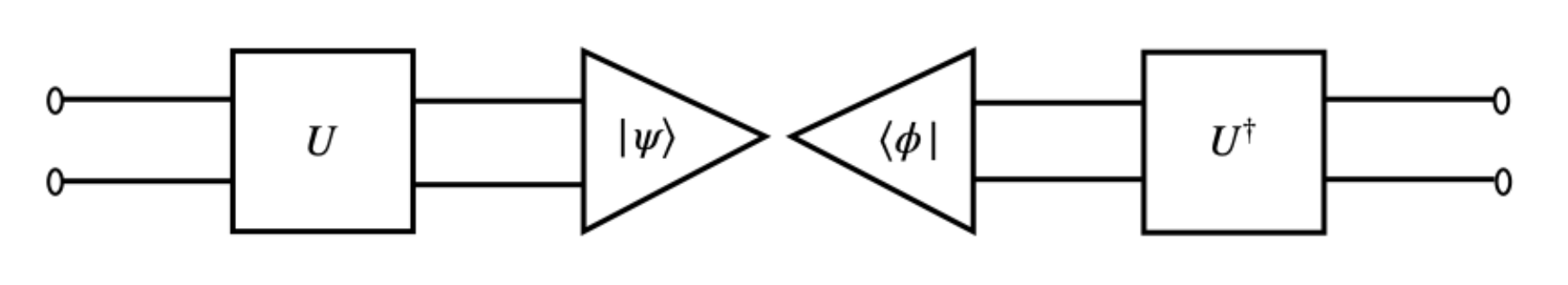} 
\caption{{\label{figMathdsT} \footnotesize{The transition matrix
      operator ${\mathds T}$ for states $|\psi\ran$ and $|\phi\ran$.}}}
\end{minipage}
\end{figure}

Diagrammatically, the time evolution operator and the
time-dependent transition matrix operator ${\mathds T}$ are represented
in Figure~\ref{figTimeEvOp} and Figure~\ref{figMathdsT}.

\bigskip

\noindent \underline{\bf Expectation values:}\ \
Picking an eigenstate basis, we can write this explicitly as
\be\label{mathds-T}
{\mathds T}^\rho\equiv
   {{\cal U}(t)|\psi\ran \lan\psi|\,{\cal U}(t)^\dag\over
     {\rm Tr} (|\psi\ran \lan\psi|)}
= {e^{-iE_{ij}t} c_{ij}\, c_{kl}^*\, e^{iE_{kl}t}
  \over \sum c^*_{ij}c_{ij}}\, |ij\ran \lan kl|
\equiv T^\rho_{ij,kl} |ij\ran \lan kl|\,.
\ee
Unlike the previous cases of weak values that involve a single
copy of the time evolution operator, the operator ${\mathds T}$
here involves two copies of the time evolution operator, acting
from the left and from the right.

With a bipartite decomposition, consider as before an operator
$A$ localized to subregion-1. Then its time-dependent expectation value
(\ref{A(t)-expecvalue}) simplifies since subregion-2 does not enter
and we obtain, using an explicit basis decomposition:
\be\label{mathds-T1-1}
{\lan\psi(t)|\, A\, |\psi(t)\ran \over \lan\psi|\psi\ran}
= {\sum_{\al=k,\, j=l,\,\beta=i}\ \lan kl| c^*_{kl} e^{iE_{kl}t}\
  A_{\al\beta}|\al\ran\lan\beta|\ c_{ij} e^{-iE_{ij}t} |ij\ran\over 
  \sum c^*_{ij}c_{ij}}\
\equiv\ {\rm Tr} (A\, {\mathds T}^{\rho,1})\,,
\ee
where
\be\label{mathds-T1-2}
{\mathds T}^{\rho,1} = {\rm Tr}_2 {\mathds T}^\rho = 
{\sum_j  c_{kj}^*\, e^{iE_{kj}t}\, c_{ij}\, e^{-iE_{ij}t} 
  \over \sum c^*_{ij}c_{ij}}\, |i\ran \lan k|
\ee
is the partial trace over subregion-2 of the ${\mathds T}_\rho$
operator in (\ref{mathds-T}).
It is worth noting that this operator ${\mathds T}^{\rho,1}$ coincides
with the operator $\rho_{t,|\psi\ran,|\phi\ran}^1$ in (\ref{A-weak-psiphi})
when $|\phi\ran={\cal U}|\psi\ran$, \ie\ when $|\phi\ran$ is
time-evolved from the state $|\psi\ran$.

\bigskip

\noindent \underline{\bf 2-point autocorrelation functions:}\ \
To explore these, we first pick an eigenstate basis and write
the ${\mathds T}$-operator (\ref{mathds-Tg}) as
\be\label{mathds-Tg-ijkl}
{\mathds T} = {{\cal U}(t)\,|\psi\ran \lan\phi|\,{\cal U}(t)^\dag\over
     {\rm Tr} (|\psi\ran \lan\phi|)}
=   {e^{-iE_{ij}t} c_{ij}\, c'^*_{kl}\, e^{iE_{kl}t}
  \over \sum c'^*_{ij}c_{ij}}\, |ij\ran \lan kl|
\equiv {\mathds T}_{ij,kl} |ij\ran \lan kl|\,.
\ee
Now consider first the 2-point autocorrelation function
in state $|\psi\ran$ of operator $A$ localized to subregion-1 at time
$t=0$ and operator $B(t)$ localized to subregion-2 at time $t$:\ with
the understanding that $\al, \beta \in$ subregion-1 and $\mu, \nu \in$
subregion-2, we have
\bea
&& \qquad\qquad
A^1 = A_{\al\beta}|\al\ran\lan\beta|\,,\quad B^2 = B_{\mu\nu}|\mu\ran\lan\nu|\,,
\nn\\
&& \lan\psi|\, A(0)\,B(t)\,|\psi\ran
= \lan kl| c^*_{kl}\, A_{\al\beta}|\al\ran\lan\beta|\
{\cal U}^\dag B_{\mu\nu}|\mu\ran\lan\nu|\, {\cal U}\ c_{ij}|ij\ran \nn\\
&& = \sum\ \lan\beta l|\ c^*_{\al l}\, A_{\al\beta}\, e^{iE_{\beta l}t}\
B_{\mu j}\, c_{ij}\, e^{-iE_{ij}t}\,|i\mu\ran
= \sum\ A_{\al\beta}\,c^*_{\al\mu}\ e^{iE_{\beta\mu}t}\, B_{\mu\nu}\,
e^{-iE_{\beta\nu}t}\, c_{\beta\nu}\,.\quad
\eea
We have implemented the index contractions
$\al=k,\ \nu=j,\ \mu=l,\ \beta=i$ to arrive at the final expression.
This is thus recast now as
\be
\lan\psi|\, A(0)\,B(t)\,|\psi\ran = {\rm Tr} \big((T\, A)\, B\big)\,,\qquad
T_{\al\beta\,\mu\nu} =  e^{iE_{\beta\mu}t}\, c^*_{\al\mu}\,
c_{\beta\nu}\, e^{-iE_{\beta\nu}t}\,.
\ee
Then we see that
\be
{\rm Tr} (T\,A)_{\mu\nu} = \sum_{\al\beta} (T_{\al\beta\,\mu\nu}\, A_{\al\beta})
= \sum\ c^*_{\al \mu}\, c_{\beta \nu}\, A_{\al\beta}\,
e^{iE_{\beta \mu}t}\, e^{-iE_{\beta \nu}t}\,.
\ee
We now see that this object can be recovered from a particular
${\mathds T}$ in the basis (\ref{mathds-Tg-ijkl}) as:
\be
(T\,A) = {\rm Tr}_1 ({\mathds T})\,,\qquad
c'^*_{\beta\mu}=c^*_{\al \mu}\,A_{\al\beta}\,,
\ee
where we have taken a partial trace over subregion-1, \ie\ the first
index. In other words, ${\mathds T}$ here comprises the specific
states $|\psi\ran$ and $\lan\phi|=\lan\psi| A$.\ With this, we then
evaluate the 2-point autocorrelation function as an effective
1-point function of $B$. As we see, this is closely related to
the $T_{AB}$ operator discussed for autocorrelation functions in
\cite{Milekhin:2025ycm}.

Diagrammatically, the manipulations above involving the time-dependent
transition matrix operator ${\mathds T}$ and the operators $A$ and $B$
localized on different subregions are represented in
Figure~\ref{figmathdsTABTr2} and Figure~\ref{figmathdsTABTr1}.
Various examples are discussed in Appendix~\ref{appnc}.

\begin{figure}[h] 
\hspace{0.25pc}
  \includegraphics[width=25pc]{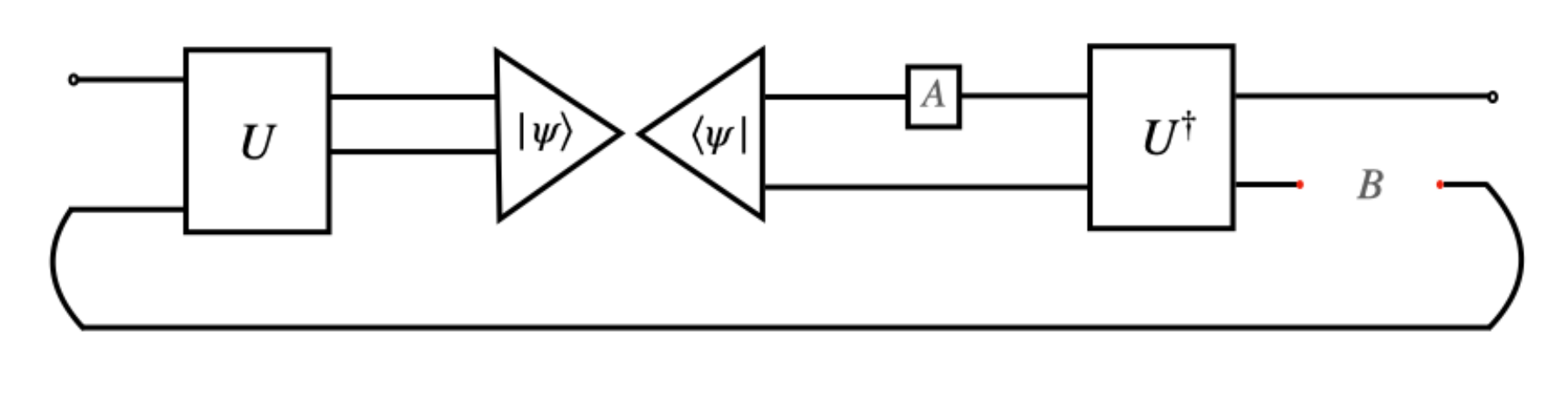} 
\hspace{0.25pc}
\begin{minipage}[b]{13pc}
\caption{{\label{figmathdsTABTr2}\!\! \footnotesize{The transition matrix
        operator ${\mathds T}$ for states $|\psi\ran$ and $|\phi\ran$
        contracted partially (\ie\ with open legs) with operators $A$
        and $B$ localized to different subregions.} }}
\end{minipage}
\end{figure}

\begin{figure}[h]
\hspace{0.25pc}
  \includegraphics[width=25pc]{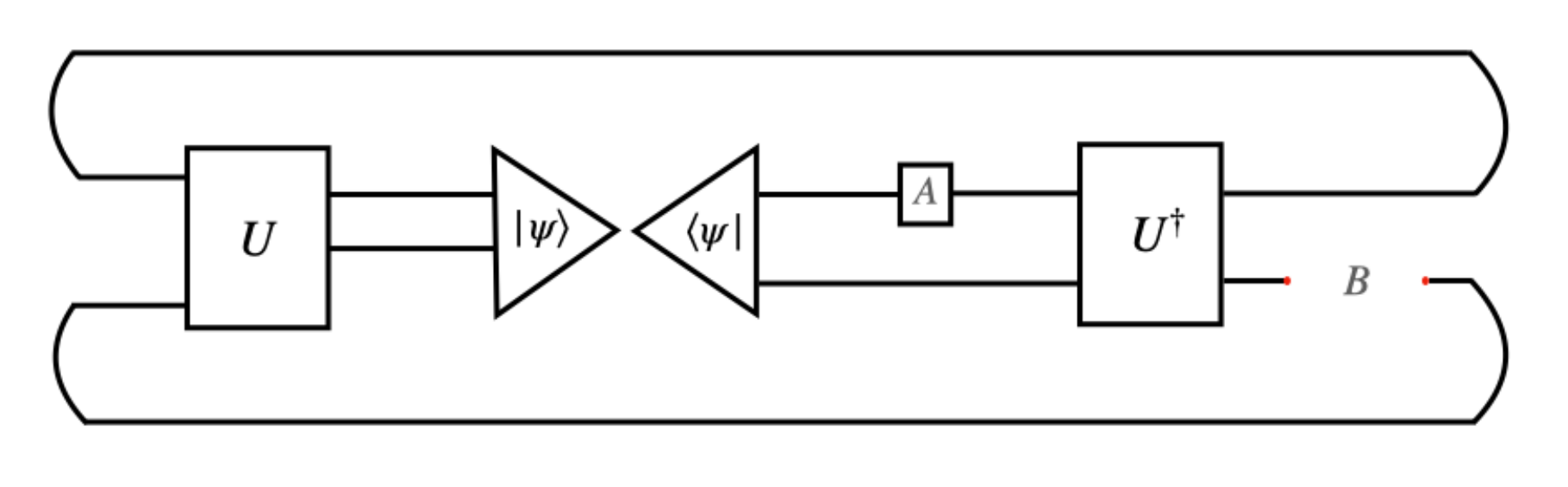}
\hspace{0.25pc}
\begin{minipage}[b]{13pc}
\caption{{\label{figmathdsTABTr1} \footnotesize{The transition matrix
      operator ${\mathds T}$ for states $|\psi\ran$ and $|\phi\ran$
      contracted with operators $A$ and $B$ localized to different
      subregions. With the operator $B$ contracted, this is the
      autocorrelation function.}}}
\end{minipage}
\end{figure}

\subsection{Future-past entangled states, time evolution}\label{sec:fp-fac}

The fact that entirely Lorentzian de Sitter necessitates entirely
timelike future-past extremal surfaces connecting the future boundary
to the past one suggests some sort of future-past entanglement between
two copies of the ghost CFT dual to $dS$ \cite{Narayan:2017xca}.
Future-past thermofield double type states were written down based on
the expectation that these are entirely positive objects in the
ghost-like $CFT_3$ with various negative norm signs cancelling, an
observation borne out in detail in toy ghost-spin models
\cite{Narayan:2016xwq,Jatkar:2017jwz} regarded as microscopic lattice
variables for such ghost CFTs.

From the bulk point of view, this sort of future-past connectedness is
a timelike version of entanglement. Some heuristic observations on
these on this bulk point of view were made in toy models in quantum
mechanics in \cite{Narayan:2022afv,Narayan:2023ebn,Narayan:2023zen}.
We will develop this further here.

Consider the future-past entangled state written in terms of basis
energy eigenstates $|I_P\ran$ at the past time slice and an
isomorphic copy $|I_F\ran$ at the future time slice. Here $I$ is a
label denoting all the qubit degrees of freedom in the system.
We have
\be
|\Psi_{fp}\ran = \sum c_{IJ} |I_F\ran |J_P\ran =
c_{IJ}\ e^{-iE_It}\ |I_P^1\ran |J_P^2\ran\,,
\ee
where we have written the explicit simple time evolution phase arising
since these are basis energy eigenstates. In the second expression we
have written explicit labels $1, 2$, to distinguish the two past
copies in this doubling: from this point of view, we have doubled
the Hilbert space and then separated one copy thereof in time to
the future time slice.

The corresponding future-past density matrix now is
\be
\rho_{fp} = |\Psi_{fp}\ran \lan \Psi_{fp}| =
c_{IJ}\,c^*_{KL}\, |I_F J_P\ran \lan K_F L_P| = 
c_{IJ}\,c^*_{KL}\, e^{-iE_It}\,e^{iE_Kt}\ |I_P^1 J_P^2\ran\ \lan K_P^1 L_P^2|\,.
\ee
The partial trace over the past copy (as written in the form with
explicit $F,P$-labels) is a partial trace over the copy-2 in the
second form with only $P$-labels. This is the index contraction $J=L$
and we have
\bea
&& {\rm Tr}_P (\rho_{fp}) = \sum c_{IJ}\,c^*_{KJ}\, |I_F\ran \lan K_F|\nn\\
&& =\, {\rm Tr}_2 \big(
c_{IJ}\,c^*_{KL}\, e^{-iE_It}\,e^{iE_Kt}\ |I_P^1 J_P^2\ran\ \lan K_P^1 L_P^2| \big)
= c_{IJ}\,c^*_{KJ}\, e^{-iE_It}\,e^{iE_Kt}\ |I_P^1 \ran\, \lan K_P^1|\,.
\eea
The second line expression is of course identical to the first one
noting the basis eigenstate time evolution\
$|I_F\ran = e^{-iE_It} |I_P\ran$. In the first line form, we see a
manifestly positive object recognizable as the reduced density matrix
at the future time slice with von Neumann entropy equal to the
entanglement entropy at the future time slice.

Consider now a bipartite system, with 2-qubits for simplicity:
the label is $I\equiv \{i,j\}$. Then (assuming the states are
normalized to avoid cumbsersome notation)
\bea
&& \rho_{fp} = \sum c_{ij\,ab}\, c^*_{kl\,cd}\,e^{-iE_{ij}t}\,e^{iE_{kl}t}\,
|ij\ran^1 |ab\ran^2\, \lan kl|^1\,\lan cd|^2 \nn\\
\xrightarrow{{\rm Tr}_2} &&
{\rho}_f = \sum c_{ij\,ab}\, c^*_{kl\,ab}\,e^{-iE_{ij}t}\,e^{iE_{kl}t}\,
|ij\ran \lan kl|\,.
\eea

In \cite{Narayan:2023zen}, some observations were made in analogy
with the discussions in \cite{VanRaamsdonk:2010pw},
\cite{VanRaamsdonk:2009ar}, of space emerging from entanglement.
In the current context, an analogy was made relating the connectedness
of time evolution from the past time slice to the future time slice
to the entangled nature of the future-past states above. In particular
a factorized future-past state with just a single nonzero coefficient
$c_{IJ}$ gives
\be
|\psi_{fp}\ran^{fac} = |\psi_F\ran\,|\psi_P\ran\quad\ra\quad
{\rm Tr}_P(\rho_{fp}) = |\psi_F\ran \lan \psi_F|
\ee
which is pure at the future time slice, with zero entropy. Note also
that the time evolution operator ${\cal U}(t)$ has disappeared in the
future-past density matrix. This calculation thus does not care
whether $|\psi_F\ran$ and $|\psi_P\ran$ belong in the same Hilbert
space: the state effectively is $|\psi_F\ran^{(1)} |\psi_P\ran^{(2)}$
with two disconnected spaces $1, 2$, unrelated in particular by
time evolution.

Note that ${\cal H}_F\times {\cal H}_P$\ (motivated by future-past $dS$
surfaces is somewhat different qualitatively from factorizing a
Hilbert space as
${\cal H}={\cal H}_L\times {\cal H}_R$ (as in the $CFT_L\times CFT_R$
dual to the eternal $AdS$ black hole \cite{Maldacena:2001kr}). In
the latter case there is no causal connection between the components
${\cal H}_L$ and ${\cal H}_R$, which are independent. In the current
case ${\cal H}_F$ is the time evolution of ${\cal H}_P$, \ie\ the
time evolution operator acts as the isomorphism.
So a state $|\Psi\ran_{FP}$ is isomorphic to a corresponding state
$|\Psi\ran_{FF}$. Since the time evolution phases cancel, the state
$|\Psi\ran_{FP}$ is future-past entangled if the corresponding
state $|\Psi\ran_{FF}$ is entangled in the ordinary sense on the
(constant) late time slice $F$.
This suggests that nonzero entanglement entropy at the future time
slice arises from non-factorizable future-past states.

The time evolution operator itself can be obtained as a reduced transition
matrix
\be\label{tE-TrTM}
{\cal U}(t)
= \sum_{i=1}^N |i\ran_F\lan i|_P  
= {\rm Tr}_2\, \big( |\psi_{FP}\ran \lan \psi_I| \big)
= \sum_{i,j} \delta_{ij}^{(2)}
\big( |i\ran_F^{(1)}|i\ran_P^{(2)}\, \lan j|_P^{(1)}\lan j|_P^{(2)} \big)\,.
\ee
with thermofield double type states\
$|\psi_{FP}\ran = \sum_i |i\ran_F^{(1)} |i\ran_P^{(2)}$
and $|\psi_I\ran = \sum_i |i\ran_P^{(1)} |i\ran_P^{(2)}$, using
$|I_F\ran = e^{-iE_It} |I_P\ran$.
The time evolution operator acts as the map from the past space to the
future one (which is not independent from the past one). Thus the
existence of the time evolution operator appears to be tantamount to
the existence of future-past entangled states of the form
$|\psi_{FP}\ran$. The existence of future-past entangled states is
thus equivalent to this time-connectedness of the future and past slices.

\subsubsection{Autocorrelation functions, factorization}

It is now natural to ask if a factorized future-past state implies
factorization of the 2-point auto-correlation function between two
operators, in other words no correlations in time.  Naively it would
appear that time-connectedness is required to encode a nontrivial
correlation with time separations. Towards exploring this, consider an
operator $A$ localized to subregion-1 (\ie\ qubit-1) and another
operator $B$ localized to subregion-2 (\ie\ qubit-2) defined as $A=A_{\alpha \beta} \ket \alpha_P \bra \beta_P$ and $B(t)=U^\dagger B_{\gamma \sigma} \ket \gamma_P \bra \sigma_P U$.

To explore the above, let us start by computing the one-point functions of the operators \( A \) and \( B \), and then proceed to evaluate the two-point function. We use the notation \( \ket{i_F j_P} =\sum_{i\ j} C_{ij}\ket{i}_F \ket{j}_P \) to represent future-past states.

The expectation value of an operator \( A \), localized in subregion-1 (at \( t = 0 \)), is given by
\bea \label{A}
&& \hskip -0.5in \langle A(0) \rangle = \bra{\psi_{FP}} A(0) \ket{\psi_{FP}}
= \sum C_{lm}^* \lan l|_F\lan m|_P\ A_{\al\beta} |\al\ran_P\lan\beta|_P\
C_{ij} |i\ran_F|j\ran_P \nn \\ 
&&= \sum_{\substack{
    i,\ j,\ l \\
    m,\ \alpha,\ \beta
}} C_{ij} C_{lm}^* e^{i(E_l - E_i)t} A_{\alpha \beta} \delta_{l i} \delta_{m \alpha} \delta_{\beta j} 
= \sum_{\alpha,\ \beta,\ l} C_{l \beta} C_{l \alpha}^* A_{\alpha \beta},
\eea
where in the third line we used \( \ket{i_F} = e^{-i E_i t} \ket{i_P} \), although we could also proceed by using \( \langle l_F | i_F \rangle = \delta_{li} \).\
Similarly, the one-point function of the operator \( B \), localized in subregion-2 (at \( t \neq 0 \)), is computed as follows:
\bea \label{B}
&& \hskip -0.5in \langle B(t) \rangle = \bra{\psi_{FP}} B(t) \ket{\psi_{FP}} \nn \\
&&= \sum_{\substack{
    i,\ j,\ l \\
    m,\ \gamma,\ \sigma
}} C_{ij} C_{lm}^* e^{i(E_l - E_i)t} e^{i(E_m - E_j)t} B_{\gamma \sigma} \delta_{l i} \delta_{m \gamma} \delta_{\sigma j}
= \sum_{l,\ \gamma,\ \sigma} C_{l \sigma} C_{l \gamma}^* B_{\gamma \sigma} e^{i(E_\gamma - E_\sigma)t}.
\eea

Now, we compute the two-point function of the operators \( A \) and \( B \):
\bea \label{AB}
&& \hskip -0.8in
\langle A(0) B(t) \rangle = \bra{\psi_{FP}} A(0) B(t) \ket{\psi_{FP}} \nn\\
&&= \sum_{\substack{
    i,\ j,\ l,\ m \\
    \alpha,\ \beta,\ \gamma,\ \sigma
}} C_{ij} C_{lm}^* \bra{l_F m_P} \left(A_{\alpha \beta} \ket{\alpha}_P \bra{\beta}_P \right) \left(U^\dagger B_{\gamma \sigma} U \ket{\gamma}_P \bra{\sigma}_P \right) \ket{i_F j_P} \nn \\
&&= \sum_{\substack{
    i,\ j,\ l,\ m \\
    \alpha,\ \beta,\ \gamma,\ \sigma
}} C_{ij} C_{lm}^* A_{\alpha \beta} B_{\gamma \sigma} e^{i(E_l - E_i)t} e^{i(E_\beta - E_j)t} \delta_{m \alpha} \delta_{l i} \delta_{\beta \gamma} \delta_{\sigma j} \nn\\
&& = \sum_{l,\, \alpha,\, \sigma,\, \gamma} C_{l \sigma} C_{l \alpha}^* A_{\alpha \gamma} B_{\gamma \sigma} e^{i(E_\gamma - E_\sigma)t}. 
\eea

From equations (\ref{A}), (\ref{B}), and (\ref{AB}), we see that \( \langle A(0) B(t) \rangle \neq \langle A(0) \rangle \langle B(t) \rangle \) for a non-factorized future-past state \( \ket{\psi_{FP}} = \sum C_{ij} \ket{i}_F \ket{j}_P \).

Let us now consider the case where the future-past state is factorized, i.e., \( \ket{\psi_{FP}} = \ket{i}_F \ket{i}_P \). In this case, the one-point and two-point functions simplify to:
\bea \label{Cij=deltaij}
&& \langle A(0) \rangle = A_{ll}, \qquad
\langle B(t) \rangle = B_{ll}, \nn \\
&& \hskip -0.3in \langle A(0) B(t) \rangle = \sum A_{l \gamma} B_{\gamma l} e^{i(E_\gamma - E_l)t} = A_{ll} B_{ll} + \sum_{l \neq \gamma} A_{l \gamma} B_{\gamma l} e^{i(E_\gamma - E_l)t}.
\eea

Even in this factorized case, we observe that in general \( \langle
A(0) B(t) \rangle \neq \langle A(0) \rangle \langle B(t) \rangle \),
from (\ref{Cij=deltaij}). Hence, a factorized future-past state does
not necessarily imply the factorization of the auto-correlation
function between the two operators. However, upon taking a long-time
average, the second term in \eqref{Cij=deltaij} vanishes, giving
factorization.

We have seen that factorized future-past states continue to lead to
non-factorized 2-point autocorrelation functions unless we perform a
long time average. The long time average reflects the fact that
cross-correlations die for large timelike separations between the
future and past time slices where the two operators are localized.
This is a bit reminiscent of ER=EPR \cite{Maldacena:2013xja} and the
picture of long wormholes, but now in the timelike direction.
To elaborate: the past time slice has a domain of dependence given by
its future lightcone (causal) wedge: the tip of this cone lies some
time to the future of the past time slice. Likewise the future time
slice has a domain of dependence given by its past lightcone (causal)
wedge. Thus the correlation function between operator $A$ localized
on the past time slice and operator $B$ on the future time slice
is expected to factorize only if the operators cannot influence
each other in time. In other words, we require that their domains
of dependence do not overlap, \ie\ the two time slices are adequately
well-separated in time. It appears that this feature reflects in the
above factorization of autocorrelation functions under long time
averaging despite the state being a factorized future-past state.
It would be interesting to explore aspects of timelike ER=EPR further.

\section{Attempting a synthesis: $\Psi_{dS}$ and time entanglement}
\label{sec:Synth-dSctm}

The previous two sections, sec.~5-6, discussed aspects of
entanglement-like structures with timelike separations in a
nongravitational quantum mechanics context. While this is independent
of de Sitter issues per se, it was motivated by those, in
\cite{Narayan:2022afv}. It is interesting to ask if we can import some
of the lessons and technology here returning to de Sitter space.
Let us recall the transition matrix operator ${\mathds T}$-operator
(\ref{mathds-Tg})\ (motivated by \cite{Milekhin:2025ycm}) and
apply it to the de Sitter Wavefunction: thus consider
\be
{\mathds T}_{dS} \equiv \Psi[I^+]\,\Psi^*[I^-]\,,
\ee
which is a ``cosmological transition matrix'' from an initial
$dS$ Wavefunction $\Psi[I^-]$ at the past boundary $I^-$ to the
late time $dS$ Wavefunction $\Psi[I^+]$ at the future boundary
$I^+$\ (this is reminiscent of \cite{Witten:2001kn} and de Sitter
space as a collection of past-future amplitudes).
This is a complicated object as such since the Wavefunction
contains all information about the history of the Universe.
In the context of de Sitter, we imagine it must have some
``boundedness'' property, since de Sitter entropy is finite
(albeit large). Let us examine this in the semiclassical context,
focussing on the simple case when both $\Psi[I^\pm]$ correspond
to the (ground state) Hartle-Hawking Wavefunction, \ie\
$\Psi[I^-]=\Psi[I^+]=\Psi^{HH}_{dS}$.\ Then using (\ref{ehactdS4L}),
App.~\ref{adsds}, we have
\be
dS_4:\qquad
\log \Psi[I^+, R_c] = {\pi l^2\over 2 G_4} + i {\pi R_c^3\over 8 G_4}
\left(-\frac{4}{l} +\frac{6 l}{R_c^2}\right) .
\ee
Using (\ref{DS-dS4replica}), we can associate to ${\mathds T}_{dS}$
an entropy (via replica) as
\be
\Psi[I^-]=\Psi[I^+]=\Psi^{HH}_{dS}\,:\qquad
S_{{\mathds T}_{dS}} =
\left(1 - \frac{R_c}{3} \del_{R_c}\right) \log {\mathds T}_{dS}
= S_{nb} + S_{nb}^* = {\pi l^2\over G_4}\,,\qquad
\ee
with\ $S_{nb} = \left(1- \frac{R_c}{3} \del_{R_c}\right) \log \Psi[I^+] =
{\pi l^2\over 2 G_4} + i\frac{\pi l^2}{2 G_4}\frac{R_c}{l}$\
the no-boundary extremal surface area (\ref{dS4renyiEE}), with
$+i$ as from the expanding branch Wavefunction. Not surprisingly,
$S_{{\mathds T}_{dS}}$ is $dS_4$ entropy arising from the Euclidean
sphere. In effect this is the calculation in App.~\ref{sec:dS-sphere},
but recast in terms of two copies of the Wavefunction (see
App.~\ref{sec:fpRep} for a similar discussion for future-past surface
areas). Likewise we can obtain $dS_{d+1}$ entropy from $S_{{\mathds
    T}_{dS_{d+1}}}$ for $dS_{d+1}$.\ 
Not surprisingly, the transition matrix in this special case is
simply a density matrix\
${\mathds T}_{dS} \equiv \Psi^{HH}_{dS}\,(\Psi^{HH}_{dS})^*$.\
More generally, ${\mathds T}_{dS}$ is a cosmological transition
matrix for general future/past $dS$ Wavefunctions, with the 
corresponding pseudo-entropy being complex-valued. For instance,
considering an entirely Lorentzian de Sitter suggests taking
the past boundary Wavefunction to be the conjugate of the future
one: this gives
\be
\Psi[I^-]=\Psi[I^+]^*\,:\qquad
S_{{\mathds T}_{dS}} =
\left(1 - \frac{R_c}{3} \del_{R_c}\right) \log {\mathds T}_{dS}
= 2 S_{nb} = 
{\pi l^2\over G_4} + i\frac{\pi l^2}{ G_4}\frac{R_c}{l}\,.
\ee
The real part here is $dS_4$ entropy as for the sphere: the
imaginary part is the future-past surface area, App.~\ref{sec:fpRep}.
In some sense, normalizing $\Psi[I^\pm]$ to be the pure phase
from the Lorentzian part alone (\ie\ removing the real hemisphere
part) gives the pure imaginary future-past area above.\ Related
quantities suggest analogs of relative entropy where \eg\
we fix $\Psi[I^-]$ but consider different $\Psi[I^+]$.\\
In this regard, it is worth noting that issues related to the Wheeler
de Witt equation for the Wavefunction and the Hamiltonian constraint
will play important roles in a deeper understanding of how this
transition matrix ${\mathds T}_{dS}$ encodes bulk time evolution
(since in the gravity case the Hamiltonian simply gives the
constraint). It would be interesting to explore these further.


The above discussion pertains to constructing a bulk transition matrix
operator from bulk de Sitter Wavefunctions. Via $dS/CFT$, this is a
transition matrix of the schematic form\
${\mathds T}_{dS} \equiv Z_{CFT}[I^+]\,Z_{CFT}^*[I^-]$, with
$Z_{CFT}=\Psi_{dS}$ the dual partition function of the Euclidean CFT.
Taking $Z_{CFT}[I^+]=Z_{CFT}[I^-]$ makes this a positive object, with
real-valued entropy. In a sense, this appears to have close
interrelations with the future-past density matrix obtained from
future-past entangled states \cite{Narayan:2017xca} in the exotic
ghost-like Euclidean CFT dual (which is spatial with no time). This
was based on various positivity properties that were explicitly
observed in toy ghost-spin models \cite{Jatkar:2017jwz}. (Relatedly it
is worth noting that a single $Z_{CFT}$ copy via replica facilitates
evaluating boundary Renyi entropies: we expect this will dovetail with
direct evaluation via $Tr(\rho_A^n)$.) It would be fascinating to
analyse this carefully to see how bulk time emerges from this dual
perspective, and we hope to report on this in the future.

\section{$dS$ extremal surfaces: other aspects and speculations}

\subsection{$dS$ extremal surfaces, timelike components,
  maximal areas}\label{sec:maximal}

In ordinary holographic entanglement in $AdS$-like spaces, we minimize
over spatial directions and maximize in the time direction (over Cauchy
slices), which then leads to expected RT/HRT extremal surfaces (and
likewise for quantum extremal surfaces) \cite{Wall:2012uf}. In the
no-boundary $dS$
context, this requires some clarification since the space has a top
Lorentzian component joined with a bottom Euclidean hemisphere. 
Surfaces anchored at the future boundary necessarily have timelike
components in the Lorentzian region but are spacelike as expected in
the Euclidean region: so we expect a maximization of the extremal
surface area for the timelike component in the Lorentzian region
but a minimization for the spatial component in the Euclidean region.
This might look reasonable in pure $dS$ naively, but
disconcerting in the slow-roll inflation generalizations where the
inflationary perturbation wiggles appear to not allow such clean
segregations. We will examine this in a preliminary way and find
things to be consistent at least in regions far from the
complexification point where the Lorentzian component of the
spacetime joins with the Euclidean one.

For simplicity let us consider a 3-dim metric of the schematic
form in (\ref{metric-atau-r}),
\begin{equation}
	ds^2 = g_{rr} dr^2 + r^2 (d\theta^2 + \sin^2\theta d\phi^2)\,.
\end{equation}
We are considering a maximal subregion (half-circle) on the boundary
Euclidean time slice given by the equatorial plane $\theta={\pi\over 2}$\,.
Now we see that the nature of the area functional changes depending
on whether we are in the Lorentzian or Euclidean region: we have
\bea\label{ALAE-maximin}
g_{rr}<0: && iA_L = i\int \sqrt{|g_{rr}| dr^2 - r^2 d\phi^2}\,, \nn\\
g_{rr}>0: && A_E = \int \sqrt{g_{rr} dr^2 + r^2d\phi^2}\,.
\eea
In the Euclidean region, we obtain a minimal area surface
stretching along the $r$-direction alone, with no excursions in
the $\phi$-direction (for maximal subregions): this is as for $AdS$
spacelike extremal surfaces.\
In the Lorentzian region $g_{rr}<0$ however, we see that minimization
leads to null surfaces with vanishing area and no bearing on boundary
entanglement entropy in the dual theory (which is admittedly exotic
and ghost-like, but allows complex-valued entanglement generalizations
at least in toy models \cite{Narayan:2016xwq,Jatkar:2017jwz}).
Interpreting this area integral with an appropriate time contour
leads to a maximum total proper time instead. Thus we find
\be
iA_L \ra i\int \sqrt{|g_{rr}|} dr\,, \qquad
A_E \ra \int \sqrt{g_{rr}} dr\,.
\ee
The proper time here being maximized is analogous to ordinary
particle trajectories as timelike geodesics.
This of course fits the structure of the $dS$ area integrals
(\ref{IRsurfAdSdS}). The $dS_4$ and other higher dimensional cases
are similar structurally.

Let us now examine this in the $dS_3$ slow-roll case near the future
boundary: then at large $r$, the area functional (\ref{ALAE-maximin})
contains inside the radical
\be
{1\over r} (1+\epsilon\log r) dr^2 - r^2d\phi^2
\ee
where we have used (\ref{grr-O(epsilon)}) and expanded the slow-roll
contribution $\beta_{>}$ in (\ref{beta>(r)dS3}).
Had this $O(\epsilon)$-correction been negative, it could have
competed with the $(-r^2d\phi^2)$-term potentially spoiling the
timelike nature of the surface and maximization. As it stands, the
small slow-roll correction is positive consistent with the expectation
that the timelike nature near the future boundary in pure $dS$ does
not change much due to the slow-roll perturbations in the Lorentzian
region. We note however that numerically examining $\beta_>(r)$
reveals non-positive behaviour in the vicinity of the complexification
point $r=1^+$: we expect that things are more murky here on the
timelike/spacelike nature of the extremal surface.

For the Euclidean part, we need to continue $\beta_>(r)\ra \beta_<(r)$
valid in the $r<1$ region: then expanding near the nbp (\ie\ far from
the complexification point) gives inside the radical
\be
{1+\epsilon r^2 \frac{1}{4} (-4-2 i \pi +\log (16)) \over 1-r^2}\, dr^2
+ r^2d\phi^2
\ee
Thus the slow roll contribution has a negative real part near $r=0$,
and does not compete with the $(r^2d\phi^2)$-term, thereby vindicating
minimization in the Euclidean region alongwith the slow-roll
corrections. In the vicinity of the complexification point $r\sim 1^-$,
the signs become murky as for $r\sim 1^+$.

Similar things happen for $dS_4$ slow-roll: near the future boundary,
$\beta_{>}(r)$ in (\ref{beta>(r)dS4}) gives a positive
contribution preserving maximization, while $\beta_<(r)$ near the
no-boundary point gives a negative contribution preserving minimization.

It would be interesting to systematize the rules for extremal
surfaces with timelike components.

\subsection{The imaginary part in no-boundary $dS$ areas}
\label{impnba}

We have recovered the no-boundary $dS_3, dS_4$ extremal surface areas
(\ref{dS3renyi}), (\ref{dS4renyiEE}), from the $n\ra 1$ limit of the
boundary Renyi entropies via replica. The imaginary part in these
contains $+i$, which can be obtained via the $AdS$ analytic
continuation $L\ra il$. The Wavefunction in the boundary Renyi
calculation (\ref{dSrenyi}) corresponds to the expanding branch of
$dS$ as we have seen. On the other hand, the areas
(\ref{IRsurfAdSdS3}), (\ref{IRsurfAdSdS4}), with the imaginary part
containing $-i$ could be understood via the $AdS$ analytic
continuation $L\ra -il$. As we saw in (\ref{dels}),
(\ref{DS-dS4replica}), these arise if we instead used the Wavefunction
with the leading divergent term of the opposite sign (which
corresponds to the contracting $dS$ branch). The technical point
is that the areas correspond to codim-2 surfaces whose leading
(imaginary) divergence is one $l^2$-factor less than the leading
divergence in the Wavefunction, which leads to the single relative
minus sign.

In the $dS_3$ case we saw that the boundary-quotient variables
(\ref{dS3replicaQuotientSdS3}) describe a Schwarzschild $dS_3$
geometry and the cosmic brane (which becomes the $dS$ extremal surface
as $n\ra 1$) is the $r=0$ singularity. There is no obvious physical
interpretation of this singularity as a ``physical'' cosmic brane,
nor is one necessary.

However say we engage in the speculation of thinking of these
(complex) areas as amplitudes for cosmic branes. Then the 
no-boundary extremal surface area (taken standalone) with the $-i$ sign
is akin to the amplitude for a cosmic brane
created from ``nothing'', \ie\ with HH no-boundary initial conditions:
but this arises from the contracting branch Wavefunction in the Renyi
calculation as we have seen. Instead if we fix the Wavefunction to
correspond to the expanding branch, then the codim-2 pure $dS_4,
dS_3$ surface areas (\ref{dS3renyi}), (\ref{dS4renyiEE}), 
contain $+i$ and interpreting these as amplitudes for creation
appears to require negative energy. To see this, let us express the
areas as (with $S_0$ half $dS_{4,3}$ entropy respectively)
\be
\psi_{cb}(R_c) = e^{S_0}\,e^{\pm i{S_0\over l} R_c}\,, \quad
S_0={\pi\,l^2\over 2G_4}\,;  \qquad
\psi_{cb}(R_c) = e^{S_0}\,e^{\pm i {2\over \pi}S_0\log(\frac{2 R_c}{l}) }\,,
\quad  S_0={\pi\,l\over 4G_3}\,,
\ee
writing both signs $\pm i$ in the imaginary part.\
Defining $R_c$ as the bulk time endpoint of the cosmic brane, the
Lorentzian timelike part is a pure phase which resembles the phase
from time evolution starting at $t=l$ and ending at $t=R_c$. The
``energy'' is $\mp {S_0\over l}$\,: this is only defined upto
numerical factors (\eg\ by rescaling $R_c\ra k R_c$).
Then the time derivative satisfies
\be
i{\del\over\del R_c} \psi_{cb}(R_c) = \mp {S_0\over l} \psi_{cb}(R_c)\,.
\ee
This reflects a Schrodinger-like time evolution for the amplitude
$\psi_{cb}(R_c)$, evolving forward in time $t_F\equiv R_c$ (of the
future boundary) but with energy being negative if we take $+i$
in the areas. Alternatively we view the cosmic brane as positive
energy but as starting at the future boundary and evolving backward
in time.

The cosmic brane evolving forward in time, from
the no-boundary point (nbp) $r=0$ to the complexification point
$r=1$ (beginning of the Lorentzian region) to the future boundary
$r={R_c\over l}$\,, is encoded in the area integrals in
(\ref{IRsurfAdSdS}) with the time contour
$\int_0^l(\ldots)+\int_l^{R_c}(\ldots)$\ (Figure~\ref{fig3}).\
The opposite sign reflects inverse time evolution.

Likewise consider the no-boundary extremal surface areas 
\cite{Goswami:2024vfl} for slow-roll inflation, with areas to
$O(\epsilon)$ given in (\ref{sr43-area-final}).
From the time contours used in defining the area integrals in the
complex time-plane (Figure~\ref{fig3}), we have the area integrals
arising as in (\ref{slowrollAreadS4dS3}): so this again corresponds
to the cosmic brane evolving forward in time, but with negative
energy in the corresponding Schrodinger evolution equation. 
To see this for the $dS_4$ case in (\ref{sr43-area-final}),
the time derivative gives
\be\label{sr4EEevoln}
i{\del\over\del R_c} \psi_{cb}^{sr4}(R_c) = \mp {S_0\over l} \left( 1 -
{\epsilon\over 6} + \epsilon\,\log{R_c\over l}
\right)\psi_{cb}^{sr4}(R_c)\ .
\ee
For these slow-roll cases there is further $R_c$-dependence, akin
to a driving potential. As $R_c$ increases, the term inside the bracket
increases, albeit suppressed by the slow-roll parameter $\epsilon$:
so this suggests external ``energy'' is being pumped into the
system, consistent with the inflaton rolling down the hill.\
This falls out of a perturbative correction when
\be
\epsilon\,\log{R_c\over l} \sim O(1)\qquad\ie\qquad
R_c\sim l\,e^{1/\epsilon}\,.
\ee
This means the future boundary cannot be taken ``too far'' in the
future to preserve the validity of this $O(\epsilon)$ expression.
Roughly speaking, going too far into the future means inflation
does not remain slow-roll.

For the $dS_3$ slow-roll case in (\ref{sr43-area-final}), the natural
time parameter is perhaps $\log R_c$, giving
\be
i{\del\over\del\log R_C} \psi_{cb}^{sr3}(R_c) = \mp {l\over 2 G_3} \left(
{1\over\pi} - {\epsilon\over 8} + {\epsilon\over 2} \log {R_c\over l} \right)
\psi_{cb}^{sr3}(R_c)
\ee
for evolution.
Qualitatively this is similar to the $dS_4$ slow-roll case, the
increase with $R_c$ suggesting ``energy'' being pumped in, but the
detailed $R_c$/time dependence is different.

It is worth noting two points: (1) the finite real part, which is half
$dS$ entropy, decreases with the slow-roll correction (always
negative; see (\ref{sr43-area-final})), which might suggest that the
number of degrees of freedom has decreased, consistent with the
thinking that the maximum amount of ``stuff'' in the space has
decreased. This stems from $l$-dependence.\ (2) The imaginary part on
the other hand evolves to increase with $R_c$ as above, so ``energy''
is being pumped in, in some sense. This stems from
$R_c$-dependence.\ In this light, it would be interesting to
understand analogs of the entropic c-function, and reinterpret the
(forward or inverse) time evolution in terms of boundary holographic
RG flows. This point of view does not require a cosmic brane
interpretation per se: we are simply trying to recast the evolution
equation for the extremal surface area as an RG evolution equation for
boundary entanglement entropy. This is reminiscent of
\cite{Strominger:2001gp} and inverse RG flows: it would be interesting
to explore this further.

\section{Discussion}\label{sec:Disc}

We have developed further previous work on de Sitter extremal surfaces and
time entanglement structures in quantum mechanics. In the first part,
we first discussed explicit quotient geometries as tools to evaluate
the extremal surface areas. Then we constructed smooth bulk geometries
with replica boundary conditions at the future boundary and evaluated
boundary Renyi entropies in $dS/CFT$.
This puts on a firmer footing previous Lewkowycz-Maldacena replica
arguments \cite{Narayan:2023zen} based on analytic continuation for
the extremal surface areas via appropriate cosmic branes.

We have seen various features in the de Sitter studies, which we
now recap and summarize.\\
(1) The quotient metrics (sec.~\ref{sec:singQuoGeom}) are adequate near
$n=1$ for calculating boundary entanglement entropy via $dS/CFT$ but
are akin to the singular spaces in \cite{Fursaev:2006ih}. In accord
however with $AdS/CFT$ logic (\eg\ for perturbations)
\cite{Headrick:2010zt}, we require smooth bulk geometries at finite $n$.
In sec.~\ref{sec:dSreplica}, we constructed smooth bulk geometries
with replica boundary conditions which we
used to evaluate boundary Renyi entropies in $dS/CFT$.
The bulk calculation via $Z_{CFT}=\Psi_{dS}$ pertains to the
semiclassical de Sitter Wavefunction and thus evaluates pseudo-Renyi
entropies as in eq.(\ref{dSrenyi}), which in general are complex-valued
(technically the semiclassical analysis uses the bulk action alongwith
the Gibbons-Hawking boundary term as well as appropriate counterterms).\\
(2) The $dS_3$ case (sec.~\ref{sec:dS3replica}) is most easily
expressed in the static coordinatization which enables a simple
visualization of the way the replica copies are glued together
(Figure~\ref{figreplicadS3}). There are close parallels with the
$AdS_3$ replica discussion in \cite{Lewkowycz:2013nqa}. Evaluating
(\ref{dSrenyi}) finally results in the boundary Renyi entropies being
pure imaginary, consistent with the imaginary central charge in
$dS_3/CFT_2$ \cite{Maldacena:2002vr}: the $n=1$ limit recovers the
boundary entanglement entropy matching the extremal surface area. In
terms of boundary quotiented variables, these are Schwarzschild $dS_3$
geometries with mass related to $n$. The cosmic brane is then
identified with the conical singularities at the North and South
poles.\\
The Renyi entropies here appear relatively simple, structurally
similar to entanglement entropies appended with simple multiplicative
$n$-factors. These however pertain to simple (maximal) subregions: it
is likely that generic subregions or multiple subregions lead to more
complicated Renyi entropies. It might be interesting to study
appropriate generalizations of the $AdS_3$ case \cite{Hartman:2013mia},
\cite{Faulkner:2013yia}, which might reveal nontrivial aspects in
the present nonunitary $dS_3/CFT_2$ context here.\\
(3) The 4-dim de Sitter geometry (sec.~\ref{sec:dS4replica}) involves
hyperbolic foliations and is a complex geometry which satisfies a
regularity criterion that amounts to requiring a smooth Euclidean
continuation (which amounts to a $-AdS$ space). This is based on a
regularity criterion that was studied for certain complex $dS_4$
branes in \cite{Das:2013mfa} (see also \cite{Maldacena:2019cbz}
for related discussions). The entry of complex metrics here is not
new: this is known from various previous studies in cosmology
including \cite{Hartle:1983ai}, as also discussed more recently in
\cite{Maldacena:2024uhs}. For instance, the regularity criteria
above (as well as in no-boundary $dS$-like geometries) are analogous
to positive frequency rather than reality. It would be interesting
to understand how these interface with \eg\ the KSW criterion
\cite{Kontsevich:2021dmb,Witten:2021nzp}.\\
In this regard, it is worth noting that \eg\ no-boundary pure $dS_4$
is a semiclassical geometry with a semiclassical picture of time,
defined by the time contour in (\ref{dS4-S1xH2}), which goes from
Lorentzian to Euclidean: in these variables, the geometry ``ends'' in
the Euclidean part where $f(r=il)=0$. The corresponding Renyi-replica
geometries (\ref{dS4H2n-rep}) also retain this semiclassical picture
of the time contour, with the geometry ``ending'' at the complex
horizon $f(r_h)=0$ as we have seen in (\ref{dS4n-c1rh}). So while
these are complex geometries, we are thinking of them with the crutch
of the semiclassical time contour and the corresponding geometric
picture of time.  This is consistent and adequate as we have seen for
the relatively simple subregions we have been discussing (as for the
$dS_3$ case also, above), with the expectation that the time contour
dovetails with the geometric picture of time in the semiclassical
regimes near pure de Sitter. However more generally when there exist
competing extremal surfaces, one expects competing complex geometry
saddles for the corresponding Renyi analysis. It would thus be
interesting to understand how to systematize the rules for complex
geometries and time contours in general. Of course these intertwine
with various related issues in quantum cosmology beginning with
\cite{Hartle:1983ai}.\\
The embedding of the hyperbolic foliations into global $dS_4$ is
nontrivial and illustrates how the $n=1$ limit recovers the known
complex-valued extremal surface area. Relatedly, it might be
interesting to understand replicas in global $dS_4$ directly. This
analysis can be also understood via analytic continuation from a
$-AdS$ framework (with close parallels to the $AdS$ hyperbolic black
holes and Renyi analysis in \cite{Hung:2011nu}), and is valid for de
Sitter in any dimension (sec.~\ref{sec:dSd+1replica}).\\
(4) In the end, our analyses of boundary (pseudo-)Renyi entropies via
$dS/CFT$ end up being analytic continuations of appropriate $AdS$
studies. This might appear boring(!) but in fact this is required
for consistency, given the fact that these are entanglement or Renyi
entropies in the boundary $Z_{CFT}$ evaluated via holography.  Indeed
we recall that the extremal surfaces were defined via subregions on
boundary Euclidean time slices and were bulk simulations of boundary
entanglement entropy
\cite{Narayan:2015vda,Narayan:2017xca,Doi:2022iyj,Narayan:2022afv}). Thus
the analytic continuation here is in accord with boundary correlation
functions in $dS/CFT$ obtainable via known analytic continuations from
$AdS$ \cite{Maldacena:2002vr}.\\
(5) We have seen that the imaginary parts of the extremal surface areas
differ depending on the $AdS$ analytic continuation being either
$L_{AdS}\ra \pm il_{dS}$. From the point of view of the codim-2 $dS$
extremal surfaces regarded standalone, choosing $L\ra -il$ gives
(\ref{IRsurfAdSdS}), which reflects the time contour
Figure~\ref{fig3}. However from the point of view of the Renyi
calculations here, we have seen $L\ra il$ arising naturally: in this
context we focus on the expanding branch of the Wavefunction with
$\log\Psi$ scaling as $-i(..)$ in the leading divergence with the
codim-2 area arising as the first subleading term with one
$l^2$-factor less, which thus gives $+i$ in Im(area). There does not
appear to be any inconsistency however, since it is unclear if one
should insist on the timelike+Euclidean cosmic brane in these $dS$
cases here admitting a physical interpretation. For instance the
cosmic brane in the $dS_3$ case arises as the conical singularity
(timelike in the Lorentzian region) in the Schwarzschild $dS_3$
boundary-quotient description (\ref{dS3replicaQuotientSdS3}).
See however the speculative remarks in sec.~\ref{impnba}.\\
(6) The timelike components in the extremal surfaces might appear
novel from the point of view of systematizing the rules for extremal
surfaces. We have seen (sec.~\ref{sec:maximal}) that in particular
they are maximal area, akin to the fact that timelike geodesics for
ordinary massive particles correspond to maximum proper time. In the
no-boundary context, this becomes more interesting since the Euclidean
part continues to be minimal area while the Lorentzian/timelike part
is maximal area.  In the case of no-boundary slow-roll inflation,
there are nontrivial corrections: as we saw, the maximal/minimal area
properties for timelike/Euclidean components continue to hold away
from the complexification point. Thinking of these areas in the
complex-time-plane requires more care towards formulating the
framework of extremal surfaces in such cases.\\
(7) Relatedly, it is worth noting that in the cases we have discussed
pertaining to maximal subregions, the codim-2 extremal surfaces are
uniquely defined with no competing saddles (this is also vindicated by
the analytic continuation from $AdS$). Thus we expect that the replica
geometries and the associated Renyi entropies are unambiguous as the
relevant saddles. The geometries and the corresponding Renyi entropies
we have described corroborate this expectation. However it is unclear
how to rigorously establish this in these sorts of cosmological
backgrounds.  These questions, and those in (6) above, in some sense
might interface with understanding covariant formulations of such
extremal surfaces (perhaps generalizing \cite{Dong:2016hjy}) and
related issues such as analogs of entropy inequalities and subregion
duality in de Sitter like spaces.

$dS/CFT$ suggests exotic nonunitary ghost-like CFT duals to de Sitter
space. Evaluating boundary Renyi entropies in such CFTs directly is an
interesting open question, perhaps developing aspects of
\cite{Headrick:2010zt}, \cite{Hartman:2013mia},
\cite{Faulkner:2013yia}, in the present context. It is also perhaps of
interest to study these in the toy ghost-spin models
\cite{Narayan:2016xwq,Jatkar:2017jwz} where entanglement studies
explicitly reveal complex-valued entropies arising naturally, the
negative norm contributions here leading to imaginary components (we
note that these are also pseudo-entropies strictly speaking since the
adjoint of a state $|\psi\ran$ being nontrivial makes
$\rho_A=Tr_B(|\psi\ran\lan\psi|)$ akin to a reduced transition
matrix).  In the present context, it is of interest to study
(pseudo-)Renyi entropies directly from $Tr(\rho_A^n)$ towards gaining
insights into nonunitary theories. Related questions arise on analogs
of (pseudo-)Renyi entropy inequalities in de Sitter and dual contexts.
Certain observations on $dS$ area/entropy inequalities (and simple
qubit pseudo-entropies) were made in \cite{Narayan:2023zen}: while
these perhaps lead to more questions than answers(!) in de Sitter,
they can be recognized via analytic continuation as encoding the
well-known $AdS$ entropy inequalities. It would be interesting to
explore similar issues with the Renyi entropies here.

In the second part (independent of de Sitter), we studied various
aspects of time entanglement in quantum mechanics, in particular the
reduced time evolution operator, weak values, operators localized to
subregions (sec.~\ref{sec:weak-rte}), and autocorrelation functions.
The transition matrix operator ${\mathds T}$ in (\ref{mathds-Tg})
containing time-evolved versions of two distinct states, and thus two
copies of the time evolution operator, was useful in recasting
correlation functions of operators localized to subregions, and
relates to the spacetime density operator in \cite{Milekhin:2025ycm}.
Finally we discussed aspects of future-past entangled states towards
exploring the way the timelike separations here intertwine with
factorization aspects of autocorrelation functions over long
timescales.

There are various questions that linger in obtaining a deeper
understanding of time entanglement structures in quantum systems, and
in particular of the emergence of time via entanglement-like
structures with timelike separations.  Part of our motivation here was
to connect aspects of time entanglement and pseudo-entropy in
\cite{Doi:2022iyj}, \cite{Narayan:2022afv}, \cite{Narayan:2023ebn},
with some aspects of the studies in \cite{Milekhin:2025ycm}: we hope
to explore and develop this further.

Finally, in sec.~\ref{sec:Synth-dSctm}, we attempted a synthesis of
the quantum mechanical time entanglement aspects with de Sitter space,
and considered a ``cosmological transition matrix''. This involves two
copies of the de Sitter Wavefunction, at the past and the future
boundary, and is reminiscent of the picture in \cite{Witten:2001kn} of
de Sitter space as a collection of past-future amplitudes. We
discussed how de Sitter entropy and the future-past surface area arise
from special cases here. This suggests tantalizing related aspects in
two copies of the dual ghost-like CFT and the emergence of bulk time:
we are engaged in ongoing work in this regard, and hope to report on
it in the future.


\vspace{5mm}

{\footnotesize {\bf Acknowledgements:}\ \ It is a pleasure to thank
  Cesar Agon, Alok Laddha, Alexey Milekhin, Onkar Parrikar, Sabrina
  Pasterski, Sunil Sake, Sandip Trivedi and especially Rob Myers
  and Ronak Soni for useful discussions and comments.  KN thanks the
  hospitality of Perimeter Institute, Waterloo, CA, as this work was
  nearing completion. This work is partially supported by a grant to
  CMI from the Infosys Foundation.  }


\appendix

\section{More on $dS$-like quotient geometries}\label{App:singQuoGeom}

\subsection{Future-past extremal surface area via quotient}\label{sec:fpRep}

Here we discuss quotient geometries for future-past timelike surfaces
stretching between $I^\pm$ \cite{Narayan:2017xca,Narayan:2020nsc} in
entirely Lorentzian de Sitter, reviewed in
sec.~\ref{sec:dSnbRev}. The (pure imaginary) $dS_4$, $dS_3$ areas 
are
\be\label{dS43rev}
(dS_4)\quad S_{fp} = -i\,{\pi l^2\over G_4} {R_c\over l}\,;\qquad
(dS_3)\quad S_{fp} = -i{l\over G_3}\log {R_c\over l}\,.
\ee
Roughly two copies of the no-boundary surfaces (\ref{nbdS43rev})
earlier glued with appropriate time-contours make up the future-past
surfaces \cite{Narayan:2022afv}: the areas of the IR extremal surfaces
satisfy
\be\label{dS43rev2}
S_{fp}=S_{nb}-S_{nb}^*\,,\qquad
{\rm Re}(S_{nb}) = {1\over 2}\cdot dS\ entropy\,,
\ee
encapsulating the fact that the future-past surface area $S_{fp}$
pertains to two copies of the Wavefunction, arising from
$I^+\cup I^-$ glued appropriately at the midslice.
These future-past surface areas can be obtained as follows.
Consider (for concreteness) global $n$-quotient $dS_4$
\begin{equation}
ds^2 = -l^2 d\eta^2 + l^2 \cosh^2\eta \left( d\theta_1^2
+ \frac{1}{n^2} \sin^2\theta_1 d\theta_2^2
+ \cos^2\theta_1 d\theta_3^2 \right),
\label{nsheetdS4fp}
\end{equation}
where the time coordinate has the range $\eta \in [-\eta_0, \eta_0]$
with $\eta_0$ the future/past boundary cutoff. The $\eta$-coordinate
being real essentially restricts the no-boundary $dS_4$ metric
(\ref{metnds3}) to the Lorentzian region automatically since
$r=l\cosh\eta>l$.  At the moment of time symmetry $\eta=0$ we have a
3-sphere of minimal size (for $n=1$).  The
global $dS_4$ geometry (with $n=1$) can be thought of as gluing an
expanding branch de Sitter with $\eta>0$ to the contracting branch
de Sitter  $\eta<0$ at the midslice $\eta=0$.\
This is entirely Lorentzian: the Euclidean hemisphere is absent. The future-past extremal surface is encoded by
a time contour \cite{Narayan:2022afv} that goes from the past
boundary $I^-$ to the future boundary $I^+$ running entirely along the
Lorentzian timelike direction. This gives the pure imaginary area:
the first relation in (\ref{dS43rev2}) implies that the Euclidean
parts of the two no-boundary time contours cancel.

The quotient calculation is along the lines of
sec.~\ref{rcdS3}-\ref{rcdS4}. The quotient space (\ref{nsheetdS4fp})
contains conical singularities emanating from the subregion boundary
as described after (\ref{metnds3}), running along the $\eta$-direction
(time) wrapping the $S^1$ parametrized by $\theta_3$. These are
desingularized by the cosmic brane source term whose action/area is
\begin{equation}
	i S_{sing} = -I_{brane} = -\frac{n-1}{n} \frac{A_{brane}}{4 G}\,,\qquad
	A_{brane} = -2 \pi l^2 i \int_{-\eta_0}^{\eta_0} d\eta \cosh\eta 
	= -4 \pi l R_c\,,
\end{equation}
using $\sinh\eta_0 \approx \cosh\eta_0 = \frac{R_c}{l}$ for
large $\eta_0$. In terms of the $r$-coordinate in (\ref{metnds3}),
we have\ $A_{brane}=2\pi l^2\int r\sqrt{dr^2\over -({r^2\over l^2}-1)}$\ \
which is the imaginary part in (\ref{Abraneds4}).
Thus, as in \eqref{sent}, the total boundary entanglement entropy
obtained is the $dS_4$ area in (\ref{dS43rev}).

We now give an alternate derivation using the on-shell action:
\begin{align}
	S &=\ {1\over 16\pi G_4}\int d^4x \sqrt{-g} \left(\textbf{R}- \frac{6}{l^2}\right) - {1\over 8\pi G_4}\int d^3\sigma \sqrt{\gamma} K  \nonumber \\
	&=\ \frac{6}{l^2} {V_{S^3} l^4\over 16\pi G_4} \int_{0}^{\eta_0} \cosh^3\eta
	- {6 l^2 V_{S^3}\over 16\pi G_4}\tanh\eta_0 \cosh[3](\eta_0) \nn\\
	&\ \qquad\qquad\qquad
	+ \frac{6}{l^2} {V_{S^3} l^4\over 16\pi G_4} \int_{-\eta_0}^{0} \cosh^3\eta
	- {6 l^2 V_{S^3}\over 16\pi G_4}\tanh\eta_0 \cosh^3\eta_0 \nonumber \\
	&=\ S_++S_- =\ -{16 \pi^2 l^2\over 16\pi G_4} \sinh^3\eta_0\
	=\ -{\pi l^2\over G_4} \left(\frac{R_c^2}{l^2}-1\right)^{3/2} .
\end{align}
where $S_+$ corresponds to the contribution coming from the expanding
branch while $S_-$ corresponds to the contribution coming from
contracting branch. In the above we used\
$K= \frac{3 \tanh(\eta_0)}{l}$\,. Also the fact that the normal at the
lower boundary points in an opposite direction relative to the upper
boundary leads to the Gibbons-Hawking-York contribution being the
same at both ends. The final equality arises from
$\cosh\eta_0={R_c\over l}$\,.

It is instructive to recast the above noting that the no-boundary
Wavefunction is
\begin{equation}
\log\Psi_{nb} = S_{hs} - i S_+\,, \qquad \log\Psi_{nb}^* = S_{hs} +i S_-\,,
\end{equation}
where $S_{hs}$ is the Euclidean hemisphere part and $-i S_\pm$ are the
Lorentzian parts. We use $-i S_+$ here (as stated in the comments
after (\ref{dels}), sec.~\ref{adsds}) correlating this sign with the
analytic continuation $L \rightarrow - i l$.
Thus we obtain
\begin{equation}\label{logPsi-Psi*}
	\log\Psi_{nb}-\log\Psi_{nb}^* = -i (S_++S_-) \equiv -i S\,.
\end{equation}
Finally using (\ref{DS-dS4replica}), we obtain the boundary
entanglement entropy,
\begin{align}
	\mathcal{S}_{fp} 
	= \left(1 - \frac{R_c}{3} \del_{R_c}\right) (\log\Psi_{nb}-\log\Psi_{nb}^*)
	\equiv S_{nb} - S_{nb}^* = -i {\pi l^2\over G_4} {R_c\over l}\,.
\end{align}
matching (\ref{dS43rev}).
In this form, using (\ref{logPsi-Psi*}), we see that the future-past
area pertains to the product amplitude
$\Psi_{nb}\, {\Psi^*_{nb}}^{\!\!-1}$ which in a sense encodes
evolving from $I^-$ to the midslice $\eta=0$ and thereon to $I^+$.

\subsection{$dS$ $\ra$ Euclidean sphere: $dS$ entropy}\label{sec:dS-sphere}

In sec.~\ref{sec:fpRep} we recovered the pure imaginary future-past
areas via quotients.  Here we compute the real part of the no-boundary
extremal surface area \eqref{DS-dS4replica}, which satisfies
(\ref{dS43rev2}), as the entropy from a quotient space corresponding
to two copies of the Euclidean hemisphere joined to a full Euclidean
sphere.\
Consider the $n$-quotient $S^4$ metric
\begin{equation}
ds^2=l^2 d\theta^2 + l^2 \cos^2\theta \left(d\theta_1^2
+ \frac{1}{n^2} \sin^2\theta_1 d\theta_2^2 + \cos^2\theta_1 d\theta_3^2\right).
\end{equation}
This can be obtained by Euclideanizing the time $\eta$-direction
in the entirely Lorentzian $dS_4$ metric (\ref{nsheetdS4fp}) as
$\eta\ra -i\theta$. However it is worth noting that the metric
(\ref{nsheetdS4fp}) is equivalent to the Lorentzian part of the
no-boundary metric (\ref{metnds3}) using $r=l\cosh\eta>l$: the above
Euclidean sphere metric then maps to the hemisphere part with
$r=l\cos\theta<l$.

Thus the locus of the conical singularity here maps to that in
the hemisphere part $r<l$ of the no-boundary surface in
(\ref{metnds3}). The cosmic brane action then becomes
\begin{equation}
I_{brane} = \frac{n-1}{n} \frac{l^2}{4 G_4} (2\pi)
\left( 2 \int_0^{\pi/2} d\theta\,\cos\theta \right)
= \frac{n-1}{n} \frac{\pi l^2 }{G_4}\quad\ra\quad
\mathcal{S} = \frac{\pi l^2}{G_4}\,.
\end{equation}
The $2\pi$-factor is from the $\theta_3$-circle while the factor
of $2$ inside the brackets is from the two hemispheres. The extremal
surface area as $n\ra 1$ is obtained via (\ref{sent}), 
which is of course the total $dS$ entropy arising from the two
hemispheres. Note that $I_{brane}$ above in terms of the $r$-coordinate
is ${n-1\over n}$ times 
${V_{S^{1}}\over 4G_{4}} (2 \int_0^l {r\,dr\over \sqrt{1-{r^2\over l^2}}})$
which is twice the Euclidean part of the area integral (\ref{Abraneds4}).
The no-boundary point $r=0$ here is $\theta={\pi\over 2}$.

Likewise $dS_3$ entropy can be obtained from the cosmic brane action
in an $n$-quotient $S^3$ metric\
$ds^2=l^2 d\theta^2 + l^2 \cos^2\theta (d\theta_1^2
+ \frac{1}{n^2} \sin^2\theta_1 d\theta_2^2)$,\ which is essentially
two copies of the hemisphere part of the no-boundary $dS_3$ metric
(\ref{metnds2}) with $n=1$ after quotienting. Similar $dS_{d+1}$ entropy
can be recovered from the $S^{d+1}$ quotient appropriately
Euclideanizing (\ref{dSd+1-nreplica}).

Pictorially, the future-past surface sec.~\ref{sec:fpRep} is a
trumpet-like geometry obtained by removing the Euclidean hemisphere
contribution in the no-boundary areas (\ref{dS43rev}),
(\ref{dS43rev2}): the time contour \cite{Narayan:2022afv} encoding
this is entirely Lorentzian going from $I^-$ to $I^+$. The real
extremal surface in the Euclidean sphere discussed here arises from
a distinct time contour. With the no-boundary surface defined by the
time contour $[nbp \ra 0\ra R_c]$\ (Figure~\ref{fig3}), the real
sphere part here is the time contour
$[nbp \ra 0\ra R_c] + [R_c \ra 0\ra nbp] \equiv
[nbp \ra 0 \ra nbp]$.


\subsection{No-boundary slow-roll inflation, extremal surfaces, quotient}
\label{sec:dSsr-rep}

\begin{figure}[h] 
\hspace{1pc}
\includegraphics[width=22pc]{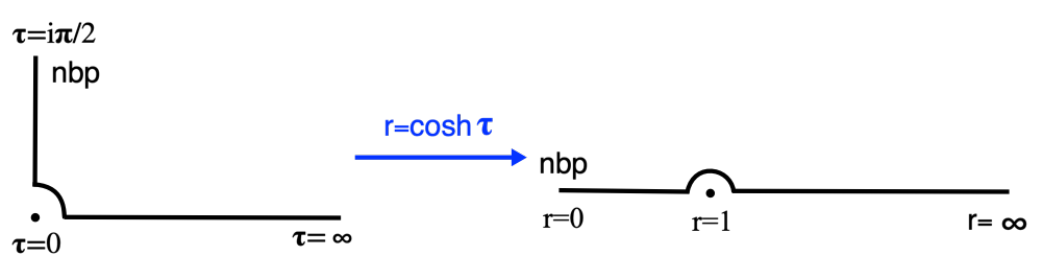}
\hspace{2pc}
\begin{minipage}[b]{12pc}
\caption{{ \label{fig3}
    \footnotesize{Time contour in the complex $r$-plane (right)
      and the $\tau$-plane (left).
        }}}
\end{minipage}
\end{figure}

No-boundary extremal surfaces in slow-roll inflation models were
studied in \cite{Goswami:2024vfl}, with certain inflationary
cosmological perturbations to no-boundary de Sitter space
\cite{Hartle:1983ai} which preserve the spatial spherical isometry of
$dS$ in global coordinates. These perturbations are described by
scalar inflaton perturbations defined imposing regularity at the
no-boundary point: this induces corresponding metric perturbations as
well. The perturbations have explicit analytic (although still
adequately complicated) expressions to $O(\epsilon)$ in the slow-roll
parameter $\epsilon$, as described in \cite{Maldacena:2024uhs} and
related previous work. The inflationary perturbations induce small
wiggles around pure de Sitter, so the metric has various interesting
real and imaginary pieces. Thus the no-boundary extremal surface areas
now have nontrivial real and imaginary pieces which now arise from
both the Euclidean hemisphere and the Lorentzian timelike regions. In
general it turns out that the corresponding area integrals must be
regarded carefully in the complex time-plane defining appropriate
contours (Figure~\ref{fig3}) that avoid extra poles at the
complexification point that arise from the slow-roll perturbations
(similar in spirit to calculations of the semiclasical Wavefunction of
the Universe).  Doing this carefully, we eventually find divergent
pure imaginary pieces from near the future boundary as well as real
and imaginary finite slow-roll corrections to the leading half de
Sitter entropy ${\pi\,l^2\over 2G_4}$ contribution from the
hemisphere. These finite $O(\epsilon)$ corrections precisely match the
finite $O(\epsilon)$ corrections in the expansion of the semiclassical
Wavefunction of the Universe (equivalently the on-shell action) in
slow-roll inflation described in \cite{Maldacena:2024uhs}.  This is
consistent with the Lewkowycz-Maldacena interpretation in
\cite{Narayan:2023zen} of these no-boundary extremal surface areas
giving the probability for cosmic brane creation but now in the
slow-roll no-boundary geometry.

The slow-roll inflation background arises in the Einstein-scalar theory
with action
\bea \label{dSd+1acslr}
& & I={1\over 16\pi G_{d+1}} \int d^{d+1}x \sqrt{g}
\left( R - \left(\del\phi\right)^2-2V(\phi)\right)
-{1\over 8\pi G_{d+1}}\int d^d x\sqrt{h}K\,.
\eea
The Einstein-inflaton equations can be solved perturbatively in
the slow-roll approximation: this has substantial literature
reviewed in \cite{Maldacena:2024uhs} which we refer to (see also
\cite{Goswami:2024vfl} for various details).
We restrict to the minisuperspace approximation with inflationary
perturbations preserving the $S^d$ spherical symmetry:
the spacetime metric near global $dS_{d+1}$ is
\be\label{metric-atau-r}
ds^2 = -dt^2+a(t)^2 d\Omega_d^2\, \equiv\, g_{rr} dr^2 + r^2 d\Omega_d^2\,,
\ee
For pure de Sitter in the above global coordinates, we have\
$a(t) = l\cosh\tau \equiv l\,r$ with $\tau={t\over l}$\ and\
$g_{aa} = {1\over 1-r^2}<0$ in the Lorentzian region $r>1$.
The region $r<1$ gives $g_{aa}>0$ and describes the Euclidean hemisphere.
The metric component $g_{rr}$ to $O(\epsilon)$ gives
\be\label{grr-O(epsilon)}
g_{rr} = {1\over 1-r^2}\big( 1 + 2\epsilon\,\beta_>(r) \big)\,,
\ee
where the slow roll correction $\beta_>(r)$ is a function of the
rolling inflaton profile $\phi(r)$, given below in (\ref{beta>(r)dS4}),
(\ref{beta>(r)dS3}).
The Hamiltonian constraint in an ADM formulation allows expressing the
metric in terms of the inflaton scalar.  Solving for and inputting the
inflaton profile gives the metric (\ref{grr-O(epsilon)}) with
$\beta(r)$ in the $r>1$ Lorentzian region for $dS_4$:
{\small
\be\label{beta>(r)dS4}
\beta_>(r) =
\frac{8-9 r^4+4 i r^2 \sqrt{r^2-1}+8 i \sqrt{r^2-1}-6 i r^4 \sqrt{r^2-1}+r^6 \big(6 \log \big(1-i \sqrt{r^2-1}\big)-1+3 i \pi \big)}{6 r^4 \left(r^2-1\right)}\,.
\ee }
To continue this to the $r<1$ hemisphere region, we note that it is
adequate to replace $-i\sqrt{r^2-1}$ by $\sqrt{1-r^2}$ in $\beta_>(r)$:
this defines $\beta_<(r)$.\ \
Likewise the $dS_3$ slow-roll case gives
\bea\label{beta>(r)dS3}
\beta_>(r) &=&
\frac{1}{32 r^2 \left(r^2-1\right)}
\Big[    -4 r^4 (1+\log 16)+4 r^2-4 \log ^2\big(r+\sqrt{r^2-1}\big)\nn\\
  && \qquad\qquad\qquad +\, 4 \big(2 r \sqrt{r^2-1} \left(2 r^2-1\right)+i \pi \big) \log \big(r+\sqrt{r^2-1}\big) \nn\\
  && \qquad\qquad\qquad -\, 4 i \pi  \big(-2 r^4-r \sqrt{r^2-1} +2r^3 \sqrt{r^2-1} \big)+\pi ^2 \Big]\,.
\eea
The IR no-boundary extremal surface area
$S_{sr_{{d+1}}} = S_{d+1}^{r<1} + S_{d+1}^{r>1}$ after expanding to
$O(\epsilon)$ gives for $dS_4, dS_3$, 
\bea\label{slowrollAreadS4dS3}
&& S_{sr_{{4}}} \simeq\ {\pi\,l^2\over 2G_4} \left(
-i \int_1^{R_c/l} {1+\epsilon\,\beta_>(r)\over \sqrt{r^2-1}}\ r\,dr
+ \int_0^{1} {1+\epsilon\,\beta_<(r)\over \sqrt{1-r^2}}\ r\,dr \right)
\nn\\ [1mm]
&&
S_{sr_{{3}}} \simeq\ {l\over 2G_3} \left(
-i \int_1^{R_c/l} {1+\epsilon\,\beta_>(r)\over \sqrt{r^2-1}}\ dr
+ \int_0^{1} {1+\epsilon\,\beta_<(r)\over \sqrt{1-r^2}}\ dr \right).
\eea
with the leading $dS_{d+1}$ piece and the $O(\epsilon)$ slow roll
correction, where $\beta_<(r)$ is to be obtained by analytically
continuing $\beta_>(r)$ in the Lorentzian region $r>1$ to the
hemisphere region where $r<1$. The $\beta$-factors lead to extra
poles at the complexification point so these areas must necessarily
be defined as complex time-plane integrals with appropriate contours
in the complex time $r$-plane (Figure~\ref{fig3}), as discussed in
detail in \cite{Goswami:2024vfl}. The time coordinate $\tau$ defines
the contour on the left, from the no-boundary point in the Euclidean
region at $\tau=i{\pi\over 2}$ going around the complexification point
at $\tau=0$ and thereon to the Lorentzian region with real $\tau$
going to the future boundary $\tau\ra\infty$. With the coordinate
$r=\cosh\tau$ the nbp is $r=0$ and the complexification point at
$r=1$.
Using the above expressions, we eventually obtain, to $O(\epsilon)$,
the $dS_4$, $dS_3$ slow-roll areas
\cite{Goswami:2024vfl} as
\bea\label{sr43-area-final}
S_{sr_4} \!&=&\! {\pi\,l^2\over 2G_4} \left( -i{R_c\over l} + 1\right)\
+\ \epsilon\, {\pi\,l^2\over 2G_4} \left( -i{R_c\over l} \log {R_c\over l}\
+ i {7\over 6} {R_c\over l}\ +\ \log 4 - {7\over 2} + i\pi  \right) ,\!\!\!\\
S_{sr_3} \!&=&\! {l\over 2G_3}
\Big( {\pi\over 2} -{i\over \pi}\log{R_c\over l} \Big) \nn\\
&&\!\!\!\!\!\!\! +\ \epsilon\,{l\over 2G_3}
\left( - {\pi\over 16} (1 + \log 16)\ +
\ \frac{i}{16} \big(2\log {R_c\over l} - 4(\log {R_c\over l})^2+3\pi ^2 
+ 4 (\log 2)^2-6 \log 2 \big) \right) . \nn
\eea

Towards defining a replica formulation for these areas, it is
instructive to note that the $S^3$ symmetric form (\ref{metric-atau-r})
of the metric here ensures that the $n$-quotient space is
essentially similar structurally to  (\ref{metnds3}): we obtain
\begin{equation}
ds^2= g_{rr} dr^2 + r^2 \left( d\theta_1^2 + \frac{1}{n^2} \sin^2\theta_1 d\theta_2^2 + \cos^2\theta_1 d\theta_3^2 \right) . \label{replica-dS4sr}
\end{equation} 
We see that the singularity structure here is essentially identical
to that in pure $dS_4$ since the replica construction and subsequent
quotienting only involves the $S^3$ part of the space leaving $g_{rr}$
unaffected. Likewise the inflaton is unaffected since it has no
$S^3$-dependence. Thus the required cosmic brane source locus is
structurally similar to that in $dS_4$: the resulting area of course
has the same complications that the unreplicated metric exhibits, and
is given by (\ref{slowrollAreadS4dS3}). The cosmic brane part of the
action\ $iS_{sing} = -I_{brane} =
-{n-1\over n} {A_{brane}\over 4G_4} = -{n-1\over n} S_{sr_4}$\
in the $n\ra 1$ limit then results in the entropy/area $S_{sr_4}$ above.

\section{$dS_{d+1}\ra -AdS$ and boundary Renyi}\label{sec:-AdS}

In this section we will consider boundary Renyi entropies in
higher dim-$dS_{d+1}$ (see sec.~\ref{sec:dSd+1replica})
\begin{equation}
  ds^2 = -\frac{dr^2}{f(r)} + l^2 f(r) d\phi^2 + r^2 dH_{d-1}^2\,,
  \qquad f(r) =\frac{r^2}{l^2} + 1 +\frac{c_1}{r^{d-2}}\,,
\end{equation}
where
\begin{equation}
	dH_{d-1}^2 = d\chi^2 + \sinh[2](\chi) d\Omega_{d-2}^2\,.
\end{equation}
As described for $dS_4$ after (\ref{dS4H2n-rep}), analytically
continuing to $-AdS$ space with $r \rightarrow i \rho$ gives
\begin{equation}
  -f(\rho) = \frac{\rho^2}{l^2}-1-\frac{\omega^d}{\rho^{d-2}}\,,
  \qquad \omega^d = -i^{-d} c_1\,.
\end{equation}
Now at the horizon $\rho_h$ the vanishing of the blackening factor
leads to the condition
\begin{equation}
	\omega^d = \frac{\rho_h^{d}}{l^2}-\rho_h^{d-2}\,. \label{Omeq}
\end{equation}
Demanding that there is no conical deficit at horizon leads to
\begin{equation}
	-\frac{l}{2} f'(\rho_h) n =1
\end{equation}
which yields the solution
\begin{equation}
	\rho_h = l \frac{1+ \sqrt{1+ d (d-2) n^2}}{n d}\ .
\end{equation}
When $n=1$ we get $\rho_h=l$ and as a result $\omega=0$.

The action is given by
\begin{equation}
  \tilde{I}_n = \frac{1}{16\pi G} \int d^{d+1} x \sqrt{g} (R-2\Lambda)
  - \frac{1}{8\pi G} \int d^dx \sqrt{\gamma} K
\end{equation}
Substituting $R = \frac{d(d+1)}{l^2},\ \Lambda = \frac{d(d-1)}{2l^2}$\,,
and
\begin{equation}
	\sqrt{\gamma} K = l \left(d \frac{\rho^d}{l^2} - (d-1) \rho^{d-2} - \omega^d \frac{d}{2} \right) \sinh[d-2](\chi)\,,
\end{equation}
we get
\begin{align}
	\tilde{I}_n =\frac{4\pi n}{l} {V_{d-1}\over 16\pi G} \left((1-d) \rho_c^d + (d-1) \rho_c^{d-2} l^2 - \rho_h^d + l^2 \omega^d \frac{d}{2}\right) .
	\end{align}
To remove the cutoff dependent terms we add the following counterterms
\begin{equation}
	I_{ct} = {1\over 16\pi G}\int d^dx \sqrt{\gamma} \left(\alpha R_b + \beta\right) \label{Ict}
\end{equation}
such that the regulated action is
\begin{equation}
	I_n = \tilde{I}_n + I_{ct}\,.
\end{equation}
It is worth noting that the above counterterms remove all divergences
upto $d=4$, while for $d=5,6$ they only remove the leading and
subleading divergences leaving extra divergences to worry about. One
then needs additional counterterms such as $R_b^2$ to remove these
subsubleading divergences (as in the well-known formulations of
holographic renormalization procedures
\cite{Balasubramanian:1999re,Myers:1999psa,deHaro:2000vlm,
  Skenderis:2002wp}). However the Renyi entropy involves the
combination $I_n - n I_1$, and these divergences get subtracted out:
as a result eq.\eqref{Ict} is enough for our purposes. 
Using
\begin{equation}
	R_b = \frac{(d-1)(d-2)}{\rho_c^2}\,,\qquad\
	\alpha = -\frac{l}{d-2},\quad \beta =2 \frac{d-1}{l}\,,
\end{equation}
we get
\begin{equation}
	I_n-n I_1 = -\frac{2\pi n}{l} {V_{d-1}\over 16\pi G} \left(\rho_h^d + l^2 \rho_h ^{d-2}-2 l^d\right)\,.
\end{equation}
To calculate the boundary Renyi entropy (\ref{dSrenyi}), 
we note now that
\begin{equation}
	\log(\Psi) = i I_{dS} = i^{d+1} I_{-AdS}\,.
\end{equation}
Substituting and simplifying we obtain
\begin{equation}
	S_n =  \frac{1}{1-n} i^{d-1} \frac{2\pi n}{l} {V_{d-1}\over 16\pi G} \left(\rho_h^d + l^2 \rho_h ^{d-2}-2 l^d\right)  
\end{equation}
In the limit $n \rightarrow 1$ we obtain
\begin{equation}
	S_{EE} = \frac{1}{1-n} i^{d-1} \frac{2\pi}{l} {V_{d-1}\over 16\pi G} 2 l^d (1-n) = \frac{1}{4G} i^{d-1} V_{d-1} l^{d-1}
\end{equation}
which is eq.\eqref{EEhighd}.
As is clear, there are various parallels with the Renyi studies
in \cite{Hung:2011nu}.

\section{The on-shell action for $dS_{d+1}$}\label{adsds} 

The $dS_{d+1}$ action is given by
\begin{equation}\label{ehact}
	I_g = \frac{1}{16\pi G} \int d^{d+1} x \sqrt{-g}
	\left(\textbf{R}- 2 \Lambda\right)
	- \frac{1}{8\pi G} \int d^d x \sqrt{\gamma} K\,,
\end{equation}
where $\textbf{R}$, $l$, $\gamma$, $K$ are the Ricci scalar,
$dS_{d+1}$ scale, the induced metric on the boundary and the trace of the
extrinsic curvature respectively.

We are evaluating the Wavefunction for no-boundary $dS_{d+1}$ space with a Lorentzian part evolving from the Euclidean hemisphere. With the (future) boundary at $r=R_c$, the outward pointing normal and the extrinsic curvature are (and $\textbf{R} =\frac{d(d+1)}{l^2}, \Lambda = \frac{d(d-1)}{2l^2}$)
\begin{equation}\label{KdS}
	n^r = \sqrt{\frac{r^2}{l^2}-1}\,,\qquad  
	K = \nabla_{\mu} n^{\mu} = \frac{1}{\sqrt{-g}} \del_r(\sqrt{-g} n^r)
	= \frac{1}{r^d} \sqrt{\frac{r^2}{l^2}-1}\, \del_r (r^d)
	\sim \frac{d}{l} - \frac{d l}{2 r^2}\,.
\end{equation}

Using these in (\ref{ehact}), we obtain the on-shell action for the Lorentzian part as
\begin{align}
	16\pi G\, I^L_g = \Omega_{d} \int_l^{R_c}  dr r^d \frac{1}{\sqrt{\frac{r^2}{l^2}-1}} - 2 \Omega_d \left(\frac{d}{l} - \frac{d l}{2 R_c^2}\right) R_c^d
\end{align}
Depending on whether $d$ is odd or even the integrand above will give different results. For example if $d$ is even there will be a log divergence which will be absent when $d$ is odd. 

Let us mention then the answers for some specific values of $d$ that were used in the text. For $d=2$
\begin{equation}\label{ehactdS3L}
	i I^L_g = {i\over 16\pi G} \left(- 8 \pi  \frac{R_c^2}{l}+4 \pi  l+ 8 \pi  l \log(\frac{2 R_c}{l}) \right)\,,
\end{equation}
while for $d=3$ one gets
\begin{equation}\label{ehactdS4L}
	i I^L_g = {i\over 16\pi G}\,2 \pi^2 R_c^3 \left(-\frac{4}{l} +\frac{6 l}{R_c^2}\right) .
\end{equation}
The Euclidean part for any $d$ is half the volume of the $d+1$-dim sphere
\begin{equation}\label{ehactdSE}
	-I^E_g = \frac{1}{2} {l^{d-1}\over 16\pi G} \Omega_{d+1}\,.
\end{equation}
The total Wavefunction above then pertains to the expanding branch
with $R_c$ the future boundary cutoff and is given by
\begin{equation}\label{dSact-LG}
	\log\Psi = i I^L_g - I^E_g\,,
\end{equation}
with the specific $I^{L,E}_g$ for $dS_3, dS_4$ as above.

\bigskip


The arguments above are intrinsic to de Sitter but they can be
obtained via analytic continuation from $AdS$.
Consider the Euclidean $AdS_3$ action with the Einstein-Hilbert and
Gibbons-Hawking terms:
\begin{equation}
	16\pi G_3\, S= -\int d^3x \sqrt{g} \left(\textbf{R} +\frac{2}{L^2}\right) - 2\int d^2\sigma \sqrt{\gamma} K\,, \label{ehact3}
\end{equation}
The Ricci scalar here in $AdS_3$ is $\textbf{R} = -\frac{6}{L^2}$.
The global $EAdS_3$ metric is given by
\begin{equation}\label{gads2}
	ds^2 = L^2 d\tau^2 + L^2 \sinh^2\tau (d\theta^2+\sin^2\theta d\phi^2)\,,
\end{equation}
where the range is $\tau=(0,\tau_0)$ with boundary at $\tau=\tau_0$.
The trace of the extrinsic curvature is $K=\frac{2 \coth(\tau_0)}{L}$\,.
The on-shell action for the $EAdS_3$ metric (\ref{gads2}), along the
same lines as for $dS_3$ earlier, is obtained as
\begin{equation}
	-S= {4\pi L\over 16\pi G_3} (2\tau_0 + \sinh(2\tau_0)),
\end{equation}
Using the analytic continuation from $AdS_3 \rightarrow dS_3$ as
\begin{equation}\label{AdSdSL-il}
	\tau_0 = \eta_0 + \frac{i \pi}{2}\,,\qquad L = -i l\,,
\end{equation}
and $l \cosh(\eta_0) = R_c$, we obtain for large $R_c$ the
semiclassical Wavefunction $\Psi_1 \sim e^{-S}$ as (\ref{Z1}), which
then gives (\ref{dels}).\ 
On the other hand, the analytic continuation $L \ra il$ gives the Wavefunction
(\ref{dSact-LG}) for the expanding branch and is apt for evaluating
correlation functions (see \eg\ \cite{Dey:2024zjx})).

Likewise $EAdS_4$ with\
$ds^2 =L^2( d\tau^2 + \sinh[2](\tau) d \Omega_3^2)$
and boundary $\tau_0$ has action
\be
	-S = \log Z = \int d^4x \left(\textbf{R}+\frac{6}{L^2}\right) +2 \int d\sigma \sqrt{\gamma} K 
	= L^2 V_{S^3}  \left(\cosh(3\tau_0)+3\cosh(\tau_0)-4\right).	
\ee

\section{Autocorrelation functions, ${\mathds T}$ operator: examples}\label{appnc}
In the following, we illustrate the explicit construction of the ${\mathds T}$ operator through a series of concrete examples. We consider a two-state system with the Hamiltonian\footnote{We define $S_{(x,\ y,\ z)} = \frac{\hbar}{2}\, \sigma_{(x,\ y,\, z)}$. For calculation simplicity, we set $\hbar = 2$ in the rest of this section. Further, superscripts ``1'' and ``2'' correspond to subregions one and two everywhere appearing in this section.}
\begin{equation}\label{ham}
    H= -J\, S_{z}^1  \otimes S_{z}^2,
\end{equation}
where $J$ is a coupling constant and $S_z$ is the spin operator. In the $S_z$ eigenbasis ($\ket{0}$, $\ket{1}$), the Hamiltonian (\ref{ham}) takes the following form
\begin{equation*}
    H = -J\left( \ket{00}\bra{00} - \ket{01}\bra{01} - \ket{10}\bra{10} + \ket{11}\bra{11} \right).
\end{equation*}

Since correlation functions involving only $S_z$ are time-independent, we focus only on those, where either of $S_x$ or $S_y$ is time-evolved. 
To compute the autocorrelation functions, we need the time-evolved operators defined as below
\bea \label{S's}
& & S_x^2(t) \equiv e^{iHt} (\mathbb{I}^1 \otimes S_x^2(0)) e^{-iHt}\, ; \quad \quad  \quad  \quad 
 S_y^2(t) \equiv e^{iHt} (\mathbb{I}^1 \otimes S_y^2(0)) e^{-iHt}.
\eea
In the $S_z$ eigenbasis, the operators $S_x$ and $S_y$ are given by
\begin{align} \label{S's-i}
  \mathbb{I}^1 \otimes S_x^2 &= \ket{00}\bra{01} + \ket{01}\bra{00} + \ket{10}\bra{11} + \ket{11}\bra{10}, \nn \\
  \mathbb{I}^1 \otimes S_y^2 &= -i \ket{00}\bra{01} + i \ket{01}\bra{00} - i \ket{10}\bra{11} + i \ket{11}\bra{10}.
\end{align}

Using standard Pauli matrix identities, we can express the time evolution operator as
\begin{equation} \label{TEE-ACF}
    e^{iHt} = (\mathbb{I}^1 \otimes \mathbb{I}^2) \cos(Jt) - i (S_z^1 \otimes S_z^2) \sin(Jt).
\end{equation}
Using (\ref{S's}), (\ref{S's-i}), and (\ref{TEE-ACF}), we compute various autocorrelation functions for a generic state $\ket{\psi} = \sum_{ij} C_{ij} \ket{ij}$, and extract the corresponding components of the ${\mathds T}$ operator. Some examples are given here

\paragraph{Case 1: }
\begin{equation}\label{sxsx}
\langle S_x^1(0) S_x^2(t) \rangle= \bra{\psi}\, S_x^1(0)\, S_x^2(t)\, \ket{\psi} = (C^*_{00} C_{11} + C^*_{11} C_{00}) e^{2iJt} + (C^*_{01} C_{10} + C^*_{10} C_{01}) e^{-2iJt}.
\end{equation}
This yields the non-zero ${\mathds T}$ components:
\[
T_{0101} = C^*_{00} C_{11} e^{2iJt},\quad T_{1010} = C^*_{11} C_{00} e^{2iJt},\quad
T_{0110} = C^*_{01} C_{10} e^{-2iJt},\quad T_{1001} = C^*_{10} C_{01} e^{-2iJt}\,.
\]

\paragraph{Case 2:}
\begin{equation}\label{szsx}
\langle S_z^1(0) S_x^2(t) \rangle=\bra{\psi}\, S_z^1(0)\, S_x^2(t)\, \ket{\psi} = (C^*_{01} C_{00} - C^*_{10} C_{11}) e^{2iJt} + (C^*_{00} C_{01} - C^*_{11} C_{10}) e^{-2iJt}.
\end{equation}
Resulting non-zero ${\mathds T}$ components:
\[
T_{0010} = C^*_{01} C_{00} e^{2iJt},\quad T_{1011} = -C^*_{10} C_{11} e^{2iJt},\quad
T_{0001} = C^*_{00} C_{01} e^{-2iJt},\quad T_{1110} = -C^*_{11} C_{10} e^{-2iJt} \, .
\]

\paragraph{Case 3:}
\begin{equation}\label{szsy}
\langle S_z^1(0) S_y^2(t) \rangle=\bra{\psi}\, S_z^1(0)\, S_y^2(t)\, \ket{\psi} = i(C^*_{01} C_{00} + C^*_{10} C_{11}) e^{2iJt} - i(C^*_{00} C_{01} + C^*_{11} C_{10}) e^{-2iJt}.
\end{equation}
Corresponding non-zero ${\mathds T}$ components:
\[
T_{0010} =  C^*_{01} C_{00} e^{2iJt},\quad T_{1011} =  C^*_{10} C_{11} e^{2iJt},\quad
T_{0001} = - C^*_{00} C_{01} e^{-2iJt},\quad T_{1110} = - C^*_{11} C_{10} e^{-2iJt} \, .
\]

\paragraph{Case 4:}
\begin{equation}\label{sxsy}
\langle S_x^1(0) S_y^2(t) \rangle=\bra{\psi}\, S_x^1(0)\, S_y^2(t)\, \ket{\psi} = (-i C^*_{00} C_{11} + i C^*_{11} C_{00}) e^{-2iJt} + (i C^*_{01} C_{10} -i C^*_{10} C_{01}) e^{2iJt}.
\end{equation}
The corresponding non-zero entries of the ${\mathds T}$ operator take the form:
\[
T_{0101} = - C^*_{00} C_{11} e^{-2iJt},\quad T_{1010} =  C^*_{11} C_{00} e^{-2iJt},\quad
T_{0110} =  C^*_{01} C_{10} e^{2iJt},\quad T_{1001} = - C^*_{10} C_{01} e^{2iJt} \, .
\]

{\footnotesize{
\bibliographystyle{JHEP} 
\bibliography{r}
}}


\end{document}